\documentclass[twocolumn,rmp,showpacs,preprintnumbers,amsmath,amssymb]{revtex4}
\usepackage{graphicx}% Include figure files
\usepackage{dcolumn}% Align table columns on decimal point
\usepackage{bm}% bold math
\usepackage{longtable}% bold math

\begin{document}

%\nofiles

\preprint{APS/123-QED}

\title{ Atomic Data for X-ray Astrophysics}

\author{T.R. Kallman}

\affiliation{NASA Goddard Space Flight Center}

\author{P. Palmeri}

\affiliation{ Astrophysique et Spectroscopie, Universit\'e de Mons--Hainaut, Belgium}

\date{\today}% 

\begin{abstract}
We review the available atomic data used for interpreting and modeling X-ray observations.
The applications for these data can be divided into several levels of detail, ranging from 
compilations which can be used with direct inspection of raw data, such as line 
finding lists, to synthetic spectra which attempt to fit to an entire observed dataset
simultaneously.  This review covers cosmic sources driven by both electron ionization
and photoionization and touches briefly on planetary surfaces and atmospheres.
We review all of this, the applications to 
X-ray astronomy, the available data, recommendations for astronomical users, and attempt to 
point out the applications where the shortcomings are greatest.
\end{abstract}

\pacs{Valid PACS appear here}% PACS, the Physics and Astronomy
\maketitle

\tableofcontents

\section{\label{intro}Introduction}

Essentially every observation in astronomy which involves spectra requires some aspect of atomic data 
for its interpretation. This can range from stellar colors as probes of age or abundance to 
high resolution spectra of the Sun and the   synthetic models which are used in their 
interpretation. The importance of atomic data to the field of X-ray 
astronomy comes from the fact that there are many atomic features in the X-ray band and  
the physical conditions in the sources are usually far from thermodynamic equilibrium, so that 
interpretation of line fluxes or ratios requires knowledge of collisional rate coefficients and radiative rates.  
Even the instruments with the greatest spectral resolving power are not capable of fully 
resolving many line blends, so that modeling is of particular importance. Furthermore,
the field has evolved rapidly, and many of the needs for new or accurate atomic data 
have only recently become apparent.  In this paper we point out the motivation for 
accumulation of atomic data, and review available atomic data used 
for interpreting and modeling X-ray observations.

A review of atomic data for an application such as astronomy spans an  audience
including specialists with tools capable of providing new data, astronomers seeking to understand 
the accuracy and range of available data, and those who can use the data to synthesize spectra
or other observables for comparison with observations.  
The subject differs from some other review topics since it is unified more by the 
application to a common problem or challenge than by a single physical 
process.  In this paper we try  to balance the needs of these communities as best we can.  

For the physicist reader, we attempt 
to describe the capabilities of current X-ray astronomy instruments, some of the astrophysical science 
challenges,  and the scope of existing work on the measurement and calculation of relevant atomic 
constants.  It appears that much of the recent calculation and experimental work
which is most useful to X-ray astrophysics 
has been done when there is a clear understanding of the  astrophysical application.  We also 
attempt to draw attention  to those who have made important or extensive contributions in the past, 
even in cases where  their work has been superseded.  In doing this, it is not feasible 
to be comprehensive. A great deal of work has been done in this area, and 
we refer the reader to the many extensive bibliographies  which provide a more 
complete set of references.  For some physical processes we do provide relatively long lists of
references to work which spans the many ions or atomic species of interest.  These are cases where
we think such lists are both relevant and useful, since they convey the volume or comprehensiveness 
of past work and also are of current interest.  

For the astronomer reader, we attempt to describe the technical challenges of calculating 
or measuring relevant quantities.  The goal is to provide an appreciation of the  uncertainties 
imposed on astrophysics by an incomplete knowledge of atomic quantities.  In addition, in many 
cases, the data cited or described here can be directly applied to observations.

For the reader interested in the synthesis of atomic data into model spectra or diagnostics which 
can be directly fit to astronomical observations, we attempt to provide pointers to data which 
is appropriate for a given application, the limitations in accuracy and comprehensiveness 
of existing data, and the prospects for future measurements or calculations.  We also hope 
to foster cooperation between modelers or astronomers and the atomic physics community.  
This has proven to be an effective way to stimulate work on problems which are relevant to 
astrophysics, while also educating astronomers and modelers about the limitations and applicability 
of atomic structure  and cross section data.

We feel that the time is right for a review such as this, because the 
accuracy and quantity of atomic constants describing many processes has progressed significantly 
in the last 10-15 years.   This is driven 
both by advances in observational data with the launch of the $Chandra$ and $XMM-Newton$ 
X-ray astronomy satellites, both statistical quality and 
spectral resolution, and by advances in the tools available to the physics community.
There now exist values for rate coefficients and cross sections which will be accurate enough to allow 
interpretation of much of the available observational data for the near future, and current 
techniques will continue to improve the situation. 
In principle, the project of calculating or measuring atomic cross sections for 
any specific application has finite scope, and can at some point be considered complete.
Although this is not imminent, we can estimate the quantity and accuracy of atomic 
data that will be needed for most applications to X-ray astronomy in the near future.

In addition, the advent of on-line databases has revolutionized the dissemination of atomic data.  
There is no longer a need for publication of rate coefficients or cross sections in obscure 
laboratory internal publications or conference proceedings, and results based on 
proprietary or fee-based databases or computational tools are receiving limited attention.
The use of the web, faster computers and essentially 
unlimited storage have made it practical to save and disseminate great quantities of 
data.   So there is no technical limitation to the sharing of data.  
In this review we therefore intentionally omit reference to unpublished data,
data which is part of a fee-based or proprietary database or code package, or which is published 
in obscure laboratory internal reports or conference proceedings which are not widely available.
We feel strongly that scientific information of all types, to be truly useful, 
must be freely and easily available, and its provenance must be transparent to all.  

Atomic data can be used in different ways according to the level of detail  needed for a
specific application. The levels of detail range from compilations used in direct inspection of
raw data, such as line finding lists, to synthetic spectra which attempt to fit to an entire
observed dataset simultaneously. The cosmic sources also differ, ranging from 
electron-ionized and photoionized plasmas to planetary surfaces and atmospheres.
Although traditional X-ray astronomy is generally agreed to span the energy range 0.1 - 10~keV,
modeling of X-ray data inevitably requires knowledge of cross sections 
which extend to the ionization potential of hydrogen.  We adopt this broader definition,
and include some discussion of cross sections and rate coefficients associated with photons emitted in 
the EUV and UV bands.  X-ray spectra thus far are only capable of detecting 
the 15 or so most cosmically abundant elements, including C, N, O, Ne, Na, Mg, Si, Al, S, Ar, Ca, 
Fe, and Ni, and we generally restrict our consideration to calculations and experiments relevant to 
these elements.  

Much of the information relevant to this topic has already been reviewed by others.  
Recent  advances in astrophysical X-ray spectroscopy  have been reviewed  by \citet{Kahn04}, 
\citet{Paer03} and some  experimental results in atomic data by \citet{Beie03b}.  The subject 
of atomic data for X-ray astronomy has been discussed in specialized 
conference proceedings, including  \citet{Silv93}, \citet{Baut00c}, \citet{Ferl01} and \citet{Smit05}.
More specialized reviews pertaining to many of the 
subtopics discussed here will be mentioned in what follows.

This review is organized as follows:  We begin with a discussion of some examples of the astronomical 
X-ray spectra, choosing from recent observations using $Chandra$ and 
$XMM-Newton$, followed by a review of the techniques used to derive atomic data, both theoretical and experimental.
We then discuss the atomic data currently available, classified according to the 
level of modeling needed in order to apply the data.  These range from 
line finding lists which can be applied by inspection, to diagnostics of physical quantities 
involving a small number of measurable parameters such as line ratios or equivalent widths,
to models which attempt to synthesize a large portion of the spectrum.  These we divide into the likely 
excitation mechanism:  coronal, for sources where the primary
energy input is mechanical; photoionized, where photon excitation, heating and ionization plays a dominant
role; and planetary, associated with neutral, molecular or solid material.
Then we further subdivide according to physical process.  

We divide the discussion of each physical process or subtopic into sections: In the first, denoted `background', 
we provide an overview of the work on the topic, dating from the early suggestions 
of the importance of the process and simple approaches to estimating rate coefficients and 
cross sections.   In the second section, denoted `Recent Developments',
we change tactics slightly, and provide an overview of the work which we consider to be
the current state of the art, or which point the way to future developments 
or improvements in the state of the art.  Inherent in this is the 
omission of some work in favor of that which we consider to be either of sufficient 
accuracy and comprehensiveness to be adequate for the needs of X-ray astronomy today, or which we 
think points the way toward such data. 

The study of X-ray spectra associated
with planets, comets, and other primarily neutral or solid objects is newer than the study of 
other emission mechanisms.  As a result, there is less work so far on interpretive tools for such spectra, and 
we do not give a separate discussion of line finding lists or discrete diagnostics for them.   
We also include a glossary in order to help keep track of the many acronyms and terms describing
computational techniques, physical effects, and approximations.
Throughout this paper we adopt chemical notation for ionic charge states, i.e. with a superscript
denoting the ionic charge.  This is in contrast to the spectroscopic notation used in 
much of the astrophysics literature, in which the charge state is denoted with a roman 
numeral: I=neutral, II=charge +1, etc. 
In the final section of the paper we discuss remaining needs for atomic data and prospects for future work. 

\section{\label{need}On the need for atomic data}

The recent impetus for atomic data applications to astronomy comes from developments in instrumentation.
Evolution of instruments has been rapid in the field of space astronomy, which has arisen
and grown to a level of sensitivity and spatial and spectral resolution which is comparable to that 
of ground-based optical instrumentation in approximately the last 40 years.  The most recent 
developments are summarized in the review by \citet{Paer03} for the X-ray band.  
Much of the current work is motivated by the launch of the grating spectrographs
on the $Chandra$ and $XMM-Newton$ satellites in the year 1999.  These are described 
in detail in \citet{Brin00} for the Low Energy Transmission Grating (LETG), in \citet{Cani05} for the 
High Energy Transmission Grating  (HETG) on the $Chandra$ satellite,
and by \citet{Rasm01} for the reflection grating spectrograph (RGS) on $XMM-Newton$.  
The gratings provide spectra with resolving power $\varepsilon/\Delta\varepsilon \sim$300 -- 1000 in the 
energy range between 0.2~keV and 8~keV, with effective areas $\sim$ 10 -- 100 cm$^2$.  In 
terms of wavelength, the resolution is $\Delta\lambda=0.01 - 0.05 $\AA\ depending on the 
instrument, and the wavelength range extends
from $\simeq$1.8 \AA\ for the $Chandra$ HETG to 170 \AA\ for the $Chandra$ LETGS.
This is in contrast with the previous generation of instruments, exemplified by the $ASCA$ satellite \citep{Tana86}, 
which carried moderate resolution CCD (Charge Coupled Device) instruments and had a spectral resolving power 
$\sim$20.   The grating instruments have discovered 
line features in spectra previously thought to be featureless, and require a significantly more
comprehensive and accurate database of atomic constants than had previously been in use.
However, both $Chandra$ and $XMM-Newton$ also carry CCDs whose use
is necessitated by cosmic sources which are too faint for adequate signal in the gratings, or
which are spatially extended and hence cannot make use of the full grating spectral resolution.

The $Chandra$ and $XMM-Newton$ gratings represent a tremendous advance in resolution,
but are not capable of unambiguously resolving thermal or natural line shapes from many 
sources.  The nominal resolution ($\sigma$) of the HETG at the energy of the O$^{7+}$ Ly$\alpha$ line
corresponds to a Doppler velocity of 170 km s$^{-1}$ \citep{Cani05}.   In a thermal plasma this would be 
obtained at a temperature of 10$^{7.5}$ K, which is significantly greater than the 
temperature where this ion is predicted to be abundant under most plausible conditions.
The natural width of the line corresponds to 0.008 eV or an equivalent Doppler velocity of 
3.7 km s$^{-1}$.  Natural widths increase rapidly with the nuclear charge, and so for iron
the corresponding velocity is 550 km s$^{-1}$.   A typical dielectronic recombination (DR)
satellite line to a hydrogenic resonance line (these terms are described in detail later 
in this review) is shifted by $\Delta\varepsilon/\varepsilon\simeq 0.002$
\citep{Boik77}.  This is at the nominal resolution of the $Chandra$ HETG.  
With good statistics, parameters describing line centroids or widths can be determined 
to precision better than the nominal resolution by factors of several.  Nonetheless, it is clear that 
the $Chandra$ spectra are not capable of measuring thermal line shapes unambiguously, 
and the wavelength precision is not comparable to that obtainable in the laboratory or in solar observations with 
Bragg crystal spectrometers, for example.  On the other hand, the HETG can resolve the important
diagnostic lines discussed in this review, and it can detect the effects of supersonic turbulence
or bulk motions at Mach numbers of $\sim$ 2 -- 3 or more.  This is of great astrophysical interest, 
since many X-ray plasmas are located in regions of strong gravity, or near shocks or other energetic 
sources.

\begin{figure*}
\includegraphics*[angle=0, scale=0.7]{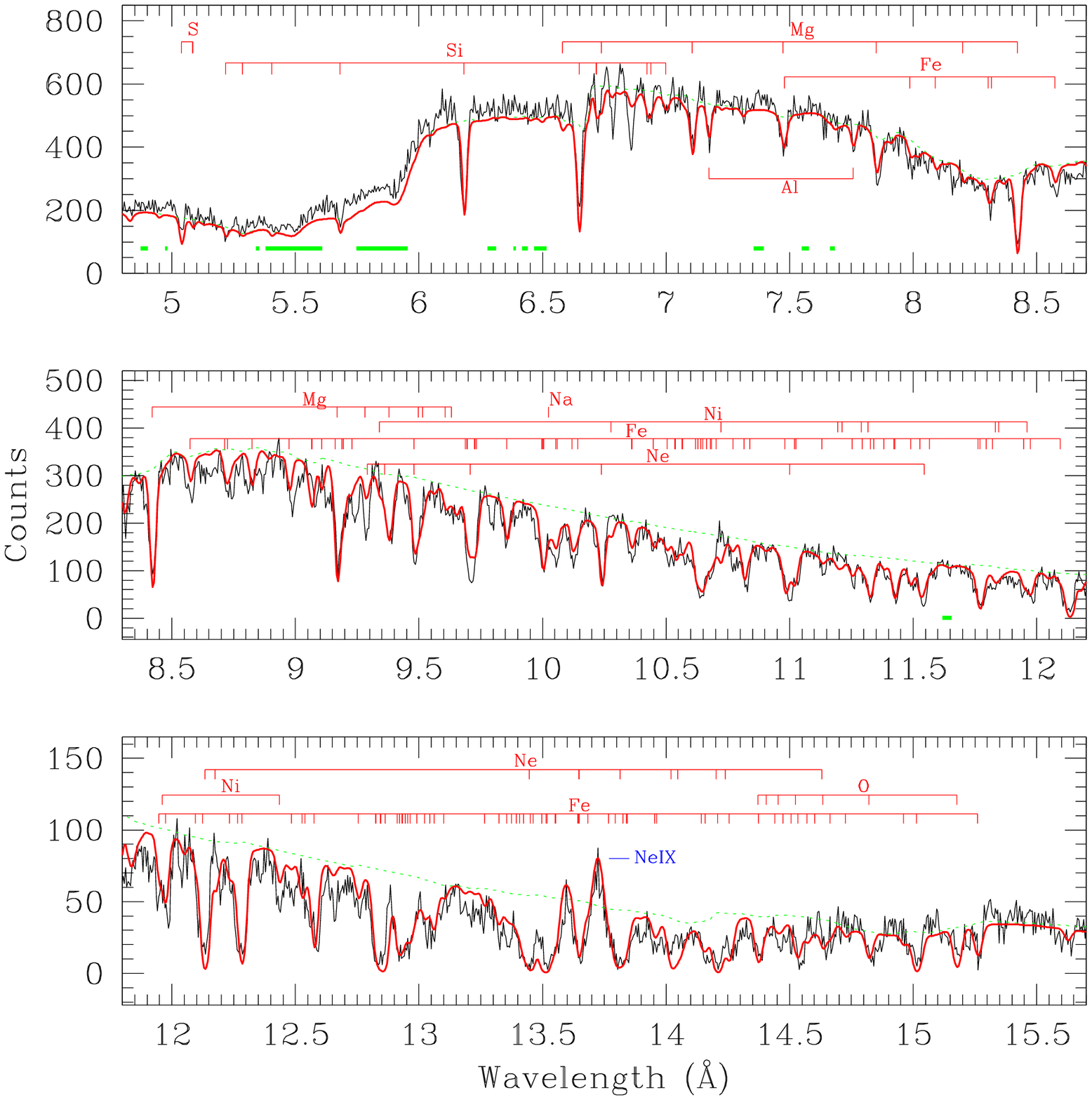}
\caption{\label{3783fig} $Chandra$ HETG spectrum of NGC 3783, from \citet{Kron03}}
\end{figure*}

As an illustration of the capabilities and limitations of the data obtained with $Chandra$ and $XMM-Newton$ 
we present spectra obtained using these instruments which test the available atomic data and 
synthetic spectra.   These are not chosen because they are typical, but rather because they 
best show the capabilities of the instruments and the available atomic data. Figure \ref{3783fig} shows the 
spectrum of the Seyfert galaxy NGC 3783 obtained with the $Chandra$ HETG.  This object is 
an active galaxy, and the spectrum shows absorption from a large number of resonance lines from
the K shells of highly ionized ions of medium-Z elements, O - S, and also from the L shell of 
partially ionized iron.  The solid curve is a model fit, described by \citet{Kron03}, showing 
general consistency but not perfect agreement for the majority of the strong lines.  In this object the 
gas is likely to be photoionized, and all the lines are shifted to the blue from their rest wavelength by an
amount corresponding to a velocity of $\sim$500 -- 1000 km s$^{-1}$.  This is interpreted as being a
Doppler shift in an outflowing wind, and similar spectra have been observed from most of the other 
objects of this class.  Discovery of such outflowing gas is surprising, since active galaxies 
are strong sources of continuum X-rays, and are thought to contain black holes which radiate 
as gas is pulled inward.  This spectrum also illustrates the fact that the data from the gratings on  $Chandra$ and 
$XMM-Newton$ can be used to constrain observed line 
wavelengths to within $\sim 0.25\%$.  This exceeds the precision of most {\em ab initio} calculations 
and therefore requires laboratory measurements to provide wavelengths precise enough to allow 
reliable line identifications and to infer line shifts.

\begin{figure*}
\includegraphics*[angle=270, scale=0.5]{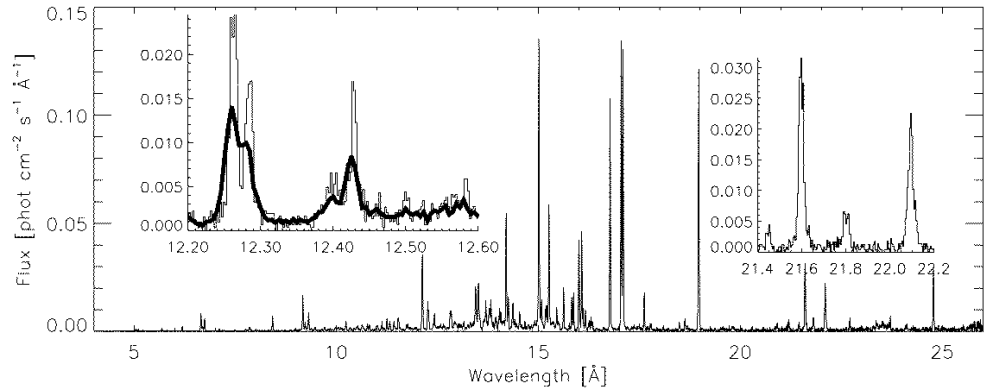}% Here is how to import EPS art
\caption{\label{capella} HETG spectrum of Capella, from \citet{Cani00}.  Insets show the 
regions of the spectrum in the vicinity of the lines of He-like Ne$^{8+}$ and O$^{6+}$.
In the left inset panel the two curves are the data from the two arms of the 
dispersed image.  The thin line is the data  from the higher energy arm (HEG) 
and the thick line is the data from the medium energy arm (MEG).}
\end{figure*}

Figure \ref{capella} shows the spectrum of the nearby active star Capella obtained with the 
$Chandra$ HETG \citep{Cani00}.  This shows rich line emission, similar to the Sun, and which is 
typical for a coronal source.  
For the most part the lines in the 10 -- 18 \AA\ range are due to
emission from L-shell iron ions. The observed iron L spectrum can be reproduced 
almost entirely by assuming a single electron temperature of $kT_e$ = 600 eV \citep{Beha01b}. 
This temperature is consistent with both the measured fractional ion abundances 
of iron and with the temperature derived from ratios of Fe$^{16+}$ lines.
However, there are some remaining discrepancies between single temperature models and the data for 
the lines of Fe$^{17+}$ near 16 \AA\ and also Fe$^{16+}$ and Fe$^{17+}$ at 
15.015 and 14.206 \AA\ which are all overestimated by current models.   The statistical 
quality of this spectrum is comparable to that obtained from the Sun,
illustrating the power of X-ray spectroscopy to study nearby objects outside
the solar system.
%Figure \ref{sun} shows the spectrum of the Sun in the same wavelength region, 
%obtained by the xx instrument \citep{Mcke80}.

%\begin{figure*}
%\includegraphics*[angle=0, scale=0.7]{Mckenziefig.ps}% Here is how to import EPS art
%\caption{\label{sun} spectrum of the Sun, from \citet{Mcke80}}
%\end{figure*}

\begin{figure*}
\includegraphics*[angle=270, scale=0.7]{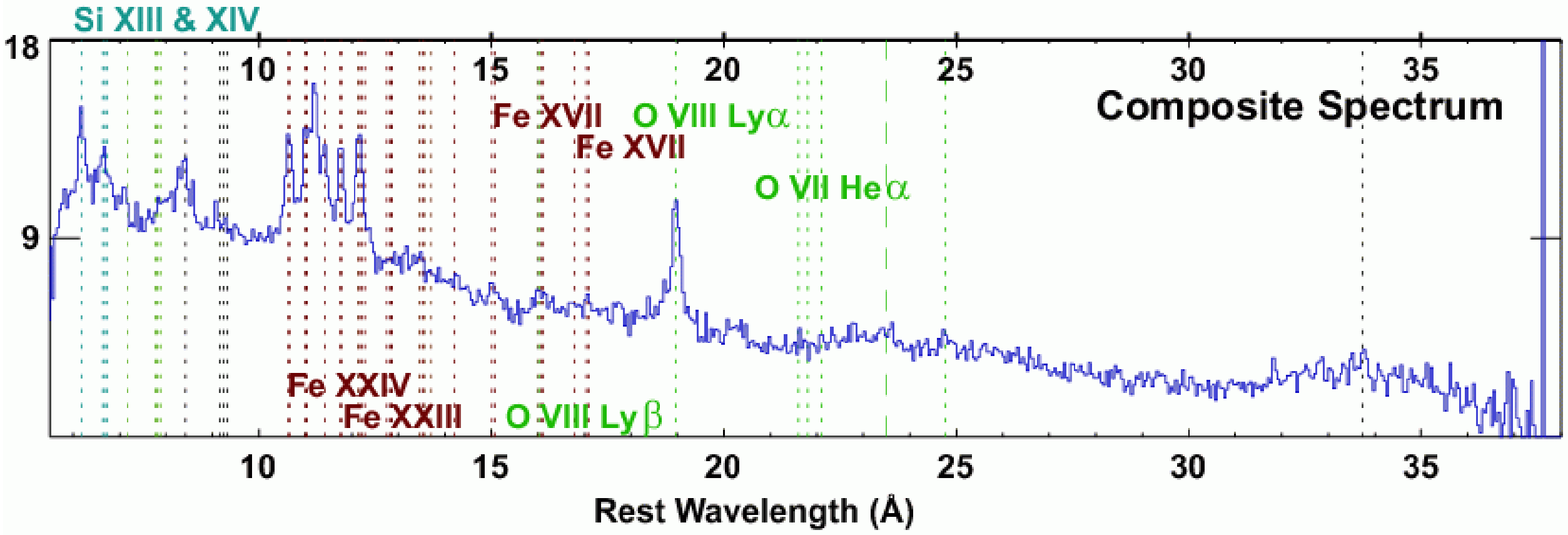}% Here is how to import EPS art
\caption{\label{cluster} Composite spectrum of galaxy clusters where cooling gas 
is expected, from \citet{Pete03}.  The solid curve shows data from the RGS on $XMM-Newton$
and dashed vertical lines show the predicted locations of various coronal emission lines.}
\end{figure*}

An illustration of the power of high resolution X-ray spectroscopy to study distant objects is 
shown in Figure \ref{cluster}, which shows the spectrum of a composite of galaxy clusters 
taken with the RGS on $XMM-Newton$ \citep{Pete03}.  The blue curve shows the data
coadded for 13 clusters, which clearly exhibit coronal line emission from Fe$^{22+}$, Fe$^{23+}$,
and O$^{7+}$ Ly$\alpha$.  This is consistent with emission from gas at an electron 
kinetic temperature of $kT_e \simeq 1.5$~keV.  
This temperature has also been derived from lower resolution observations of 
these and similar clusters.  The gas densities can also be derived from the 
X-ray luminosities and sizes.  Based on these quantities, the time for 
the gas to cool radiatively is predicted to be short compared with the probable 
cluster age, hence the lines from lower ionization species are expected as the 
gas cools to $\sim$ 1~keV and below.
The vertical dashed curves in the Figure denote the positions of the 
lines which are predicted from such cooling flow models, such as O$^{6+}$ and Fe$^{16+}$. 
Their absence requires a major reexamination of the prevalent model for cluster X-ray gas.
One possibility is that the gas, rather than cooling steadily over cosmological time,
is reheated by active galaxies in the cluster.  If so, it would provide an additional
example of the influence of black holes on their environments and on their likely 
influence on the growth of cosmic structures \citep{Rusz04}.

\begin{figure*}
\includegraphics*[angle=90, scale=0.7]{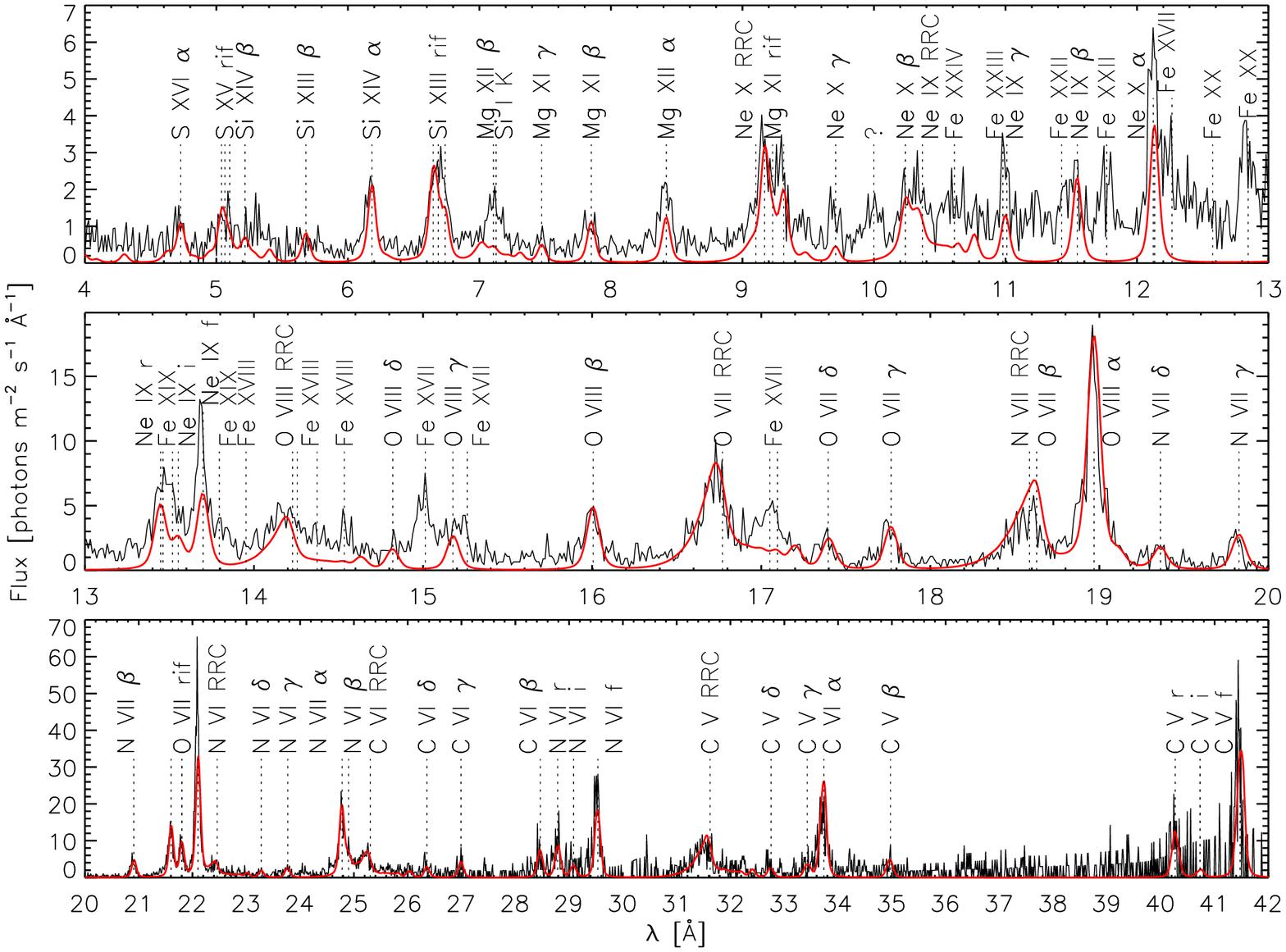}% Here is how to import EPS art
\caption{\label{ngc1068} Spectrum of the Seyfert 2 galaxy NGC 1068, taken 
with the LETG on the Chandra satellite, from \citet{Brin02}.  Black curve shows the data and red 
curve shows a fit to a photoexcited and recombining model \citep{Kink02}.}
\end{figure*}

Figure \ref{ngc1068},  shows the spectrum of a Seyfert galaxy NGC 1068 taken with the LETG on $Chandra$ 
\citep{Brin02}.  The extended long wavelength coverage is a unique feature of this instrument.  
The black curve shows the data taken of the nuclear region of the galaxy, 
and the red curve shows a model.  The model consists of emission due to gas which is emitting 
primarily as a result of photoionization, photoexcitation and scattering, as evidenced by the radiative 
recombination continuum (RRC) features due to O$^{6+}$ near 16.5 \AA\ and C$^{4+}$ near 31.5 \AA\, and their 
strengths relative to the resonance lines from these ions.  
This is interpreted as being due to the fact that the galaxy contains a strong source of 
EUV and X-ray continuum radiation, likely a black hole, but that this source is hidden from our direct view 
by cold obscuring gas.  This appears to be typical of type 2 Seyfert galaxies, which  are likely intrinsically 
similar to the type 1 Seyfert galaxy NGC 3783 (Figure \ref{3783fig}).  Both types 
contain a black hole partially surrounded by torus of nearly opaque cold gas, but their spectra reveal the 
fact that they are viewed at different orientations \citep{Anto85}.  Thus X-ray spectra 
such as these provide evidence for the presence 
of hidden black holes via their scattered radiation.

These examples illustrate the statistical quality of recent high resolution X-ray spectra,
and, crudely, the extent to which current synthetic spectra are able to match the 
observations.  They also provide a glimpse of the astrophysical issues which can be addressed 
using X-ray spectra and the potential importance of atomic data to address these issues.
They also illustrate the unique nature of the information provided by X-ray spectra; the scientific 
insights provided by these observations are not accessible to any other waveband, or to observations
using lower resolution X-ray instruments.  

\subsection{\label{sensitivity}Atomic Data Accuracy Requirements}

Spectra such as the ones shown so far in this section can be used as a practical guide 
for X-ray atomic data accuracy requirements.  
Simple estimates can be obtained by examination of the spectra.  
The strongest emission line (the  Ly$\alpha$ line of O$^{7+}$) in the spectrum of Capella, 
as shown in Figure \ref{capella} for example, 
has flux 0.02 photons cm$^{-2}$ s$^{-1}$, 
corresponding to a total of $\simeq$1300  counts in a 3$\times 10^4$ s observation.  
The dominant error for such an observation is counting statistics, and assuming 
Gaussian errors this corresponds to a statistical error on the line intensity of $\simeq$3$\%$.  This may be viewed 
as the implied accuracy requirement for 
a single atomic rate coefficient which affects the line flux linearly, if all other 
quantities which can affect the line flux are known or not of interest.   Such rate 
coefficients, when they describe collisional processes, are derived by averaging the energy dependent 
cross section for the process over the Maxwellian distribution of electron velocities.  So an accuracy 
requirement on a rate coefficient translates approximately linearly into the requirement on the 
energy dependent cross section if interpreted as applying uniformly over all energies.  On the other 
hand, a greater uncertainty in the cross section can be tolerated if it applies over a narrow energy range.
 An observed line flux can be
used to infer the total number of emitting ions if the excitation rate for the line is known.
The accuracy of such a determination will be limited by the atomic data if the uncertainty in the 
excitation rate exceeds the statistical error  in the observed spectrum.

\begin{figure*}
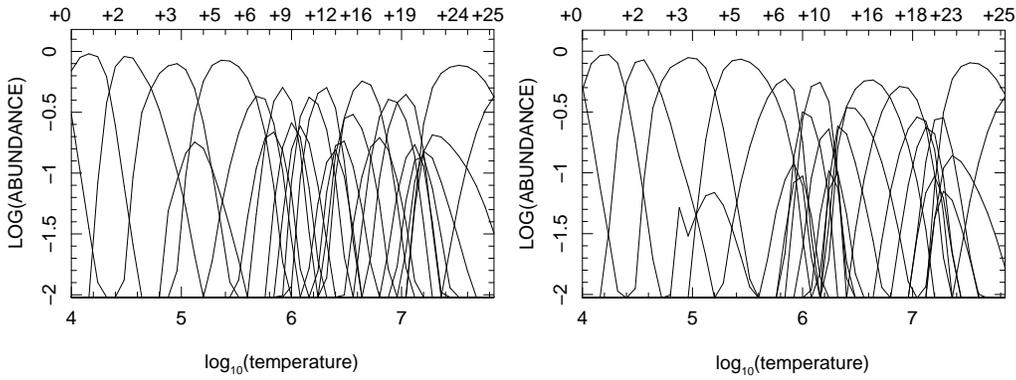

\includegraphics*[angle=270, scale=0.3]{f5a.ps}% Here is how to import EPS art
\includegraphics*[angle=270, scale=0.3]{f5b.ps}% Here is how to import EPS art
\caption{\label{ionbal} Left Panel: Ionization balance for iron in a coronal gas, calculated 
using rate coefficients described in the text. Right panel:  Ionization balance calculated 
using perturbed rate coefficients as described in the text. Curves correspond to various ions of 
iron:  in the left panel Fe$^+$ dominates at log(T)=4, Fe$^{+5}$ dominates at log(T)=5.2,
Fe$^{+8}$ dominates at log(T)=5.9, Fe$^{+12}$ dominates at log(T)=6.3,
Fe$^{+16}$ dominates at log(T)=6.6, Fe$^{+24}$ dominates at log(T)=7.5.
In the right panel, the temperature of maximum abundance of 
Fe$^{+8}$ -- Fe$^{+23}$ is displaced to lower temperature by $\simeq$0.1 in log(T).}
\end{figure*}

On the other hand, inferring physical quantities such as 
gas temperature, density or elemental abundance from a line or a spectrum in many cases requires 
comparison with models.  Model results depend on various atomic rate coefficients:
those affecting the ionization balance, those affecting excitation and those affecting photon emission.
The dependence of  observables such as line strengths on rate coefficients affecting ionization balance or excitation
can be non-linear, and are not well suited to  analytic estimates for atomic data accuracy requirements.
Numerical experiments can illustrate the sensitivity of line strengths to the rate coefficients affecting ionization
and excitation.  As an illustration, Figure \ref{ionbal} shows the ionization balance in coronal equilibrium 
for iron.  Ion fractions are plotted as a function of temperature T in the range 4$\leq$log(T)$\leq$8 
and the dominant ionization stage ranges from Fe$^+$ at low temperature to Fe$^{26+}$ at the highest temperature.
The left panel shows these quantities calculated using rate coefficients from \citet{Arna92} for iron.  We denote these 
as our baseline ionization balance (but note that the work by \citet{Brya06} represents a significant update).  
The right panel has been calculated using the baseline rate coefficients, except that the rate coefficients for 
recombination for each ion have been perturbed, i.e. multiplied by a randomly chosen factor ranging between 
0.79 and 1.28 (i.e. $\Delta\log[{\rm rate coefficient}]=\pm 0.1$).  This range is 
only slightly greater than the magnitude of the effect which is likely to be associated with  
plasma microfields, cf., Section \ref{disec} and \citet{Badn03}, and so reflects a realistic limit to the 
level of accuracy needed 
from zero-external-field rate coefficients for this process.   Within this range, the multiplicative factor affecting recombination 
is randomly chosen and evenly distributed.  The recombination 
rate coefficients are stored as analytic functions of temperature and, for each ion, the same multiplicative factor 
is applied at all temperatures.  Comparison of the two panels of Figure \ref{ionbal} 
indicates the sensitivity of the ionization
distribution to changes in the recombination rate of this magnitude:  at a given temperature 
ion fractions can differ by factors $\sim$2--4, although closed shell ions such as Ne-like 
and He-like ions are relatively unaffected.  The temperature where the maximum fraction occurs 
can also change by $\sim$0.1 -- 0.2 in log(T).  

\begin{figure}
\includegraphics*[angle=270, scale=0.33]{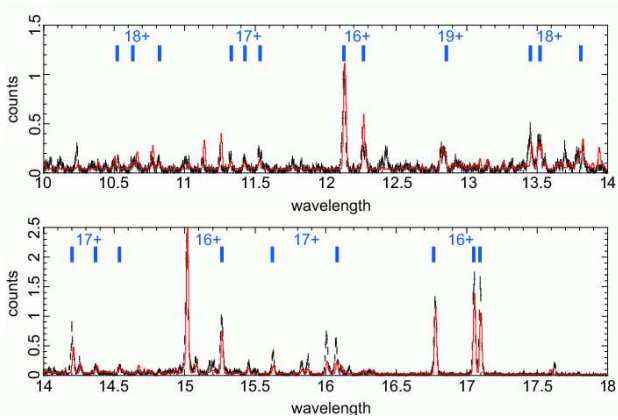}% Here is how to import EPS art
\caption{\label{capellafit} Fit to a 3$\times 10^4$ s 
$Chandra$ HETG observation of the active star Capella in the 10 -- 18 \AA\ region
using the baseline ionization balance shown in the left panel of figure \ref{ionbal}.}
\end{figure}

\begin{figure}
\includegraphics*[angle=270, scale=0.33]{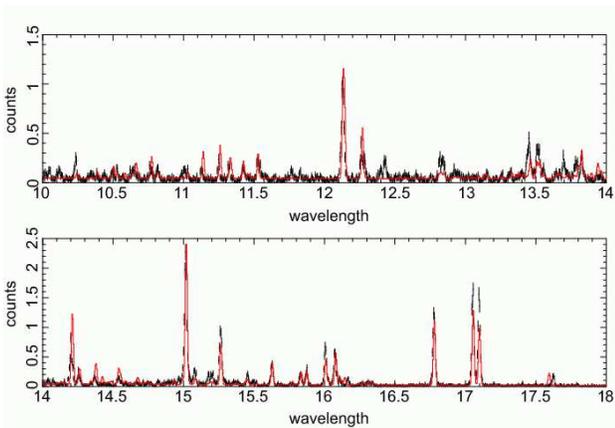}% Here is how to import EPS art
\caption{\label{capellafake} Same as figure \ref{capellafit} except using ionization balance 
shown in figure \ref{ionbal} derived from perturbed rate coefficients as described in the text and 
displayed in the right panel of figure \ref{ionbal}.}
\end{figure}

The effects of the ionization balance change on the synthetic spectrum, and on the inferred 
distribution of gas temperatures in a real astrophysical source, are illustrated in Figures 
\ref{capellafit} and \ref{capellafake}.  Figure \ref{capellafit} shows
 an observed spectrum from the active star Capella taken with the HETG grating on the $Chandra$ 
satellite, the same source shown in Figure \ref{capella}.  The observed spectrum is shown as the 
data points with error bars.  The solid (red) curve in Figure \ref{capellafit} is a synthetic 
spectrum calculated using the baseline ionization balance shown in the left panel of Figure \ref{ionbal}.  This spectrum is 
calculated using the ionization balance at two temperatures, log(T)=6.9 and log(T)=7.1.  The element 
abundances and the relative normalizations of the two components  were allowed to vary in order to 
achieve a fit which has $\chi^2$=3267 for the 1602 independent energy channels shown in this figure. 
This is not in a range of $\chi^2$ which permits application of  traditional statistical measures of 
probability of random occurrence \citep{Bevi69}, and is primarily of illustrative value.  Similar values of 
$\chi^2$ are found when these data are fit to two temperature component models 
calculated with other modeling packages which are 
in widespread use, such as {\sc apec} \citep{Smit01} and {\sc mekal} \citep{Mewe95b}.

In the region of the spectrum shown most of the features are due to iron, the exception being the 
Lyman lines of Ne$^{9+}$.  We find a best fit value of [Ne/Fe], the abundance ratio relative to solar
\citep{Grev96}, of 0.66.
The strongest lines in this spectrum are indicated in Figure \ref{capellafit}: Fe$^{16+}$ between 12 and 17 \AA, 
and the Ly$\alpha$ line of Ne$^{9+}$ at 12.13 \AA, which is blended with the corresponding Fe$^{16+}$ line.
Higher ionization lines include lines from Fe$^{18+}$ at 10.5 -- 10.8 \AA,  
Fe$^{17+}$ between 11 and 16 \AA, 
Fe$^{18+}$ at 13.5 -- 13.8 \AA\ and Fe$^{19+}$  at 12.85 \AA.  
Lower ionization lines include O$^{7+}$ at 16.00 \AA,  Fe$^{17+}$ blended with Ne$^{9+}$ L$\beta$ at 
10.24 \AA\  and Ne$^{8+}$  at 11.03 \AA\  .  The baseline model accounts for the strengths of many of the strongest lines,
but under predicts the Fe$^{16+}$  15 and 17 \AA\ lines; this is a manifestation 
of problems with rate coefficients affecting this ion which are in widespread use and which are discussed 
in section \ref{fe17sec}.  

 The emission 
line spectrum radiated by the plasma is calculated using rate coefficients for electron impact collisional excitation
culled from the references discussed later in this paper.  These are enumerated in \citet{Baut01}, with the 
addition of rate coefficients from the recent {\sc chianti} version 5 compilation \citep{Land05d}.  
It is well known that a source 
such as Capella is best fit using a broad distribution of temperatures, ranging from log(T)$\simeq$6 -- 7.4
\citep{Cani00}, but we adopt the two component model shown here for illustrative purposes.

The effects of perturbing the ionization balance on the fit to the spectrum can be seen in Figure \ref{capellafake},
in which the analogous procedure is carried out using the ionization balance shown in the right panel of 
Figure \ref{ionbal}.  In making this fit we have not attempted to reoptimize any of the parameters describing 
the fit, i.e. the normalization, temperatures or elemental abundances in the two temperature components.
Rather, we have left them the same as in Figure \ref{capellafit} in order to illustrate the 
sensitivity to the ionization balance.   Comparison with the 
fit shown in Figure \ref{capellafit} shows that the fits to the strongest lines, i.e. those of Fe$^{16+}$ and 
Ne$^{9+}$, are essentially identical.  This is a reflection of the fact that the ionization balance of the 
closed-shell Fe$^{16+}$ is little affected by the perturbation in recombination rate coefficients.  However, significant 
differences are apparent in the fit to the ions in adjacent ionization stages, notably Fe$^{17+}$ and Fe$^{19+}$.
These show up in the lines near 14.2-14.5 \AA\ for Fe$^{17+}$ and 12.8 \AA\ for Fe$^{19+}$, which are under predicted 
by the model shown in Figure \ref{capellafake}.  The lines of Fe$^{18+}$ are not strongly affected by 
the perturbation in the recombination rate coefficients.  The fit between data and model in Figure \ref{ionbal} 
has $\chi^2$=3610 for 1602 energy channels and is clearly worse than for the baseline model.   
However, when the perturbed model is iteratively fitted to the $Chandra$ data a better fit is found when the 
high temperature component increases to log(T)=7.2.  The resulting fit is improved, although still inferior to 
the baseline model:  $\chi^2$=3522 for 1602 energy channels as compared with 
 $\chi^2$=3267 for 1602 energy channels for the baseline model.

These experiments illustrate some of the effects of changes in the rate coefficients affecting ionization balance
on fitting to observed astrophysical X-ray spectra  and on the inferred temperature distribution.  They indicate
that changes of $\sim$25$\%$ in these rate coefficients can affect the model strengths of the lines of the more 
delicate ions by  factors $\sim$3 -- 10, while the strong lines from stable ions are more robust.  
Such changes in model line strengths lead to changes in the inferred physical conditions 
derived by iterative model fitting which are significant.  In this case we find changes of 0.1 in log(T) from 
such a procedure, which is at the resolution of the grid of models used to calculate the spectra.

In a coronal plasma there is an exact symmetry between recombination and electron impact 
collisional ionization, so that perturbations to recombination rate coefficients can be interpreted as corresponding
inverse perturbations to the ionization rate coefficients.  The inferences from these numerical experiments are 
reciprocal between the two types of rates.  At the same time, it is clear that the above model fits have
been applied to only one spectrum, and only one numerical experiment at perturbing the rate coefficients has been 
performed.  In order to adequately characterize the effects of perturbing various rate coefficients, this experiment 
should be repeated in order to truly randomize the perturbations, and experiments should 
be carried out with different amplitudes and with fits to other astrophysical X-ray spectra.
More extensive experimentation of this type has been carried out with application to solar 
EUV and X-ray line emission by \citet{Savi02c} and \citet{Gian00}.  \citet{Masa97} examined the effects 
of uncertainties in rates affecting ionization balance on the results of fitting 
to moderate resolution X-ray spectra obtained using CCD instruments.

\section{\label{theory}Theoretical Techniques}

Theoretical calculation of atomic processes begins with the solution to the multi-electronic Schr\"odinger or Dirac equation,
\begin{equation}
\left \{ \sum_i h_i + \sum_{i<j} V^{e-e}_{ij} \right \} \Psi = E \Psi.
\label{schrod}
\end{equation} 
where $h_i$ are the one-electron Hamiltonians and $V^{e-e}_{ij}$ the electron-electron interaction
potentials. In the fully relativistic case,  $h_i$ are Dirac Hamiltonians, $V^{e-e}_{ij}$ 
include the direct and exchange part of the electrostatic interaction
 (the interaction energy due to the positional correlation
of parallel-spin electrons according to the Pauli exclusion principle)
 and the Breit 
interaction (the two-body magnetic interaction), and $\Psi$ is a four-component wave function. 
In the non-relativistic case, $h_i$ are
Schr\"odinger Hamiltonians and $V^{e-e}_{ij}$ include only the electrostatic interaction. 
In the Breit--Pauli (BP) relativistic approximation,  the non-relativistic multi-electronic 
Hamiltonian is corrected by adding terms resulting from the reduction of the one-electron 
Dirac Hamiltonian and the Breit interaction to the Pauli form \citep{Betsal72}. 
The relativistic correction terms are then the mass-correction, Darwin, and spin-orbit 
terms that are added to $h_i$, and the spin-other-orbit, spin-spin, orbit-orbit, 
spin-contact, and two-body Darwin terms that are added to $V^{e-e}_{ij}$.   
Higher-order relativistic interactions like QED have also to be treated in the case of highly-ionized heavy atoms.
The angular part of the solution of the Schr\"odinger equation is known exactly and is found using
the Racah algebra \citep{Raca42, Raca43, Fano59, Edmo60}. With regards to the radial part,
the exact solution cannot be found due to the $V^{e-e}_{ij}$ terms. One of the simplest approximation
is the independent-particle or central-field model where 
\begin{equation}
\sum_{j \ne i} V^{e-e}_{ij} \sim V_i(r)
\end{equation}
where $V_i(r)$ are local central potentials in which each electron moves independently.
The atomic state function $\Psi$ is then a product of the atomic orbitals $\phi_i$ which are 
eigenstates of the mono-electronic Hamiltonians,
\begin{equation}
\left \{ h_i + V_i(r)  \right \} \phi_i = \epsilon_i \phi_i,~i=1, ... , N
\label{indpar}
\end{equation}
where $N$ is the number of electrons.
Different forms of the functions $V_i(r)$ can be considered \citep{Cowa81}. Forms that depend
upon the orbitals themselves lead to the use of the self-consistent-field (SCF) iterative procedure \citep{Cowa81}.

A more elaborate solution of Eq.~(\ref{schrod}) comes from the application of the variational principle.
Using the trial atomic state function defined by 
\begin{equation}
\Psi = (N!)^{-1/2} det~\phi_i (x_j)
\label{sladet}
\end{equation}
where $x_j$ are the electron space and spin coordinates
and the determinant insures the Pauli exclusion principle, the $\phi_i$ are obtained by
requiring that the expectation value of the multi-electronic Hamiltonian is minimum
\begin{equation}
\langle \delta \Psi | \sum_i h_i + \sum_{i<j} V^{e-e}_{ij} | \Psi \rangle= 0
\label{varprin}
\end{equation}
where the variation is taken with respect to the radial parts of the orbitals $\phi_i$.
This leads to the Hartree--Fock equations \citep{Hart57}
\begin{equation*}
h_i \phi_i (x) + \sum_j \int dr' \phi_j (x') {\frac{1}{|r-r'|} } \phi_j (x') \phi_i (x)
= \epsilon_i \phi_i (x) 
\end{equation*}
\begin{equation}
+ \sum_j \int dr' \phi_j (x') \frac{1}{|r-r'|}  \phi_i (x') \phi_j (x),~i=1, ... , N. 
\label{hf}
\end{equation}
These non-homogeneous coupled integro-differential equations are solved iteratively in a SCF procedure.
An important property of the Hartree--Fock approximation follows from the Brillouin's
theorem which implies that the diagonal matrix elements of single-particle operators
are given correctly to first order by Hartree--Fock atomic state functions. This approach
essentially takes into account the part of the electron correlations within an electronic 
configuration. 

Further improvement can be made using trial atomic state functions $\Psi$ that
are expansions of configuration state functions $\Phi$,
\begin{equation}
\Psi = \sum_k c_k \Phi_k
\label{ci}
\end{equation}
where $\Phi_k$ are themselves expansions of Slater determinants (Eq.~(\ref{sladet})) and $c_k$
are the mixing coefficients. The summation in Eq.~(\ref{ci}) can be systematically extended in principle 
to yield results of arbitrarily high accuracy or to achieve convergence. This approach 
is referred to as the configuration interaction (CI) approximation. When the variational principle (Eq.~(\ref{varprin})) 
is applied to Eq.~(\ref{ci}) varying the $c_k$ and the orbitals $\phi_i$, one talks 
about the multi-configuration approach such as  in the multi-configuration Hartree-Fock (MCHF) \citep{Fisc96} 
and multi-configuration Dirac-Fock (MCDF) \citep{Parp96} 
methods.  Otherwise one talks rather about the superposition of configurations approach 
where Eq.~(\ref{indpar}) or Eq.~(\ref{hf}) is used to determine the orbitals $\phi_i$ 
and the multi-electronic Hamiltonian matrix is diagonalized afterward to obtain 
the $c_k$ and the eigenvalue $E$ as, for example, in Cowan's code \citep{Cowa81}. 
In practice, the CI expansion in Eq.~(\ref{ci}) is limited
and in order to include more correlation in $\Psi$, the multi-electronic Hamiltonian 
matrix elements are corrected to reproduce the few available experimental energy levels. The approximation becomes then
a semi-empirical one.
 
A different approximation which is also capable of systematic improvement of the atomic state function
is based on the many-body perturbation theory (MBPT) of \citet{Brue55} and \citet{Gold57}. 
In this method, the multi-electronic Hamiltonian is written as a sum of a zero-order Hamiltonian and a perturbation term
\begin{equation}
H = H_0 + H_{pert}
\end{equation}
where
\begin{equation}
H_0 = \sum_i \left \{  h_i + V_i(r) \right \}
\end{equation}
and
\begin{equation}
H_{pert} = \sum_{i<j} V^{e-e}_{ij} - \sum_i V_i(r).
\end{equation}
A complete set of orbitals $\phi_i$ are then obtained by solving Eq.~(\ref{indpar}) and used in the
order-by-order Rayleigh--Schr\"odinger perturbation expansion of the atomic state 
energies $E$ and other properties such as  oscillator strengths. An example of
the reduction of the perturbation expansion to MBPT formulas using Feynman diagrams and 
second-quantization method is given in \citet{Avgo92} for closed-shell atoms.

For processes involving the continuum, the continuum atomic state function $\Psi(\varepsilon)$ for the
$(N+1)$-electron system has to be evaluated, 
\begin{equation}
\Psi(\varepsilon) = \sum_{\alpha} \chi_{\alpha} \phi_{\alpha}(\varepsilon)
\label{conpsi}
\end{equation}
where $\varepsilon$ is the energy of the free electron, $\phi_{\alpha}$ are free-electron orbitals
and $\chi_{\alpha}$ are atomic state function of the $N$-electron target Hamiltonian.
The free-electron orbitals are solutions of the following Schr\"odinger equations
\begin{equation}
\left \{ h^{K}_{\alpha} + \varepsilon \right \} \phi_{\alpha} = \sum_{\alpha'} U_{\alpha \alpha'} \phi_{\alpha'}
\label{conh}
\end{equation}
where $h^{K}_{\alpha}$ are one-electron kinetic Hamiltonians and the matrix potential 
$U_{\alpha \alpha'}$ is defined by
\begin{equation}
\begin{array}{cc}
U_{\alpha \alpha'} =  \int dx_1 ... dx_N ~\chi_{\alpha}(x_1 ... x_N) \\ 
 \times~ U(x_1 ... x_N x_{N+1} ) ~\chi_{\alpha'}(x_1 ... x_N)
 \end{array}
\end{equation}
where $U$ is the sum of the nuclear and electron-electron interaction potentials 
acting on all the $(N+1)$ electrons of the target-plus-free-electron system.
 
 The widely used distorted wave (DW) approximation consists in neglecting the coupling between
 the different channels or the interaction between continuum states, i.e.
\begin{equation}
U_{\alpha \alpha'} = 0, \alpha \ne \alpha', 
\end{equation}
by using the central-field model defined in Eq.~(\ref{indpar}). It becomes a good approximation
for both the non-resonant background and resonance contribution.  However, many DW excitation rate 
coefficients omit the resonance contribution entirely.

At sufficiently large ionization or free-electron energies, DW orbitals approximate closely 
to Coulomb orbitals calculated using
\begin{equation}
U_{\alpha \alpha} \sim -{\frac{(Z-N)}{r}}
\end{equation}
where $Z$ is the atomic number.
 It is then called the Coulomb--Born (CB) approximation. The exchange part of the electrostatic 
interaction is neglected in CB approximation but included in the Coulomb--Born--Oppenheimer (CBO) 
and Coulomb--Born--Exchange (CBE) approximations.

The simplest approximation consists in considering plane waves to describe the continuum orbitals.
It is equivalent to switching off the matrix potential, i.e. $U_{\alpha \alpha} \sim 0$.  
This technique called the Born approximation is most likely to be accurate 
for energies far above threshold, and for an electron colliding a neutral atom.
It is a good approximation for the background cross section over a wide range
of energies for positive ions, because the infinite energy limit for scattering of a positive
ion is exactly the Born cross section and this limit controls the finite energy cross section
over a wide range of energies due to the slowly varying behavior of the cross section
with energy.

The three above-mentioned approximations make assumptions about the 
wave function of the continuum electrons which are only appropriate  
when the energy of the projectile electron (or free 
electron) is much greater than the binding energy of the target 
electrons (or bound electrons) of interest.   In contrast, approximations such as the
more elaborate and widely used $R$-matrix method \citep{Burber93} can be invoked 
for  free electrons at low energies and for low ionization charges up to negative ions where 
the coupling between channels is generally strong.  

Moreover, the importance  of resonant enhancement of rate coefficients for collisional 
excitation (see e.g. Section~\ref{discretediagnostics}.A.1)  means that the R-matrix method is widely used for 
highly-charged ions as well, as it is a very efficient way of determining the resonance 
contribution. Distorted-wave cross sections are traditionally non-resonant. There
are two ways in which resonances can be taken account of subsequently. Firstly,
(as exemplified by HULLAC) autoionizing states can be included in the collisional-radiative 
equations when calculating level populations and effective recombination and ionization 
rate coefficients. This can become very demanding computationally but it does allow 
for collisional redistribution of autoionizing states in dense plasmas. Secondly, 
the low-to-medium density approach, appropriate to the solar atmosphere, magnetic fusion 
plasmas, is to include the resonances in the excitation calculation, either via  
R-matrix (always present), or perturbatively (as has been done for some ions using FAC)
and so omit autoionizing states from the collisional-radiative population rate 
equations (otherwise they would be double-counted). This hugely simplifies 
the collisional-radiative problem but means there is a density range below that
for collisional equilibrium, where partial collisional redistribution of autoionizing
states takes place, and where non-autoionizing level populations are not well described
\footnote{The authors thank Dr. N. Badnell for pointing this out and providing the text for this 
paragraph.}.

The R-matrix  method consists of dividing the space 
into an internal and external regions. In the internal region where the interaction 
between the free electron and the target electrons is strong, the $(N+1)$-system 
atomic state function $\Psi(\varepsilon)$ is expanded in terms of target eigenfunctions $\chi$,
\begin{equation}
\Psi(\varepsilon) = {\cal A} \sum_{\alpha} \bar{\chi}_{\alpha} \frac{u_{\varepsilon,\alpha}(r)}{r} + 
\sum_i c_i \Phi_i
\label{closec}
\end{equation}
where ${\cal A}$ is the anti symmetrization operator insuring the Pauli exclusion principle, 
the bar sign over $\chi$ indicates the coupling with the angular and spin parts of the free-electron
orbital $\phi(\varepsilon)$, $u_{\varepsilon}(r)/r$ represents the radial part of the 
latter, and $\Phi_i$ are bound states of the $(N+1)$-system that are constructed with 
target orbitals to ensure completeness of
$\Psi(\varepsilon)$ and improve short-range correlations. The functions $u_{\varepsilon}(r)$ and 
the coefficients $c_i$ are obtained applying the Kohn variational principle to Eq.~(\ref{closec}) 
giving rise to a set of
coupled integro-differential equations.

These are solved by determining the $R$-matrix which is a real, symmetric matrix
whose rows and columns are labeled according to the possible channels of the
scattering problem.
At this point, one should notice by looking at Eq.~(\ref{closec}) that resonances (quasi-bound states)
are automatically included either in terms of closed channels in the first expansion or as linear
combinations of bound states in the second expansion.
In the external region,
$\Psi(\varepsilon)$ is represented by Eq.~(\ref{conpsi}) and Eq.~(\ref{conh}) is solved with the
following long-range matrix potential,
\begin{equation}
U_{\alpha \alpha'} (r) = \sum_{\lambda} C^{(\lambda)}_{\alpha \alpha'} / r^{\lambda + 1}.
\end{equation}

The MBPT method described above is also a powerful technique to compute continuum processes.
Here, the perturbation expansion is reduced using Feynman diagrams involving,
for instance, the photoelectron of the photoionization process.

We now describe some specific codes for computations of atomic structure and scattering.
Codes employing similar assumptions and computational algorithms are grouped together.

-{\sc hullac}:

The Hebrew University Lawrence Livermore Atomic Code \citep{Bars01} is a full package 
for structure and scattering calculations. It solves the multi-electronic Dirac equations 
and includes QED corrections. An analytical parameterized central-field potential 
is used to obtain the analytical Slater-type spin-orbitals. The atomic states $\Psi$ 
are constructed using CI expansions. The parameters appearing in both the central-field 
potential and the spin-orbitals are varied in a MC approach in order to minimize 
the energies of any set of levels or, in its semi-empirical mode, to minimize the 
rms deviation between a set of experimental energy levels and their corresponding 
theoretical values. The continuum is treated in a DW approximation. Using interpolation 
techniques for the radial part of collision strengths, it can calculate large quantities of data.

-{\sc fac}:

The Flexible Atomic Code \citep{Gu03}  is also a fully relativistic program computing 
both structure and scattering data. It uses a modified 
Dirac--Fock--Slater central-field potential which includes an approximate treatment 
of the exchange interaction. The orbitals are optimized in a SCF iterative procedure during which the average
energy of a fictitious mean configuration with fractional orbital occupation numbers 
is minimized. This mean configuration represents the average electron cloud of the 
configurations retained in the CI expansion. This is an intermediate approach between 
MC and SC.  The DW method is used to calculate
continuum atomic state functions $\Psi(\varepsilon)$.

-{\sc mcdf/grasp/grasp92}:

The Multi-Configuration Dirac--Fock structure code \citep{Gran80} and its more recent versions, 
{\sc grasp} \citep{Dyal89} and {\sc grasp92} \citep{Parp96}, use the elaborated fully relativistic 
Hartree--Fock MC variational principle to obtain the $c_i$ and four-component $\phi_i$ in the
CI expansion defined by Eq.~(\ref{ci}). One energy level or the average of a group 
or all energy levels is minimized in the orbital optimization. The QED interaction 
is treated as corrections to the energy
levels. 

-{\sc superstructure/autostructure/autolsj}:

{\sc superstructure} \citep{Eiss74} solves the Breit--Pauli Schr\"odinger equation. CI atomic
state functions are built using orbitals generated from the scaled 
Thomas--Fermi--Dirac central-field potential that is based on the free-electron gas with exchange 
approximation. The scaling factors for each orbital are optimized in a MC
variational procedure minimizing a $LS$ term energy or weighted average of $LS$ term
energies. Semi-empirical corrections can be applied to the multi-electronic Hamiltonian 
in which theoretical $LS$ term energies are corrected in order to reproduce 
the centers of gravity of the available experimental multiplets. Other versions are
{\sc autostructure} \citep{Badn86, Badn97} and {\sc autolsj} \citep{Duba81}.
These include the treatment of the continuum in a DW approach. In {\sc autostructure},
analytical Slater-type orbitals can be also used, and non-orthogonal orbital
basis sets can be considered for the calculations of inner-shell processes.

-{\sc civ3}:

The Configuration Interaction Version 3 code of \citet{Hibb75} uses analytical Slater-type orbitals in
CI expansions to represent atomic state functions and solves either the non-relativistic or the 
Breit--Pauli Schr\"odinger equations. The orbital parameters are varied in a MC optimization 
procedure to minimize one or several energy levels using an analytical central-field potential. The atomic
state function can be improved in a semi-empirical procedure called fine tuning in 
which the multi-electronic Hamiltonian matrix elements are corrected in order to 
reproduce the available experimental
energy levels.

-{\sc r-matrix/bprm/darc}:

The {\sc bprm} scattering package described in \citet{Burber93} implements the elaborated 
$R$-matrix method in the Breit--Pauli relativistic approximation.  In the
latter, only the one-body Darwin, mass-correction and spin-orbit terms 
are considered.  The scattering package can use either 
{\sc superstructure}, {\sc autostructure} or {\sc civ3} target orbitals. The radiation 
and Auger dampings which are important in
processes involving inner-shells have been included using an optical potential \citep{Gorba96, 
Gorba00}. A fully relativistic implementation of the $R$-matrix approach using {\sc 
grasp} target spin-orbitals is also available ({\sc darc}; \citet{Ait96}).

-{\sc mchf}:

In the Multi-Configuration Hartree--Fock structure code \citep{Fisc97}, 
the non-relativistic multi-electronic Schr\"odinger equation is solved using  the elaborated 
Hartree--Fock MC variational principle to obtain the orbitals and the mixing coefficients 
utilized in the CI expansion of the atomic state function. The relativistic corrections 
are included by diagonalizing afterward the Breit--Pauli multi-electronic Hamiltonian. 

-{\sc hfr}:

The pseudo-relativistic Hartree--Fock code of \citet{Cowa81} solves the Hartree--Fock
equations for the spherically averaged atom for each electronic configurations. These
equations are the result of the application of the variational principle to the configuration
average energy. Relativistic corrections are also included in this set of equations, i.e.,
the Blume--Watson spin--orbit, mass-variation and one-body Darwin terms. The
Blume--Watson spin--orbit term comprises the part of the Breit interaction that can
be reduced to a one-body operator. CI is taken into account in an SC approach.
The radial parts of the multi-electronic Hamiltonian can be adjusted to reproduce the
available energy levels in a least-squares fit procedure. These semi-empirical corrections are
used to allow the inclusion of higher-order correlations in the atomic state functions.
Continuum orbitals are calculated in a DW approach using HXR central--field potential
which uses an approximate exchange part of the Hartree--Fock non-local potential.

-{\sc mz}:

This code is base on the Z expansion technique \citep{Safr80} in which a MBPT is applied using 
screened hydrogenic orbitals. The perturbation expansion then contains powers of $Z$, the atomic
number. The Breit interaction and QED are also treated in this expansion. Convergence 
is best with this method for highly ionized ions, and all members of an isoelectronic 
sequence are treated simultaneously.

\section{\label{exp}Experimental Techniques}

Experiments have the potential to directly simulate in the laboratory the same processes
observed remotely using telescopes, and so can provide the most accurate and appropriate 
data for use in interpreting astronomical observations.  This cannot always be 
achieved, owing to practical challenges of achieving the densities or radiation environments
occurring in some sources, and also because the scope of the data needs exceeds the available
experimental resources.  So in many cases we rely on experimental results as crucial checks 
on computations, while for many other quantities of interest experiments represent the only 
means to determine atomic quantities with sufficient accuracy for comparison with 
observations.  This accuracy is needed in the case of line 
wavelengths or energy levels, where inaccurate values can lead to line misidentification or misleading 
redshifts or radial velocities.  Experiments are also crucial for checks of the accuracy of  
DR rate coefficients at low temperature ($\sim 10^4$ K), 
owing to the fact that this process requires energy levels which are accurate to 
$\simeq$1 eV.

Experimental techniques have progressed at a rate comparable with the 
development of theoretical tools and computer platforms.  
Notable in their application to X-ray processes are the electron beam ion trap
(EBIT), which has been described in detail by \citet{Beie90}, and the 
storage rings described by \citet{Abra93}, \citet{Schi98}, and \citet{Gwin01}.  These 
developments have been reviewed recently by 
\citet{Beie03b} and we will not repeat them here.  In what follows we will describe, for 
each process, which experiments have contributed to the available database and 
what future developments might include.

\section{\label{lowlevel}Line Finding Lists and Related Tabulations}

In order to organize the many different types and applications for atomic data, we 
divide them according to the level of computation required for their use.  
Atomic data which is simplest to use, in principle, includes things like line finding lists,
which can be utilized with a minimum of 
detailed fitting or model calculation. Such tabulations are crucial to interpretation of datasets taken at 
high resolution.  They allow a crude determination of the likely conditions in the 
gas under study, and may be able to address simple questions about element abundances and the likely
range of plasma conditions.  In addition, they are needed for successful application of the 
tools described in the later sections.  These differ according to the physical excitation mechanism,
and we discuss them separately.

\subsection{ Coronal Plasmas}

\noindent{\bf Background}

Coronal plasmas are those in which the ionization balance is determined primarily by collisions 
between ions and thermal electrons.  They are common to situations involving mechanical heating, 
such as supernova remnants, stellar coronae and clusters of galaxies.  For our purposes 
this includes objects such as supernova remnants which may not be in ionization equilibrium \citep{Shkl68}.
A feature of coronal plasmas which are in or near equilibrium 
is that the ionization balance adjusts so that the ionization potentials of the most 
abundant ions are comparable to the gas temperature.  As a result, the electrons have sufficient
energy to excite many atomic bound states, and so the spectra are generally rich in line emission.
For coronal plasmas, the most useful low level tools are likely to be finding lists for 
emission lines. This typically includes resonance lines, lines which have the ionic ground term as the 
lowest level, since these are often the strongest and therefore most easily measured.
It also includes satellite lines, which resemble resonance lines but with the addition 
of a spectator electron, since these can be used to infer temperature.  Finally, it includes 
electric-dipole forbidden lines and lines connecting excited levels (subordinate lines)
which are potential diagnostics of various plasma 
parameters such as electron density and the effects of cascades.

Prior to the launch of $Chandra$ and $XMM-Newton$ the need for accurate X-ray line  wavelengths 
in astronomy was driven by the study of solar X-rays.  This led to the development of line lists 
in close cooperation with laboratory spectroscopists.
Early work originated from data taken by rocket flights and the OSO satellites.
These include rocket observations of first solar detection of O$^{6+}$ lines and others 
in the 14 -- 22 \AA\ range by \citet{Free70}.
\citet{Walk70} presented observation of the forbidden lines in the He isosequence.
\citet{Meek68} presented flare spectra from crystal spectrographs on OSO-4 in the 
wavelength ranges 0.5 -- 3.9 \AA\ and 1.0 -- 8.5 \AA.  
An analysis of the solar spectrum in the range 33 -- 110 \AA\ was presented by \citet{Widi68},
and a rocket UV spectrum of the Sun from 60 to 385 \AA\ was presented by \citet{Behr72}.
These and other solar X-ray observations were reviewed by \citet{Culh74}.
The launch of the second generation of solar instruments, the NRL spectroheliograph on Skylab and $SMM$
led to lists by \citet{Widi82} of solar flare lines in the 170-345 \AA\ range, 
and by \citet{Phil82} of the  flare lines in the  5.7 -- 19 \AA\ range. 
\citet{Leme84} published a compilation of observations of the iron K line from solar flares.  
A compilation of solar spectra of the K line of Fe$^{17+}$ -- Fe$^{23+}$ was performed by 
\citet{Seel86}, along with a comparison theoretical values calculated with the Z expansion
technique.
\citet{Dosc84} compiled a line list based on observations of the Sun for the 10-200 \AA\ range.

Motivated at least in part  by solar observations, laboratory measurements of spectra 
using plasma machines or sparks provide spectra in the EUV and X-ray for 
many ions.  
Examples include study of S,  Ar and Ca ions by \citet{Deut66, Deut67}.
\citet{Feld68} published a list of lines from iron in the 10-18 \AA\ range from both laboratory and solar flare 
spectra.  
\citet{Kunz68} measured He-like line intensities from a theta pinch machine for C$^{4+}$ and O$^{6+}$ ions. 
\citet{Conn70} compared laboratory plasma measurements with Hartree-Fock structure calculations for
Fe  between 10 and 17 \AA.  
\citet{Fawc70} classified lines between 240 -- 750 \AA\ produced in a laser plasma, and 
\citet{Fawc71}  classified lines of Fe$^{10+}$ -- Fe$^{14+}$ from a theta pinch machine.
\citet{Gold72} classified  Li-like spectra of K - Mn in the EUV.
\citet{Fawc72b} identified  Ni$^{10+}$ -- Ni$^{16+}$ and Co$^{9+}$ -- Co$^{16+}$ $3s-3p$
and $3p-3d$ lines from laboratory spectra. 
\citet{Fawc74} made {\em ab initio} calculations of $2p-3d$ line wavelengths   and oscillator strengths in 
Fe$^{18+}$ - Fe$^{23+}$, and compared the results with experimental values.
\citet{Gold73} classified $2s-2p$ transitions in laboratory spectra of Ca$^{16+}$ 
and Ti$^{18+}$.  \citet{Feld73} classified transitions of Fe$^{17+}$ and Fe$^{18+}$ observed in laser produced plasmas.
\citet{Fawc73} classified $n$=2 -- 3 transitions from S$^{9+}$ -- S$^{13+}$ and Ar$^{11+}$ -- Ar$^{15+}$.
Other examples include study of Fe by \citet{Fawc72a} and by \citet{Boik78}.  Comparison of theoretical and 
experimental $n$=2 energy levels for the C, N, Li Be, O and F isoelectronic sequences 
was carried out by \citet{Edle83,Edle84,Edle85}.
Identifications of EUV lines of Fe$^{9+}$ -- Fe$^{23+}$ 
based on solar and laboratory data were performed by
\citet{Jupe93}.

\noindent{\bf Recent Developments}

A commonly accepted standard set of line lists for all applications have been compiled from 
laboratory spectra and energy levels for various elements by  
NIST.  These include compilations for:  Mg \citep{Kauf91}
Al \citep{Mart79, Kauf91b},  Si \citep{Mart83},
and S \citep{Mart90, Kauf93}.  Collections of energy levels and transition probabilities for iron were made by  
\citet{Suga79}, \citet{Corl82}, \citet{Suga85} and by 
\citet{Shir90}.  The data are defined as NIST Standard Reference Data, 
implying known or estimated accuracy.  An exception is the energy level data compiled by  \citet{Kell87} for 
some ions, which also includes some inner shell energies.  
Together with older data on spectra of light ions \citep{Wies66, Wies96} this has been incorporated 
into the NIST spectroscopic database \citep{Fuhr99} at: http://physics.nist.gov/PhysRefData/ASD/index.html.
The accuracies of the transition strengths are evaluated according to the NIST convention:  A=3$\%$, B=10$\%$, 
C=25$\%$,D=50$\%$,E=$\>$50$\%$.  Many of the lines in the X-ray band have ratings of C or below.
A consequence of the restriction to critically evaluated data is that the data included 
in this database lag behind the production of data available for such a compilation. 
This does not greatly affect the selection or accuracy of data for strong lines in well studied 
wavelength bands, but it does result in incompleteness of some line lists for use in X-ray astronomy.

Recent work using the EBIT instrument at Lawrence Livermore National Laboratory (LLNL) represent the beginning of an effort to fill the need for 
line wavelengths which are both comprehensive and sufficiently accurate to be used in fitting high resolution
space astronomy data.  \citet{Brow01a} have measured the spectrum of ions of Fe$^{16+}$ -- Fe$^{23+}$ in 
the 10 -- 20 \AA\ wavelength range.  \citet{Leps03} have measured the spectra of ions of Ar in the 
range 20 -- 50 \AA, and \citet{Leps05a, Leps05b} summarize measurements for Ne-like through Li-like Ar, S and Si.

Using  many-body perturbation theory approaches, \citet{Gu05} has 
calculated the level energies of excited L shell complexes in iron and nickel ions. 
Comparison  with experimental results shows that the line wavelengths are generally in agreement to within $\leq$10 m\AA. 
Energy levels with comparable accuracy have been obtained by \citet{Koto05} also using a many-body 
perturbation method.
This is a significant improvement over that obtainable with {\em ab initio} calculations using standard 
configuration interaction methods, and is adequate for many purposes for fitting to observed spectra
using $Chandra$ and $XMM-Newton$. 

Attempts to update and extend the NIST database for use in analyzing $Chandra$ data include the lists for 
Ne, Mg, Si, S in the 10 -- 30 \AA\ wavelength range by \citet{Podo03}. 
Lists specific to the X-ray band are given as part of the {\sc chianti} database \citep{Dere97b, Land99, Land05}.
Other compilations widely used in analyzing X-ray and EUV spectra are  {\sc apec} \citep{Smit01}
and {\sc mekal} \citep{Mewe95b}.  These generally represent compilations from diverse sources, including 
both experimental and theoretical work, in order to be sufficiently  comprehensive to be  used in 
analyzing $Chandra$ and $XMM-Newton$ data.  Effort has been devoted to updating these to the accuracy required 
for interpreting high resolution observations, since most of these compilations were begun 
before the launch of these instruments.  These updates are available as part of the corresponding 
data analysis packages in widespread use for analyzing X-ray spectral data
(cf. http://heasarc.gsfc.nasa.gov/docs/software/lheasoft/, http://cxc.harvard.edu/ciao/).

\subsection{Photoionized Plasmas}

\noindent{\bf Background}

In astrophysical sources containing a strong source of ionizing continuum radiation 
and where mechanical heating is negligible, the ionization balance in the gas will 
be determined by photoionization.  Electron-ion collisions play a role in 
recombination and cooling the gas, and in equilibrium the temperature is determined by 
balancing the heating by photoelectrons with collisional cooling \citep{Tart69}.  A key distinction between 
photoionized and coronal plasmas is that the electron temperature in a photoionized plasma is 
$\sim 5 - 10 \%$ of the ionization potential of the dominant ions.  Line emission is due to 
recombination, inner shell fluorescence, or collisional excitation of low-lying levels.  
Emission line equivalent widths are defined as the ratio of the line flux to the 
average continuum level in the vicinity of the line.  These are generally smaller 
in photoionized plasmas than for coronal plasmas, owing partly to the 
contribution from the  continuum source and also 
because of the reduced energy available from electron collisions.  

The interpretation of photoionized plasmas shares with coronal plasmas the need for 
line lists.   Photoionization by a radiation field which is non-thermal can lead
to a situation where there is a substantial flux of ionizing photons with energies 
far above the first ionization potential of the abundant ions.  This is in contrast to 
the typical situation in coronal plasmas.  It greatly increases the importance 
of inner shells in opacity and ionization.  As a result, spectra can have strong 
fluorescence line emission and inner shell lines seen in absorption as the source 
of photoionization is viewed through the transmitted spectrum of intervening gas.  
Fluorescence emission and inner shell absorption are unique X-ray signatures which can 
be observed from gas under a wide range of conditions, including neutral or near-neutral gas.
Few inner shell transitions, other than satellite lines excited by DR, 
are included in the compilations of lines for coronal plasmas.  
Theoretical wavelengths for K lines 
were calculated by \citet{Hous69} using a single configuration Hartree-Fock method.
\citet{Kaas93} compiled wavelengths and branching ratios for many inner shell lines based on 
energy levels of \citet{Lotz67}. 
A compilation of solar spectra of the K lines of Fe$^{17+}$ -- Fe$^{23+}$ was performed by 
\citet{Seel86}, along with a comparison with theoretical values calculated with the Z expansion
technique. \citet{Shir00} compiled experimental energy levels for Fe ions.  

\noindent{\bf Recent Developments}

Wavelengths and oscillator strengths of inner shell 
$n$=2 to $n$=3  transitions in iron ions (Fe$^0$ -- Fe$^{15+}$) have been calculated by 
\citet{Beha01} using {\sc hullac}.  These features form a broad unresolved transition array (UTA) 
near 16 -- 17 \AA.   Calculations of transition probabilities for a subset of these lines, 
using {\sc civ3} and an extensive selection of configurations in the CI expansion was carried out by \citet{Kise03}.  
Extensive calculations of the UTA transitions, including transitions to $n$=4 states, 
have been carried out by \citet{Duba03}  using {\sc autolsj}.
Wavelengths, A-values and Auger rate coefficients for K lines of all the ions of iron have been published by 
\citet{Palm02}, \citet{Palm03}, \citet{Palm03b} and \citet{Baut03}.  These 
represent a compilation of experimental results where available, supplemented by
calculations using {\sc autostructure} and {\sc hfr} for the lower ionization states.

An additional observable which is associated primarily with photoionized gases is the photon spectrum
due to radiative recombination (RR).  The continuum radiation associated with  radiative recombination 
(RRC) is emitted as a  feature at the threshold energy which has an 
exponential shape, i.e. $j(\varepsilon)\propto {\rm exp}(-(\varepsilon-\varepsilon_{Th})/kT_e)$, above 
threshold $\varepsilon_{th}$, with a width proportional to the electron temperature $T_e$.  RRC features in coronal 
plasmas are broadened by the higher characteristic temperature and therefore are 
not easily observable. There is no published database of experimental RRC threshold energies as 
such, although energy level databases such as the NIST database can be used to derive them. 

Lists of resonance lines, which can be applied to spectra seen in absorption in front 
of a strong continuum source, were  compiled by \citet{Vern94}.  This includes
lines from all wavelengths longward of the He$^+$ Lyman limit at 
227.838 \AA\ and all the ion states of all elements from hydrogen to Bismuth ($Z$=83) 
and includes experimental and  critically evaluated wavelengths.
At shorter wavelengths, \citet{Vern96b} compiled  experimental and critically evaluated energy level
and derived wavelengths for lines originating from ground-term multiplets  in the spectral region 
1--200 \AA. 

{\em Ab initio} wavelengths for some K lines of medium-Z elements have been calculated by 
\citet{Beha02} using {\sc hullac}. Comparison with $Chandra$ spectra suggests a likely accuracy of 
$\geq$ 20 m\AA.  Wavelengths for the K lines of oxygen have been calculated 
by \citet{Prad00} and \citet{Prad03} using {\sc bprm}.   Using a
multiplatform approach ({\sc autostructure, hfr} and {\sc bprm}), \citet{Garc05} 
calculated the atomic data relevant to the photoabsorption near the 
K-edge of all the oxygen ions. The accuracy of the K line wavelengths was estimated 
to be better than 20 m\AA\ by comparison to the available experimental data.  
Experimental measurements of inner shell lines include 
K lines of Fe$^{16+}$ -- Fe$^{23+}$  measured from tokamak 
plasmas by \citet{Beie93}, measurements of K lines of Fe$^{9+}$ --Fe$^{15+}$ by \citet{Deca97} using 
EBIT, and the EBIT measurements of the K lines from O$^{4+}$ and O$^{5+}$  by \citet{Schm04}.  
\citet{Gu05b}
combined EBIT measurements of the K lines of oxygen ions 
with FAC calculations of population kinetics to provide 
line wavelengths and identifications with accuracies ranging from  5 to 20 m\AA\  .
Wavelengths of $n=1-2$ inner-shell transitions in the Be-like up to 
F-like ions of
magnesium, aluminum, silicon, phosphorus and sulfur were measured in a
laser-produced plasma and were interpreted using the Z expansion 
technique \citep{Faen94}. \citet{Biem00} 
observed the K$\alpha$ lines of Ar$^{9+}$--Ar$^{16+}$ in a plasma focus 
discharge  and modeled
the spectra using {\sc hfr} and {\sc mcdf}.
K shell binding energies and line wavelengths can also be derived from Auger electron spectroscopy.
Examples of this are given by \citet{Bruc79, Bruc87, Bruc92} and
\citet{Lin01a, Lin01b}.  Experimental techniques and results have been reviewed by
\citet{Stol87}.

\section{\label{discretediagnostics}Discrete Diagnostics}

Beyond the detection and identification of spectral features,  physical parameters 
of the gas can be derived from the shapes and strengths of the features.
Line ratios can be used to infer density or temperature. These require the accurate
measurement of a subset of lines or spectral features, and  
can only be applied in situations where the measurement is clean enough to make these features
unambiguous. If so, they serve to allow accurate determination of some physical quantities describing the 
gas under study often without requiring extensive numerical calculation or use of spectral deconvolution. 

\subsection{Coronal Plasmas}

Use of line ratios has been extensively applied to coronal plasmas, where they have been used
to infer density, temperature, elemental abundance, departures from ionization equilibrium, and non-Maxwellian
electron velocity distribution.  Density diagnostics rely on the competition between collisions 
and radiative transitions for two or more lines with greatly differing radiative decay rates.
Ratios of lines with differing excitation energies provide temperature diagnostics.  
Examples of the use of these diagnostics in the context of classical nebulae such as 
H II regions and planetary nebulae are discussed in detail by \citet{Oste06}.
In the Sun the conditions in X-ray gas vary both in time and with position, and the use of 
discrete diagnostics provides a means to study a limited spatial region, corresponding to the 
atmospheric zone where a given set of lines is emitted most efficiently.
This may provide insight which is easier to interpret for objects which are clearly 
inhomogeneous, than  global modeling, which involves modeling 
the entire atmosphere at once \citep{Ziri88, Mari93}.  Use of such diagnostics must be applied with care, however, 
since there may be neighboring regions which may have lower emissivity but larger 
emission measure and which can thereby contaminate the diagnostic line measurements.

\subsubsection{He-like Diagnostics}

\noindent{\bf Background}

He-like ions are stable against ionization over a comparatively wide 
range of conditions, and they emit strong $n$=1 -- 2 lines which  
are relatively free from confusion with other lines.   The three 
characteristic lines  are the  resonance r ($1s^2~^1S_0 - 1s2p~^1P_1$), intercombination i 
($1s^2~^1S_0 - 1s2p~^3P_{2,1}$), and forbidden f ($1s^2~^1S_0 - 1s2s~^3S_1$), forming the  ``triplet'' lines.
The i line actually consists of two components corresponding to the two different total angular 
momentum states in the upper level, but these are blended at the resolution of all available
astronomical X-ray observations.
The relative strengths of these lines are studied through conventional line ratios R=f/i and G=(f+i)/r.  The ratio R
is sensitive to gas density via collisions or the effects of a strong UV radiation field driving the 
transition $1s2s~^3S_1 - 1s2p~^3P_{2,1}$.  The ratio G is sensitive to temperature owing to 
the fact that the excitation energies for the $^3P$ and $^3S$ levels are lower that for the 
$^1P$ levels.  The G ratio is also affected strongly by  recombination, 
which increases the populations of the $^3S$ and $^3P$ upper levels relative to the $^1P$ level.  Radiative 
excitation of the  $^1P$ from ground can have the opposite effect on G.

The importance of the He-like lines as diagnostics was pointed out by \citet{Gabr69}, who 
discussed the density dependence of R and made a density determination for the solar corona.
\citet{Blum72} discussed the processes populating the upper levels of these lines and showed 
that R is also temperature dependent. 
The dependence of  G and R on temperature and density for many elements, including the effects of satellite lines 
from DR into the Li-like ion, 
resonances and  inner shell ionization for coronal equilibrium conditions were calculated by \citet{Prad81}.
\citet{Mewe75b} calculated population densities of all levels with principal quantum number $n=2$ in 
several He-like ions with Z ranging from 6 to 20. 
Calculation of level populations including the effects of photoexcitation within the $n=2$ manifold 
were performed by \citet{Mewe78}.  \citet{Mewe78b} examined the effects of time-dependent ionization 
appropriate to solar flares.  \citet{Prad85} showed that recombination drives up 
the value of G in a plasma which is  recombination dominated, i.e. photoionized, owing to the fact 
that RR preferentially populates the triplet states.

\noindent{\bf Recent Developments}

Diagnostic use of He-like line ratios depends on rate coefficients and cross sections
populating the $n$=2 levels, and on the line wavelengths needed for accurate 
identifications. 
Other rate coefficients needed for complete modeling of He-like diagnostics  include radiative rate coefficients for the 
non-dipole allowed transitions, which have been calculated by \citet{Lin77}, and  DR satellite 
line intensities, which have been calculated by \citet{Safr01} and others as discussed in section \ref{disec}.  
Theoretical wavelengths for He-like ions up to $Z=100$ using 
non-relativistic variational calculations were calculated by \citet{Drak88}. 
Rate coefficients for 2 photon decay of metastable He-like ions were calculated by \citet{Dere97}. 
R-matrix calculations of transition probabilities for He-like ions were presented by
\citet{Fern87} and have been  incorporated into TOPbase \citep{Cunto92}.  
Wavelengths for the He-like lines in solar flares have been measured for  the $1s2s~^3S_1 - 1s2p~ ^3P_{0,2}$ lines 
in He-like Ne, Mg, Na and Si to within 20-30 m\AA\ using  the $SOHO$ SUMER instrument \citep{Curd00}. 
Lifetime measurements have been carried out for the excited states of many ions, not just 
for the He-like isosequence, and many of these are reviewed by
\citet{Trab02}.  Experiments using ion traps can measure lifetimes for 
forbidden and intercombination transitions, and these  provide an important check on transition probabilities for use 
in modeling He-like lines.  

The rate coefficients for  collisional excitation of He-like ions were reviewed by 
\citet{Duba94} and  \citet{Kato89}, 
who showed the importance of  resonances in the  $1s2s$ manifold  for 
high Z elements, i.e., Ca and Fe. 
These have been taken into account in the  modified relativistic DW calculations
of \citet{Zhan87}. Adoption of these rate coefficients is recommended for high Z elements, and 
$R$-matrix calculations for low Z elements.   These rate coefficients, together with radiative 
recombination coefficients from \citet{Pequ91}, calculations of  $f$-values and energy levels
from \citet{Fern87} and DR satellite calculations of \citet{Bely82,Bely82b}
were used by \citet{Porq00} to model He-like line ratios,  including a discussion of recombining 
plasma compared with coronal plasma.  \citet{Baut00b} calculated the line emissivities 
and level populations using similar collisional rate coefficients plus a recombination cascade  treatment, 
including the effects of 3-body recombination and suppression of DR 
at high density.   The calculations by \citet{Porq00} and \citet{Baut00b}
provide consistent physically reasonable values for density when applied to $Chandra$ spectra
from both recombination and collision-dominated objects.
\citet{Harr96} calculated rate coefficients for the emission of the lines from S$^{14+}$, 
bench marked these against observations of tokamaks, and then used them to 
set limits on the density in solar flares.

\subsubsection{\label{satsec} Dielectronic Satellite Lines}

Lines emitted during dielectronic recombination (DR) are in principle sensitive temperature diagnostics 
because the emission process involves collisional excitation of a recombining ion, so the rate 
depends on the fraction of electrons capable of surmounting the core excitation energy barrier.
DR is a two-step recombination process that begins when a free electron approaches an ion, collisionally excites
a bound electron of the ion, and is simultaneously captured.  This process is discussed in more detail in 
Section \ref{disec}; here we limit ourselves to discussion of satellite emission.
Satellite lines are emitted during the stabilization of the doubly excited state, when the 
core electrons relax to their ground level in the presence of the additional recombined electron.
This spectator electron is responsible for shifting the line wavelength away from the wavelength of the
resonance line in the parent ion.  Detection of satellite lines requires spectral resolution 
$\varepsilon/\Delta\varepsilon\geq 500$.  Interpretation of satellite intensities 
is simplified by the lack of ambiguity about the excitation mechanism, although satellites can 
also be emitted following inner shell collisional or radiative excitation.  
Even if they are not resolved, satellites can contribute significantly to the 
intensity of the adjacent resonance line.  Thus they can affect the use of discrete diagnostics, 
such as the H-like diagnostics, and must be taken into account for accurate treatment
of discrete diagnostics.

A basic description of the calculation of intensities of DR satellites and 
their diagnostic value was first presented by \citet{Gabr72} and by \citet{Gabr72b}.  Most DR proceeds through 
high $n$ ($n >$ 50) for ions in isoelectronic sequences beyond H and He, and satellites from this are indistinguishable from 
resonance lines.  For satellites, only states with $n <$ 4 produce lines which can be resolved from the adjacent 
resonance line.  The low $n$ satellites require different assumptions to calculate compared 
with the high-$n$ states, and the importance of low-$n$ states increases with Z. 
\citet{Gabr72} showed that satellites are temperature diagnostics, largely independent of ionization balance.
The satellite intensity depends on Z$^4$, and so they are relatively unimportant for Z$<10$.  The formalism for treating 
satellite intensities makes the assumption of LTE between satellite upper level and the relevant autoionizing
continuum.    The satellite intensity then depends on energy separation and branching ratio for 
line emission.  
%For the purposes of scaling arguments it can be 
%assumed that the autoionization probability is independent of Z, 
%so the Z dependence comes only from the radiative probability.  
%The Boltzmann factor is generally small so that temperature dependence is $T_e^{-1}$.  
The resonance line  intensity must also 
be corrected for unresolved satellites, and satellite intensities must include the effects of 
direct excitation of inner shells in the adjacent ion.  Enhancement of 
resonance line intensities  by unresolved satellites  has been treated by \citet{Ansa75}.
\citet{Gabr79} showed that satellites can be a diagnostic of non-Maxwellian electron energy 
distributions.  This is because of the relative importance of inner shell excitation to DR. 
The resonance line can be excited by all electrons with energies greater than threshold, while DR is a resonant process,
selectively excited by electrons at an energy corresponding to the particular resonance.  Thus, 
with 2 or more measured satellites plus the resonance line  it 
is possible to determine if the excitation rate is consistent with a Maxwellian electron velocity 
distribution, and to constrain the form of the distribution if not.

Calculations of satellite intensities using non-relativistic Hartree-Fock orbitals 
were reported in a series of papers by Karim and Bhalla.  
\citet{Bhal75} calculated intensity factors for DR satellite 
spectra for highly charged He-like ions.  This includes 
more accurate calculations of autoionization rate coefficients and more accurate 
calculations of inner shell collisional excitation rate coefficients than the previous work 
by \citet{Gabr72}.  This results in 
smaller emissivities for lines arising from $1s2p(^1P)2s ^2P$ levels.
These calculations used non-relativistic Hartree-Fock-Slater orbitals in $LS$ coupling with CI
and exchange. \citet{Kari86} and \citet{Bhal86d} used this technique to calculate intensity factors for the 
satellite to the  Ly$\alpha$ line of Ne$^{9+}$.
\citet{Kari88b} calculated DR for selected hydrogen like ions 
($Z$=10, 14, 18, 20, 22, 26, and 28) using the Hartree-Fock-Slater orbitals.
The effects of configuration interaction and spin-orbit coupling were included for
$n\leq 4$.  Configuration 
average rate coefficients were used to calculate DR rate coefficients for $n=5$, and 1/$n^3$ scaling was assumed for all higher
states.  This shows that the  maximum DR rate decreases with Z
and the position of maximum shifts to higher temperature with increasing Z.  At lower temperatures
DR rate coefficients for light elements are greater than those for heavy elements while
at higher temperature the trend is reversed.  Satellite line intensities were compared
with calculations using Thomas-Fermi \citep{Duba81} and Z-expansion \citep{Bitt84},
and $\sim$10$\%$ difference is found.   \citet{Kari88b} found an $\sim$40$\%$ 
difference in the maximum DR rate for Fe compared with \citet{Duba81}, possibly due to neglect of 
some high $nl$ states ($2lnl'$, $n > $4).  
\citet{Kari88c} calculated X-ray and Auger transition rate coefficients from doubly excited 
states of He-like ions for $Z$=10, 
14, 18, 20, 22, 24, 26, and 28, along with intensity factors of satellites originating 
from DR of ground-state hydrogenic ions via the autoionizing states. 
\citet{Bhal88} calculated satellite intensity factors for H-like Fe$^{25+}$, showing good agreement 
with Thomas-Fermi calculations for $n=2$, but discrepancies for $n >$ 2.
DR satellite spectra for high-lying resonance states of H-like Fe and Ni
were calculated by \citet{Kari95}, and for H-like Si, Ca, and Fe \citep{Kari92}.
DR rate coefficient for some selected ions in the He isoelectronic sequence 
were calculated by \citet{Kari89b} and \citet{Kari89}.  These authors presented a comparison with 
other calculations, including that of \citet{Nils86} which employs relativistic wave functions
but which uses rate coefficients which are extrapolated for levels above $n=4$.  They argued that accurate
treatment of the high-$n$ levels is important for ions with Z$< 20$.
Effects of radiative cascades on H-like DR  satellite spectra were studied by \citet{Kari88}, 
in an attempt to address conflicting claims by \citet{Gau80} and \citet{Duba81}.  Possible systematic
discrepancies are discussed between results obtained using Hartree-Slater, Thomas-Fermi and Z-expansion 
techniques.
Calculations of DR satellites arising from levels with $n >$4 for He-like Cr and Ni were presented 
by \citet{Kari90}.  The validity of the commonly used $1/n^3$ scaling for high-n satellites was discussed, 
and it was pointed out that high-$n$ satellites are likely to be unresolvable, but can appreciably affect 
total line intensity and should be taken into account.
The dependence of DR satellite intensity factors on $n$ for $1s \epsilon l \rightarrow 2l'nl" \rightarrow 1snl" $
in selected hydrogen-like ions was studied by \citep{Kari91b}, and for 
$1s^2 \epsilon l \rightarrow 1s2pnl \rightarrow 1s^2nl$ in He-like ions by \citet{Kari91}.

Calculations of satellite intensities using wave functions based on a Thomas-Fermi potential 
were carried out by Dubau and coworkers.
\citet{Duba80b} calculated the emission from DR satellites to the Mg$^{11+}$ resonance lines 
including CI, intermediate coupling and wave functions calculated with {\sc superstructure}.  
This showed that the relative importance of 
satellites increases up to $n=4$, and decreases for higher $n$.  Also included were the cascade
contribution to the lower $n$ satellites.  The results are consistent with previous work 
of \citet{Vain78} for wavelengths and radiative transition 
probabilities, but differ systematically for the autoionization probabilities.
%refer to \citet{Gabr72, Gabr72b} as basic work on this subject.
\citet{Duba80} reviewed calculations and solar observations of satellite line intensities.
\citet{Volo87}  calculated DR satellites to Ca$^{19+}$ Ly$\alpha$.  This resolved a problem 
dating from the work of \citet{Duba81} and \citet{Blan85} in which the extrapolation of the contribution of 
the low-$n$ satellites does not properly converge to the resonance line for $n >$ 4, and may
explain why the calculated intensity for Ly$\alpha$ was lower than observed from solar flares
in \citet{Duba81}.
\citet{Bely82} calculated rate coefficients for production of satellite lines by DR
and inner-shell excitation in Ca, as well 
as the production of He-like spectra by excitation, radiative and DR  and through cascades. 
\citet{Bely82b} calculated $n$=1 -- 2 spectra of Fe$^{23+}$ and Fe$^{24+}$ due to inner 
shell direct excitation, cascade, RR and DR.  The results were compared with solar spectra.
\citet{Bely83} compared results of these calculations with spectra from the PLT tokamak.
The Fe$^{24+}$ DR rate coefficient associated with the $1s-2p$ core excitation was measured for temperatures in the 
range 0.9 -- 3 keV and good agreement was obtained with contemporaneous calculations.

The use of DR satellites as temperature diagnostics in the solar corona was discussed for iron by 
\citet{Bely79}, \citet{Bely79b} and for Ca by \citet{Bely82}.
\citet{Dosc87} discussed temperature determinations from solar flares.   
These authors presented model temperature distributions and studied their effect on the satellite line 
strengths and on the temperatures which would be inferred from them based on available calculations.
A review of the X-ray emission from solar flares was provided by \citet{Dosc72}, who also discussed 
temperature determinations  using  H/He like ratios for S, Si, Mg and  satellites due to DR onto He-like  ions.
The He-like rate coefficients for Mg$^{10+}$ were calculated by \citet{Keen86b}, who 
showed that the $1s-3p$ line, when compared with the $n$=1--2 lines, 
can be used as a temperature diagnostic in solar flares and active regions.

\citet{Phil83} used {\sc hfr} to calculate inner shell spectra of Fe$^{18+}$ - Fe$^{21+}$ 
under conditions of solar flares.
The excitation was found to be primarily due to DR rather than direct excitation.  
\citet{Chen86}, \citet{Nils87} and \citet{Nils88} also carried out calculations 
of satellite spectra of H- and He-like ions using relativistic multi-configuration wave functions.  
Calculations of DR coefficients and satellite spectra for He-like ions was 
carried out by \citet{Vain78} and by \citet{Safr00},  using the Z-expansion technique and including 
Breit-Pauli operators.  The systematic dependence with Z is examined and compared with previous work.
An extensive discussion and comparison of different computational methods for 
satellite spectra for iron was presented by \citet{Kato97}.

Calculations of satellite line wavelengths and of the cross section for satellite emission by
DR have been benchmarked by the EBIT experiment.  
\citet{Beie92} measured the satellite spectrum of 
Fe$^{24+}$ and Fe$^{23+}$  in order to measure the cross section 
for DR capture and stabilization.  They find reasonably close agreement with the 
theoretical calculations of \citet{Vain78}, \citet{Bely79}, and \citet{Bely82}.
\citet{Gu01} measured the EBIT spectra for the other ions of 
iron  Fe$^{20+}$ -- Fe$^{23+}$
over a wide range of incident electron energies, 
and showed that unresolved DR satellites 
can contribute as much as 15$\%$ to the intensities of strong resonance lines.
Satellite spectra of He-like Fe and Ni obtained from a tokamak plasma were studied by \citet{Smit93}.
Experimental measurements of satellite lines from He-like ions of Ne and 
heavier elements  using EBIT were carried out by \citet{Warg01}, \citet{Smit96} and \citet{Smit00b}.

\subsubsection{Other Diagnostics}

Ions in the Be -- Ne isoelectronic sequences have ground terms with many levels which 
can be mixed collisionally at high density.  When this occurs it opens 
channels for emission in X-ray lines whose strength is therefore a  density diagnostic.
This occurs for densities greater than $\sim 10^{13}$ cm$^{-3}$ for lines in the $2s^22p^k-2s2p^{k+1}$ transition array of  
the ions of iron Fe$^{17+}$ -- Fe$^{20+}$ in the in the  80 -- 140 \AA\  range \citep{Stra84}.
\citet{Mauc05} has modeled the effect of the ground term mixing on lines in the 10 -- 20 \AA\ wavelength 
range  from these ions.
The lines from the ions Fe$^{15+}$, Fe$^{16+}$ and Fe$^{17+}$ in the 14 -- 19 \AA\ range have been modeled by 
\citet{Corn94b}.
They calculated the structure, radiative and collisional rate coefficients using {\sc autolsj}.  
These can be used to identify lines in the spectrum of solar active regions. 
The intensities of the Fe$^{15+}$ DR satellites to Fe$^{16+}$ at 15 \AA\ and of the
lines of Fe$^{17+}$ at 14.2 and 16 \AA\ are shown to be sensitive to temperature.

Calculations of level populations affecting emission in strong UV, EUV and X-ray lines
demonstrating density dependence due to the collisional mixing of levels in the ground configuration,
have been carried out in a series of papers by Bhatia and coworkers:
Si$^{6+}$\citep{Bhat03b}, Si$^{7+}$\citep{Bhat03e}, S$^{8+}$\citep{Bhat03d}, 
S$^{9+}$\citep{Bhat03c}, S$^{10+}$\citep{Bhat87b}, Ca$^{14+}$\citep{Bhat86b}, 
Ca$^{16+}$\citep{Bhat83}  , Fe$^{9+}$\citep{Bhat95b}, Fe$^{10+}$\citep{Bhat96,Bhat02}, 
Fe$^{13+}$\citep{Bhat93,Bhat94}, Fe$^{14+}$\citep{Bhat80c,Bhat97}, 
Fe$^{16+}$\citep{Bhat03}, Fe$^{18+}$\citep{Bhat89b}, Fe$^{19+}$\citep{Maso83, Bhat80}, 
Fe$^{20+}$\citep{Maso79}, Fe$^{21+}$\citep{Maso80}, Fe$^{22+}$\citep{Bhat81,Bhat86}, 
Ni$^{12+}$\citep{Bhat98}, Ni$^{15+}$\citep{Bhat99}, Ni$^{16+}$\citep{Bhat80c}, 
and Ni$^{20+}$\citep{Bhat03f}.
These include energy levels based on CI calculations using {\sc  superstructure},
transition probabilities and collision strengths in the DW approximation with the addition 
of a correction calculated in the Coulomb-Bethe approximation \citep{Burg74} to take into account high partial waves.
These references are only a representative sample of available
work on such density diagnostics.  Other references can be found in the extensive bibliographic tabulations
contained in the  AMBDAS (http://www-amdis.iaea.org/AMBDAS/) and  ORNL (http://www-cfadc.phy.ornl.gov/bibliography/search.html)
databases.  Other relevant data for application to fusion plasmas is collected in the ALADDIN (http://www-amdis.iaea.org/ALADDIN/) 
database.

\subsubsection{\label{fe17sec} Fe$^{16+}$}

The strong lines of Fe$^{16+}$ near 15 and 17 \AA\  are among the  most prominent 
in the X-ray spectra of many coronal sources.  The relative strengths of the $2p^6 - 2p^53s$ lines 
near 17 \AA\, the $2p^6 - 2p^53d$ lines near 15 \AA\ and the 
$2p^6 - 2p^54d$ lines near 12 \AA\, are temperature sensitive owing to the differing rate coefficients 
for electron impact ionization (EII) \citep{Raym86}.
The ratio of the lines $2p^6 - 2p^53d$ lines near 
15 \AA\ and the $2p^6 - 2p^53s$ lines at 17 \AA\ 
can also be an indicator of recombination 
\citep{Lied95}.  

Calculation of  transition probabilities in intermediate coupling with CI have been carried out 
by \citet{Loul71}.   The wavelengths of the 3s-3p  subordinate lines 
for several Ne-like ions have been calculated  by \citet{Kast83}.  
\citet{Bhat85} discussed the use of the  the $2p^6~^1S_0 - 2p^53s~^3P_2$ forbidden line as a density 
diagnostic.  They also presented observability diagrams which serve as a convenient 
overview of the known and unobserved lines, both for Fe$^{16+}$ and for other ions in the 
neon isoelectronic sequence.  
 Cross sections for excitation of the Fe$^{16+}$ lines have been calculated by
\citet{Mann83}, \citet{Hage87}, \citet{Zhan89} and for Fe$^{15+}$ by \citet{Zhan89b}. 
The lines have been modeled by \citet{Loul73, Loul75}, 
\citet{Smit85}, \citet{Bhat92}, \citet{Corn92, Corn94} and \citet{Phil97}.
Calculations of excitation for Fe$^{14+}$ -- Fe$^{16+}$ were reviewed by  \citet{Badn94}.
Cross sections for excitation of the Fe$^{16+}$ lines have been calculated by
\citet{Gupt00} and \citet{Chen03}

It has long been known that the ratios of the $2p^6 - 2p^53d$ components observed from the sun \citep{Park73, Park75} differ from 
calculations.   This was emphasized by \citet{Beie04}, who pointed out that 
calculations of Fe$^{16+}$ excitation cross sections 
differ systematically from the ensemble of laboratory and astrophysical 
data.  Earlier suggested explanations for this discrepancy include the effects of cascades 
\citep{Gold89}, inner shell ionization of Fe$^{15+}$ \citep{Baut00} and optical depth effects 
\citep{Bhat85}.  The effects of cascades were further explored by \citet{Loch06}, who point out 
the importance of correcting experimental measurements for polarization.
\citet{Smit85}  have shown that the Fe$^{15+}$ DR satellites, such as the 15.226 \AA\ satellite to 
the Fe$^{16+}$ $2p^6~^1S_0 - 2p^53d~^3D_1$ line at 15.261 \AA\, can be comparable to 
the intensity of the Fe$^{16+}$ $\lambda$ 15.01 \AA\ lines.  

The strongest lines 
from Fe$^{16+}$ under coronal conditions are the $2p^6~^1S_0 - 2p^53d^1P_1$ resonance and 
$2p^6~^1S_0 - 2p^53d~^3D_1$ intercombination 
lines at 15.01 and 15.26 \AA.   The relative strengths are not highly 
sensitive to temperature or density \citep{Loul73}, and they have been observed and modeled 
extensively  \citep{Fawc79}.   The ratio of the strengths of these two features, as measured 
in the laboratory \citep{Brow98} is found to be independent of conditions and excitation process, including 
radiative cascades, resonance excitation, and blends with unresolved DR satellites.
The laboratory ratio is measured to be 3.04, and is greater than that measured in many astrophysical sources.
EBIT measurements and modeling of the lines of Ni$^{19+}$ \citep{Gu04} 
confirm the results for Fe$^{17+}$.

\subsubsection{Optical Depth Diagnostics}

Multiple resonance scatterings will affect conveniently measured ratios 
of lines with differing thermalization properties.   Thermalization depends on 
the line optical depth and on the probabilities of destruction or conversion per 
scattering.  Such ratios, whose 
excitation is relatively insensitive to other conditions, include the relative strengths 
of the Fe$^{16+}$ 15 \AA\ lines.   The effects of optical depth on the ratios of  X-Ray and  EUV lines of this ion have been studied 
by  \citet{Bhat99b}.  Although optical depths have been suggested as the origin of 
ratios of the 15 \AA\  lines which are discrepant with calculations, the discrepancy now appears 
to be affected by blending and possibly with omission of important physical 
effects in the calculations \citep{Beie04}.  Other line ratios which are affected by optical 
depth include the Lyman series of hydrogenic ions, and the G ratio of He-like $n=1-2$ lines.
In both of these cases the ratio in question also depends on other factors such as density or  
temperature, and this complicates diagnostic measurement of optical depth using these lines.

\subsubsection{Abundance Diagnostics}

Elemental abundances cannot be directly derived from line intensity ratios, owing to the dependence of 
emissivities on temperature and density.  Line equivalent widths have been used 
to derive abundances in coronal plasmas, since the continuum in many spectral regions is 
dominated by electron-proton bremsstrahlung, and so is only weakly dependent on abundances
of metals \citep{Sylw98}.  This requires that the temperature be derived from measurement of the 
continuum shape, for example, since both the line and continuum emissivities 
depend on temperature.    Abundances can also be derived through full spectral fitting, as 
described in the next section.

\subsubsection{Non-equilibrium Diagnostics}

Departure from ionization equilibrium shifts the ionization balance,
at a given electron temperature, away from the equilibrium value.
This has the effect of increasing the importance of processes associated with adjacent ions in 
emitting lines from a given species, such as inner shell ionization (excitation-autoionization or 
direct ionization)
when the ionization balance shifts to the low side, or recombination (DR or RR)
when the balance shifts to the high side.  An example was studied by 
\citet{Mewe78b} for He-like lines, who showed that non-equilibrium effects can 
enhance the inner shell contribution from ionization of the Li-like species, if the temperature 
is high and the ionization is low, or the recombination from the H-like species if the temperature is low
and the ionization is high. \citet{Oelg04} studied  time-dependent recombination dominated plasmas.
\citet{Oelg01} showed that the DR satellites dominate the emission of the  Fe$^{24+}$ at temperatures 
below that of maximum abundance in collisional ionization equilibrium. Owing to their extreme 
temperature sensitivity, the satellites are excellent spectral diagnostics for such temperatures in 
photoionized, collisional or hybrid plasmas, whereas the forbidden, intercombination and resonance lines 
of Fe$^{24+}$ are not.  Similar effects occur with Fe$^{16+}$ \citep{Baut00}, and other ions.
The importance of iron K lines as diagnostics of non-equilibrium, and detailed modeling of the 
excitation and emission of these lines, was discussed by \citet{Deca03}.

\subsubsection{Non-Maxwellian Diagnostics}

\citet{Gabr79} first suggested the use of DR satellites as indicators of non-thermal electron velocities.
They showed that the intensities of two satellite lines $1s^2 nl - 1s2pnl$ with $n$= 2,3 relative to the 
Fe$^{24+}$ resonance line $1s^2-1s2p$ can be described by a single temperature only 
if the emitting plasma has a Maxwellian distribution 
of electrons. This effect was detected in the laboratory by \citet{Bart85} and in solar 
flares by \citet{Seel87}.  
The effects of non-Maxwellian velocity distributions on the coronal ionization balance
have been examined by \citet{Dzif92} and by \citet{Dzif98}, following on work by \citet{Owoc83}, 
who showed that the oxygen ionization balance in the solar corona is more sensitive to 
non-thermal effects than is iron, and that these effects can cause apparent differences
in temperatures inferred from the two elements.
The effects of non-Maxwellian velocity distributions on the He-like line ratios 
have been examined by \citet{Dzif02}.  They parameterize the electron velocity distribution
according to the formulation of \citet{Owoc83}, and show that at large departures from Maxwellian the
value of G can greatly exceed that of an equilibrium coronal plasma at the same 
value of R.  This corresponds to the appearance of greater temperature in the G ratio, 
for a given value of R.

\subsection{ Photoionized}

RRCs have an exponential shape with characteristic 
width equal to the electron temperature, and only in photoionized plasmas are they likely 
to be narrow enough to be clearly detected. This was pointed out by 
\citet{Lied95}, and has been used to infer the 
temperature in X-ray binaries \citep{Wojd01}, and in Seyfert 2 galaxies 
\citep{Sako00, Kink02, Brin02}.  This procedure relies on the determination 
of the background continuum level, and so is subject in principle to blending and confusion 
with other features.    The influence of bremsstrahlung continuum emission cannot be ignored
\citep{Mewe86}.  Blending with lines of iron complicates use of the features from Ne, for example, but is 
less important for the RRCs from Si and S.  Use of relative strengths of RRCs to determine  
abundances requires atomic data for the photoionization cross section. 
The atomic data needed for this is discussed in the next section.

Metastable levels in ions of the Be isoelectronic sequence can provide diagnostics 
of the combined effects of gas density and UV photoexcitation.
These can be applied to gases which are photoionized via their effect 
on the absorption spectrum.  As shown by \citet{Kaas04}, the absorption
from metastable $2s2p^3P$ in the ion O$^{4+}$ provides a constraint on density 
in the range $10^8$ -- $10^{13}~{\rm cm}^{-3}$.  They show a marginal detection 
of lines arising from this level from an active galaxy observed with the LETG 
on $Chandra$.

Owing to the presence of a strong continuum source, spectra of photoionized sources in the X-ray 
band can be viewed in transmission, and if so exhibit primarily absorption due to bound-free 
continuum and resonance lines.  Line strengths, as measured by the line equivalent width, 
are therefore diagnostic of the gas column density and the line intrinsic width.  Curve of growth 
analysis, familiar from the analysis of stellar spectra \citep{Miha78}, has been applied 
to analysis of $Chandra$ spectra of active galaxies by, e. g. \citet{Lee01}.  An important difference 
between the optical/UV and  X-ray case is that the damping parameter $a$, which is 
typically $\leq10^{-8}$ for a line such as Ly$\alpha$ of hydrogen,  can be 
as great as $\sim 1$ for X-ray lines such as  Fe K$\alpha$ lines owing to the typically 
very short lifetimes of line upper levels \citep{Lied03, Masa04}.

Broadening of bound-free absorption features is dominated by the intrinsic properties of the 
transition, i.e., level or resonance structure near the threshold and the phase space above the 
threshold.  So such features serve as  diagnostics of the presence of certain ions or atomic species.
Examples include the resonance structure and edge position in ions of oxygen.  These have been 
calculated for O$^{5+}$ \citep{Prad00}, 
neutral oxygen, O$^+$ and  O$^{2+}$ \citep{Prad03} and for all ions of oxygen by \citet{Garc05}.
The position and shape of the neutral oxygen edge, and the resonance structure have been calculated by 
\citet{Gorc03}  using the R-matrix code package.
Comparison with experimental measurements \citep{Schm04} using the EBIT device at 
Lawrence Livermore National Lab shows that the 
calculations of \citet{Garc05} are accurate to within $\simeq$ 10 m\AA\ for O$^{4+}$ and O$^{5+}$.
O$^{2+}$ -- O$^{5+}$ K line wavelengths were 
measured with an accuracy ranging from 5 to 20 m\AA\ in an EBIT experiment and were 
interpreted in a collisional-radiative model using the FAC code \citep{Gu05b}.
These lines were observed by \citet{Paer01}, and 
have been used by \citet{Juet04} to infer the mean ionization balance in the 
interstellar gas along  the lines of sight to several X-ray binaries.

\section{\label{highleveltools}Spectrum Synthesis and Global Fitting}

In many situations the most detailed information about the conditions in astrophysical 
plasmas can be gained by modeling the microphysical processes affecting the excitation/decay, 
and ionization/recombination in sufficient detail to synthesize the effects on the spectrum
from many ions and elements simultaneously.  
This entails calculating the ion fractions and level populations 
and also the emitted spectrum or opacity or both.
This procedure, sometimes described as spectrum synthesis or global modeling, 
is necessary in cases where the observed spectral resolution or signal-to-noise
is not sufficient to cleanly resolve individual spectral features to be used 
as discrete diagnostics.  This is the realm of data analysis in X-ray astronomy 
prior to the gratings on $Chandra$ and $XMM-Newton$, when the instrumental resolution was  
capable of resolving only the strongest lines in uncrowded spectral regions, such as
the Lyman-$\alpha$ line from O$^{7+}$ in coronal sources. 
Much of the data obtained by $Chandra$ and $XMM-Newton$ still is obtained with 
the CCD instruments alone, for sources which are too faint for gratings or which 
are spatially extended, and even grating spectra or bright sources are unable to resolve or unambiguously detect
weak lines in crowded spectral regions.
Global modeling is also useful when there is strong coupling expected 
between the ions responsible for spectral formation.  An example is in a photoionized gas, 
where in equilibrium the temperature will couple the emission and absorption properties 
of all the ions in the gas.  

\subsection{\label{coronal}Coronal Plasmas}

Early work on coronal ionization balance was carried out by 
\citet{Jord69, Jord70}, who first included a  
treatment of DR and autoionization and the suppression of these processes at high densities. 
Early work on the cooling function of coronal plasma was presented by  \citet{Cox69} and 
\citet{Dalt71}.  Other calculations of coronal ionization and emission include 
\citet{Land72}, \citet{Jain78}, \citet{Alle69} and \citet{Tuck71}. 
The work of Mewe and coworkers served to better characterize many of the rate coefficients needed for 
coronal emissivity calculations.  These began with calculations of X-ray  
\citep{Mewe72, Mewe72b, Mewe85} and EUV  \citep{Mewe75} lines from the solar corona, 
He-like lines  \citep{Mewe78}, exploration of abundance dependences on lines \citep{Mewe81} and 
continuum \citep{Gron78},  and non-equilibrium ionization \citep{Gron82}.
Many of these rate coefficients and cross sections were incorporated into
 the {\sc mekal} code \citep{Mewe85, Mewe86, Kaas92, Mewe95a, Mewe95b}, and 
updated to the  {\sc spex} code \citep{Kaas96}, a unified plasma model/analysis environment. 

Current ionization balance calculations in widespread use are those of \citet{Arna85} 
who calculated the  ionization balance for most elements of interest to X-ray astrophysics, 
and \citet{Arna92} who  evaluated and compiled
recombination and collisional ionization rate coefficients for all ions of iron, and the 
ionization balance of \citet{Mazz98}.  Both  \citet{Arna85}  and  \citet{Mazz98} make 
use of many of the rate coefficients of \citet{Shul82b, Shul82c}, which in turn use 
the DR rate coefficients of \citet{Jaco77}.  As pointed out 
in Section \ref{disec}, the rate coefficients of \citet{Jaco77} have been shown to overestimate 
the effect of cascades to autoionizing levels.  \citet{Brya06} have 
published ionization balance calculations which make use of the most recent 
rate coefficients (summarized later in this section) and which have 
now been extensively bench marked and fitted to experiments.
The  {\sc chianti} database \citep{Dere97b, Dere01} contains  
evaluated collisional excitation and radiative data, augmented to include wavelengths 
shorter than 50 \AA\ \citep{Land99} and excitation by protons \citep{Youn03},
and including much of the data appropriate to coronal plasmas reported in this review.  
The most recent update to {\sc chianti}, version 5 \citep{Land05} 
also includes transitions to the $n \geq 4$ levels of iron ions 
based on calculations using {\sc fac}, and new UV lines from C-, N-, and O-like isosequences,
and updated level energies from \citet{Land05b}.  

Other calculations of coronal ionization balance and spectra include studies of time-dependent 
ionization \citep{Suth93,Shap76}, exploration of the  
effects of ionization balance on plasma diagnostics \citep{Gian00}, calculation of 
non-equilibrium ionization in supernova remnants \citep{Hami83}, and 
spectra of supernova remnants in the adiabatic phase \citep{Itoh79}. 
Comprehensive calculations of coronal emission spectra for the X-ray band, including 
both ionization balance and spectrum synthesis by \citet{Raym77} remain in widespread use. 
These have recently been updated and greatly expanded by \citet{Bric95} and 
\citet{Smit01} to form the core of the {\sc apec} code, which is in use in analyzing
high resolution astrophysical X-ray spectra. 

\subsubsection{\label{eiion} Electron Impact Ionization (EII)} 

\noindent{\bf Background}

Electron impact ionization (EII) is of fundamental importance to coronal plasmas, 
in which electrons are energized by some mechanical agent such as 
shocks, acoustic waves or MHD dissipation.
It can be  divided into direct ionization (denoted DI), 
which we discuss in this subsection, and other processes involving an intermediate state which 
are addressed in the succeeding subsections.
It is challenging to compute owing to the fact that it requires 
a treatment of two continuum electrons in the final state.  
Therefore experimental cross sections play a key role in the determination of accurate rates.
Measurement techniques include crossed beams, merged beams, traps, plasmas, or indirect methods.  
Of these, crossed beam measurements have been most widely applied.  Storage 
ring methods have the potential to eliminate the greatest systematic uncertainty, namely the 
metastable states in the target beam.

A convenient point for comparison comes from the classical treatment of electron scattering
\citep{Seat62}.  In a collision between two free electrons, one initially at rest, 
the energy transferred is $\epsilon=E/(1+(RE/e^2)^2)$, where E is the kinetic energy 
of the incident electron and R is the impact parameter.  If this energy transfer is 
instead interpreted as the energy available for ionization of a bound electron, 
then the cross section is determined by the maximum impact parameter such that the transferred
energy is the ionization potential, $I$:

\begin{equation}
\sigma_{classical}(E)=4\left(\frac{I_H}{I}\right)^2 \left(\frac{I}{E}\right) \left(1-\frac{I}{E}\right) \pi a_0^2
\end{equation}

\noindent This formula is not accurate enough for quantitative work; at low energies it overestimates 
the cross section and at large energies measured cross sections decrease 
$\propto$log($E)/E$. The functional behavior derived from this formula, with modifications, has been used 
to parameterize the cross section in many tabulations of cross sections and collision strengths 
derived from both experiment and theory.

The semi-empirical formula of \citet{Lotz67} was developed at a time when few accurate experimental 
measurements were available.
Selected experimental rate coefficients and isoelectronic interpolation, based on 
the compilation of experimental rate coefficients by \citet{Kief66},  were fitted by hand 
for each subshell to a formula 

\begin{equation}
\sigma(E)=\zeta \frac{a~{\rm ln}(E/\chi)}{\chi E}
\end{equation}

\noindent where $\chi$ is the ionization potential of the subshell. 
This formula captures the correct behavior in the asymptotic (Born) limit, 
while allowing semi-empirical adjustment of the cross section near threshold.
It was found that a good fit to the available experimental data was obtained if 
$a$ is approximately constant and $\zeta$ is the average number of electrons per subshell.  
This was done for the lowest two charge states of H, He, Li, N, Ne, Na,  K 
and for neutral Ar, Kr, Rb, Xe, Cs and Hg.  Based on this, \citet{Lotz68} 
derived the energy dependent cross section and ionization rate coefficients for all 
ions of elements up to Ca, estimated to be accurate to $^{+40}_{-30}\%$.

\citet{Summ74} developed calculations for direct ionization using the semi-classical 
Exchange Classical Impact Parameter (ECIP) method.  This combines 
a classical binary treatment of close collisions, which gives accurate cross sections 
at low energies, with a treatment which has the proper asymptotic behavior \citep{Burg76b}.
This method was applied by \citet{Summ72, Summ74} to calculations of ionization rate coefficients.  
\citet{Burg77} evaluated these rates by comparison with experiment, and found 
better agreement than other calculations available at the time for most ions.
Calculations using the  CBO approximation,
were carried out by \citet{Gold77}, \citet{Gold78}, \citet{Gold80} and \citet{Moor80}. 
Born approximation cross sections  were computed  for ions of Al and Na \citep{Mcgu77,Mcgu82}. 
The importance of excitation-autoionization 
and direct ionization of Na-like ions was pointed out by  \citet{Samp82}.  
\citet{Shev83} used the CBE approximation to calculate the rate coefficients for EII for
ions belonging to the isoelectronic sequences from H to Ca.
DW calculations in $LS$ coupling were carried out for the isosequences: 
H and Li \citep{Youn80},
He \citep{Youn80b}, Ne \citep{Youn81}, Na \citep{Youn81b},
Cl \citep{Youn82}, and Ar \citep{Youn82b}.  These remain in widespread use for 
astrophysical modeling.

DW calculations are likely to be accurate for ions with charge greater than a few, 
but in order to provide calculations for less ionized species techniques such 
close-coupling must be used.
Close-coupling calculations for the electron-impact ionization include the time-dependent 
close-coupling method, which was applied  by \citet{Pind00b}, and \citet{Pind00} to calculations of 
ionization of He, C and Ne.  In this technique the time-dependent Schrodinger equation is 
solved for the radial wave functions.  
These calculations used a configuration-averaged potential due to the core electrons and 
resulted  in poor agreement with experiment for Ne, suggesting the need for a full Hartree-Fock
treatment of the interaction with the core electrons.
Comparison of time-independent and time-dependent close-coupling methods
was carried out by \citet{Badn98} for Na-like Mg, Al, and Si.  The time-independent methods were
R-matrix and convergent close-coupling solutions,  based on a total wavefunction 
constructed using antisymmetrized products of Laguerre pseudo-orbitals and 
physical bound orbitals.  General agreement was found between the results of the 
methods and with the experiment of \citet{Pear91} at the 10$\%$ level.
Calculation of ionization from metastable Ne was carried out by \citet{Ball04}, and 
for C$^{2+}$ by \citet{Loch05}, using
the R-matrix with pseudostates method.
Relativistic methods have not been extensively applied to calculations of collisional ionization.
The use of MCDF methods has been demonstrated by \citet{Moor90} and by \citet{Pind89}. 

\noindent{\bf Experimental Measurements}

Experimental measurements of collisional ionization have been carried out 
for many ions of interest to astrophysics, and these are summarized in 
Table \ref{table1}.  Measurements using crossed beams include those of \citet{Broo78} for 
He, C, N and O, \citet{Cran82} for Na-like ions of Mg, Al, and 
Si,  \citet{Dise84} for Ne$^+$ , \citet{Greg83} for  Ne$^{3+}$ and Ar$^{3+}$,
\citet{Greg86b} for Fe$^{5+}$, Fe$^{6+}$, Fe$^{9+}$, \citet{Greg87} 
for Fe$^{11+}$, Fe$^{13+}$, and Fe$^{15+}$, \citet{Mont84} for  Fe$^+$.
Measurements of single and multiple ionization of sulfur atoms by EII
were made by \citet{Zieg82}. 

Experiments by \citet{Sten99} and by \citet{Link95} illustrate the complications 
inherent in the measurement of collisional ionization.
These include the effects of metastable levels in the initial state ions, which can have 
greater cross section for collisional ionization than the ground state, and 
excitation-autoionization (EA) in which an ion is collisionally 
excited to a level which autoionizes.  
\citet{Sten99}  made crossed beam measurements of ionization of iron ions, 
Fe$^+$ -- Fe$^{9+}$, showing the influence of metastables
as evidenced by the ionization at energies below the ground state threshold, 
and also excitation-autoionization on the total cross section.
Metastables are not likely to be present in the lowest density 
astrophysical plasmas, and so must be separated from the total cross section 
for use in astrophysics.  However, their inclusion in a self-consistent way 
is a goal for the realistic simulation of finite-density astrophysical plasmas.
\citet{Falk83} used crossed beams to measure the influence of metastables on the 
EII for Be-like ions of B, C, N, and O by varying the 
metastable fraction in the target beam, demonstrating that metastable levels can 
dominate the collisional ionization in this isoelectronic sequence.

Experimental and theoretical data have been reviewed by several groups.  Early
reviews include those of \citet{Kief66} and \citet{Bely70}.  A review
by the Belfast group includes   \citet{Bell83} for  elements lighter 
than fluorine, and \citet{Lenn88} for  elements from fluorine to Ni.  These result in a set 
of recommended  data utilizing the scaling formula of \citet{Lotz68}, normalized to the 
distorted wave calculations of  \citet{Youn80,Youn80b, Youn81, Youn81b, Youn82, Youn82b} and, 
where available, experimental measurements.   These rate coefficients have been further evaluated 
by \citet{Kato91}, who find the rate coefficients too low for neutral and near-neutral species, and also for 
Na-like ions due to neglect of excitation-autoionization.  \citet{Voro97} 
has compiled rate coefficients based on the Belfast compilation, but which attempts to correct these 
problems by adopting the  rate coefficients of \citet{Lotz68} for many of the species in question.
An extensive review of the physical principles and many experimental and theoretical results is 
by \citet{Mull91}.
In addition, rate coefficients for collisional ionization have been reviewed as part of all the  
previously mentioned ionization balance calculations: \citet{Arna85, Arna92} and \citet{Mazz98}.
Sources include \citet{Kato91} and corrections to errors in the Belfast collections
in  http://dpc.nifs.ac.jp/aladdin/.  
Bibliographies of measurements and calculations 
of collisional ionization  are given by \citet{Burg83} and by \citet{Itik84}, \citet{Itik91}, 
and \citet{Itik96}.   Both theoretical cross sections and extensive bibliographic data  have 
been collected at the ORNL collisional database website,
http://cfadc.phy.ornl.gov/astro/ps/data/home.html.

\begingroup
\squeezetable
\begin{table}
\caption{\label{table1}Experimental measurements of 
collisional ionization cross sections. }
\begin{ruledtabular}
\begin{tabular}{ccc}
Ion&Reference&Compilation\\
\hline
H$^0$&\citet{Mcgo68}&Bely\\
H$^0$&\citet{Roth62}&Bely\\
H$^0$&\citet{Fite59}&Bely\\
H$^-$&\citet{Mcdo63}&Bely\\
H$^-$&\citet{Inok68}&Bely\\
H$^-$&\citet{Danc67}&Bely\\
H$^-$&\citet{Tiso68}&Bely\\
He$^0$&\citet{Mont84}&Belfast\\
He$^0$&\citet{Dold61}&Bely\\
He$^0$&\citet{Broo78}&Belfast\\
C$^0$&\citet{Broo78}&Belfast\\
C$^+$&\citet{Hamd78}&Belfast\\
C$^+$&\citet{Aitk71b}&Burgess\\
C$^{2+}$&\citet{Falk83}&ORNL\\
C$^{2+}$&\citet{Wood78}&Burgess\\
C$^{3+}$&\citet{Greg85}&ORNL\\
C$^{3+}$&\citet{Cran79b}&ORNL\\
C$^{4+}$&\citet{Cran79b}&ORNL\\
C$^{5+}$&\citet{Aich98}& \\
N$^0$&\citet{Broo78}&Belfast\\
N$^{2+}$&\citet{Greg85}&ORNL\\
N$^{2+}$&\citet{Aitk71b}&Burgess\\
N$^{3+}$&\citet{Falk83}&ORNL\\
N$^{3+}$&\citet{Greg85}&ORNL\\
N$^{4+}$&\citet{Cran79b}&ORNL\\
N$^{4+}$&\citet{Defr90}& \\
N$^{5+}$&\citet{Cran79}&ORNL\\
N$^{6+}$&\citet{Aich98}& \\
O$^-$&\citet{Tiso68}&Bely\\
O$^0$&\citet{Roth62}&Bely\\
O$^0$&\citet{Broo78}&Belfast\\
O$^0$&\citet{Fite68}&Bely\\
O$^+$&\citet{Aitk71}&Burgess\\
O$^+$&\citet{Loch03}& \\
O$^{2+}$&\citet{Greg85}&ORNL\\
O$^{2+}$&\citet{Aitk71}&Burgess\\
O$^{2+}$&\citet{Loch03}& \\
O$^{3+}$&\citet{Cran79}&ORNL\\
O$^{3+}$&\citet{Loch03}& \\
O$^{4+}$&\citet{Falk83}&ORNL\\
O$^{4+}$&\citet{Loch03}& \\
O$^{5+}$&\citet{Defr90}& \\
O$^{5+}$&\citet{Tref63}&Bely\\
O$^{5+}$&\citet{Cran86}&ORNL\\
O$^{5+}$&\citet{Cran79}&ORNL\\
O$^{5+}$&\citet{Cran79b}&ORNL\\
O$^{5+}$&\citet{Rinn87}&ORNL\\
O$^{7+}$&\citet{Aich98}& \\
Ne$^0$&\citet{Nagy80}&Belfast\\
Ne$^0$&\citet{Bann96}&ORNL\\
Ne$^+$&\citet{Dold63}&Bely\\
Ne$^+$&\citet{Dise84}&Belfast\\
Ne$^{2+}$&\citet{Mats90}& \\
Ne$^{2+}$&\citet{Danj84}&Belfast\\
Ne$^{2+}$&\citet{Bann96}&ORNL\\
Ne$^{3+}$&\citet{Greg83}&ORNL\\
Ne$^{4+}$&\citet{Bann96}&ORNL\\
Ne$^{4+}$&\citet{Dupo97}& \\
Ne$^{5+}$&\citet{Dupo97}& \\
Ne$^{6+}$&\citet{Bann96}&ORNL\\
\end{tabular}
\end{ruledtabular}
\end{table}
\endgroup

\setcounter{table}{0}

\begingroup
\squeezetable
\begin{table}
\caption{\label{table1b} continued: Experimental measurements of 
collisional ionization cross sections. }
\begin{ruledtabular}
\begin{tabular}{ccc}
Ion&Reference&Compilation\\
\hline
Ne$^{6+}$&\citet{Dupo97}& \\
Ne$^{7+}$&\citet{Defr90}& \\
Ne$^{7+}$&\citet{Dupo97}& \\
Ne$^{8+}$&\citet{Dupo97}& \\
Mg$^0$&\citet{Kars78}&Belfast\\
Mg$^+$&\citet{Beck04}& \\
Mg$^+$&\citet{Pear91}& \\
Mg$^+$&\citet{Mart68}&Bely\\
Al$^+$&\citet{Hayt94}& \\
Al$^{2+}$&\citet{Cran82}&ORNL\\
Al$^{3+}$&\citet{Aich01}& \\
Al$^{4+}$&\citet{Aich01}& \\
Al$^{5+}$&\citet{Aich01}& \\
Al$^{6+}$&\citet{Aich01}& \\
Al$^{7+}$&\citet{Aich01}& \\
Si$^+$&\citet{Djur93}&ORNL\\
Si$^{2+}$&\citet{Djur93}&ORNL\\
Si$^{3+}$&\citet{Cran82}&ORNL\\
Si$^{4+}$&\citet{Thom94}&ORNL\\
Si$^{5+}$&\citet{Thom94}&ORNL\\
Si$^{6+}$&\citet{Zeijl93}&ORNL\\
Si$^{7+}$&\citet{Zeijl93}&ORNL\\
S$^{4+}$&\citet{Howa86}&ORNL\\
Ar$^0$&\citet{Wetz87}&Belfast\\
Ar$^0$&\citet{Step80}&Belfast\\
Ar$^+$&\citet{Wood78}&Burgess\\
Ar$^+$&\citet{Dise88}&Belfast\\
Ar$^+$&\citet{Man87}&Belfast\\
Ar$^+$&\citet{Muel85c}&\\
Ar$^{2+}$&\citet{Man93}& \\
Ar$^{2+}$&\citet{Mats90}& \\
Ar$^{2+}$&\citet{Muel85}&ORNL\\
Ar$^{3+}$&\citet{Greg83}&ORNL\\
Ar$^{4+}$&\citet{Cran79}&ORNL\\
Ar$^{4+}$&\citet{Pindz84}&ORNL\\
Ar$^{5+}$&\citet{Greg85}&ORNL\\
Ar$^{6+}$&\citet{Howa86}&ORNL\\
Ar$^{7+}$&\citet{Zhan92}&ORNL\\
Ar$^{7+}$&\citet{Rach91}& \\
Ar$^{8+}$&\citet{Zhan91}&ORNL\\
Fe$^{2+}$&\citet{Muel85}&ORNL\\
Fe$^{5+}$&\citet{Greg86b}&ORNL\\
Fe$^{6+}$&\citet{Greg86b}&ORNL\\
Fe$^{9+}$&\citet{Greg86b}&ORNL\\
Fe$^{9+}$&\citet{Sten95b}& \\
Fe$^{11+}$&\citet{Greg87}&ORNL\\
Fe$^{13+}$&\citet{Greg87}&ORNL\\
Fe$^{15+}$&\citet{Greg87}&ORNL\\
Fe$^{15+}$&\citet{Link95}& \\
Fe$^{23+}$&\citet{Wong93}& \\
Ni$^{2+}$&\citet{Sten95}& \\
Ni$^{3+}$&\citet{Sten95}& \\
Ni$^{3+}$&\citet{Greg86}&ORNL\\
Ni$^{4+}$&\citet{Sten95}& \\
Ni$^{5+}$&\citet{Wang88}&ORNL\\
Ni$^{5+}$&\citet{Sten95}& \\
Ni$^{6+}$&\citet{Wang88}&ORNL\\
Ni$^{7+}$&\citet{Wang88}&ORNL\\
Ni$^{8+}$&\citet{Wang88}&ORNL\\
Ni$^{12+}$&\citet{Wang88}&ORNL\\
Ni$^{14+}$&\citet{Wang88}&ORNL\\
\end{tabular}
\end{ruledtabular}

\end{table}
\endgroup

The extent of the available experimental data is illustrated in Table \ref{table1}, 
which list the ion stage and reference for various experimental papers appropriate to 
astrophysically abundant elements, along with the referring compilation in cases where these 
have been adopted by a compilation such as Belfast, ORNL (http://cfadc.phy.ornl.gov/astro/ps/data/home.html) 
or \citet{Bely70}.    
It is apparent that there are multiple measurements for many ionic species, and 
that there is little overlap between the adopted datasets used by various compilations.
This is a manifestation of the fact that the ionization database needs to be 
thoroughly reexamined because different data bases give different rate 
coefficients \citep{Savi05}.  There is no consensus among the widely used calculations of 
ionization balance as to which is most accurate.  In fact, \citet{Arna85} and \citet{Mazz98} 
primarily make use of distorted wave calculations, such as those by \citet{Youn80},
rather than experimental results directly.  This is due to the 
remaining uncertainties with regard to the applicability of experimental results 
to low density environments, as well as convenience.   Experimental results 
such as those of \citet{Sten99} and \citet{Bann96}, 
in which the beam composition can be thoroughly characterized 
provide a hopeful step in this direction.  

\noindent{\bf Excitation Autoionization}

Excitation-autoionization (EA) is a process 
in which an ion is collisionally excited to a multiply excited level which then autoionizes.  
This process can dominate the total collisional ionization cross section
for many ions at energies above the threshold for direct ionization.  
Its importance was pointed out by \citet{Gold65} and by \citet{Bely68}.  
An experimental demonstration of the importance of EA was performed by \citet{Mart68}. 

Calculations of the effect in Fe$^{15+}$ were carried out by 
\citet{Cowa79}.  \citet{Burg83} performed functional fitting similar to that of Lotz for 
EA.  DW calculations include  rate coefficients from ground and excited levels of Ar L shell ions by \citet{Cohe98},
and by \citet{Grif82} and \citet{Grif87}.  Also in this category are the calculations for Na-like and Mg-like ions
and Fe$^{5+}$ -- Fe$^{13+}$ by \citet{Pind86a,Pind86b,Pind98}, and \citet{Mitn98b},  for Fe$^{0+}$ by \citet{Pind95} 
and for all ions of Ni by \citet{Grif88} and \citet{Pind91}.
CBE calculations of EII with inclusion of 
EA in the H -- C  isoelectronic sequences were performed by \citet{Samp79,Samp81}, and  \citet{Samp82}
made similar calculations for  Na-like ions

Close-coupling calculations including
inner shell excitation and EA were made for Li-like ions of C, N, and O \citep{Henr79}, 
Na-like ions of Al and Si \citep{Henr82},
and for Fe$^{23+}$ using R-matrix \citep{Butl85}.
R-matrix with pseudostates calculations include pseudo-orbitals
in order to allow for an accurate treatment of the bound and continuum wavefunctions.
Ions which have been treated in this way include 
H \citep{Bart96}, He \citep{Huds96}, Li$^+$ \citep{Brow99}, Be$^+$ \citep{Bart97, Pind97}, B \citep{Marc97b},
B$^{2+}$ \citep{Marc97,Woit98},  C$^{3+}$ \citep{Mitn99} and Na-like Mg, Al and Si \citep{Badn98}, Al$^{2+}$
\citep{Teng00b}.

\begin{figure*}
\includegraphics*[angle=270, scale=0.6]{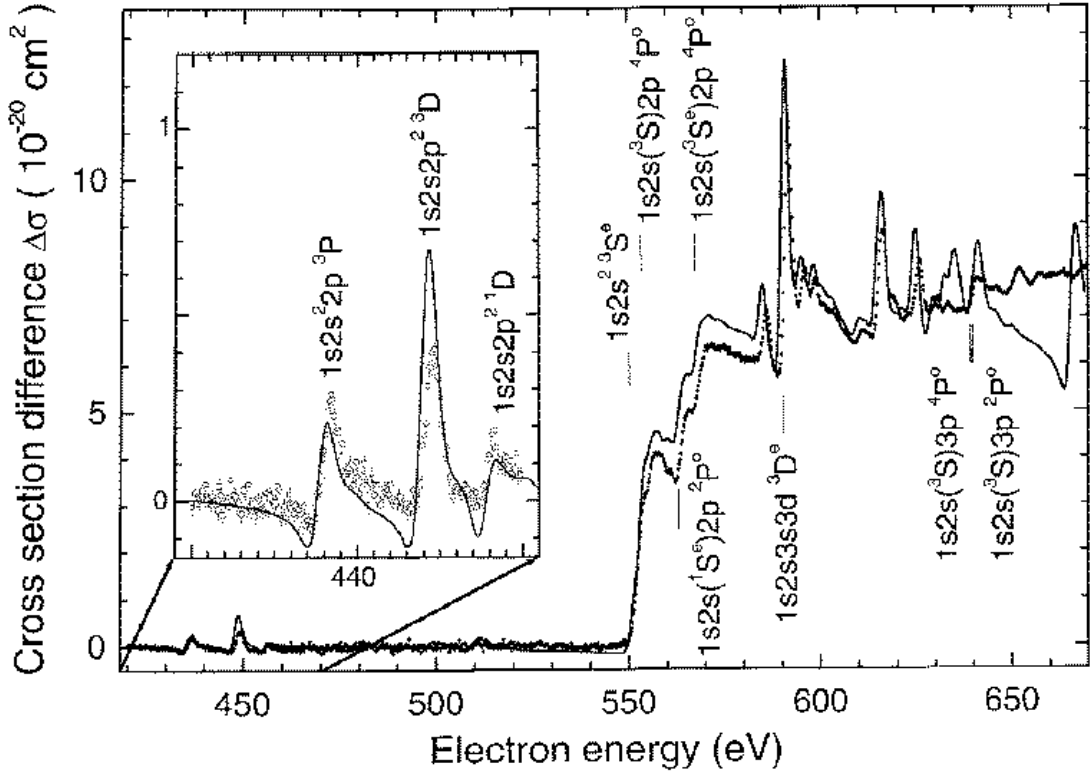}% 
\caption{\label{Teng} Cross section for electron-impact ionization of O$^{5+}$ 
in the vicinity of the $1s2s^2~^3S^o$ EA threshold.  The DI contribution has been 
subtracted off.
Data points are crossed beam measurements \citep{Muel00}, solid curve is R-matrix calculation. }
\end{figure*}

An additional process which affects EII is 
resonance excitation double autoionization (REDA), which is  the first stage of dielectronic
recombination, dielectronic capture (cf. Section \ref{disec}), followed by double autoionization.
A related process is resonant excitation  auto-double-ionization (READI), in which the autoionization
occurs as a single event resulting in the ejection of two Auger electrons.
The net result is an effective ionization event.
\citet{Link95} have studied the effects of EA and REDA in Fe$^{15+}$ using the Heidelberg storage 
ring, and compared  their results with distorted wave calculations by \citet{Badn93} and 
\citet{Chen90}.  EA dominates the cross section by a factor $\sim$5 above 800 eV for 
this ion, and REDA can contribute $\sim$20--30$\%$ to the total ionization
rate.  Theoretical cross sections reproduce the 
magnitude of the experimental cross section, but they do not 
accurately reproduce the complex resonance structure.  
Possible reasons for this are the isolated resonance approximation which omits 
interacting resonance effects.
Experimental measurements of multiple ionization have been carried out by \citet{Muel85b} for 
Ar$^+$ and Ar$^{4+}$,  \citet{Teng00} and \citet{Knop01} for C$^{3+}$, and \citet{Aich01b} for Ne$^{7+}$.
Semiempirical formulae for electron impact double ionization cross sections 
were presented by \citet{Shev05, Shev06}.
R-matrix calculations can reproduce the fine structure in the EA/REDA/READI cross section 
for some ions, 
as evidenced by the comparison shown in Figure \ref{Teng}. 
This shows the R-matrix calculation and 
crossed-beam measurements of excitation-autoionization for the K shell of 
O$^{5+}$  \citep{Muel00}.  

\subsubsection{\label{disec} Dielectronic Recombination (DR)} 

\noindent{\bf Background}

In coronal equilibrium ionization balance is determined by the relative rates of collisional ionization and recombination, 
where the most important recombination processes at low density are radiative recombination (RR), discussed 
in the next section, and dielectronic recombination (DR), which we discuss in this section.  
We offer first a note about terminology.
It is most common to refer to recombination of ion $j$ when discussing recombination 
from ion with charge state $j$ to charge state $(j-1)$, and we will adopt this convention in most of our 
discussion of DR.  In the less common case where DR is discussed in terms of the final state ion, $(j-1)$,
we will refer to it as recombination $into$ ion $(j-1)$.

DR dominates over RR in coronal equilibrium for many ions and at a wide range of temperature,
but not universally.  The H- and He-like ions are an exception, in which RR dominates at the 
temperatures characteristic of equilibrium.  Accurate ionization balance calculations require that
both processes be included.  A description of the process is provided by \citet{Savi03}, which we 
quote with minor modification:
`` DR is a two-step recombination process that begins when a free electron approaches an ion, 
collisionally excites a bound electron of the ion, and is simultaneously captured. 
The electron excitation can be labeled $nl_j \rightarrow n'l'_{j'}$, where $n$ is the principal quantum number of 
the core electron, $l$ is its orbital angular momentum, and $j$ is its total angular momentum. 
This intermediate state, formed by the simultaneous excitation and capture, may autoionize. 
The DR process is complete when the intermediate state emits a photon which reduces the 
total energy of the recombined ion to below its ionization limit. Conservation of energy 
requires that for DR to go forward $E_k = \Delta E - E_b$. Here $E_k$ is the kinetic energy 
of the incident electron, $\Delta E$ is the excitation energy of the initially bound electron, 
and $E_b$ is the binding energy released when the incident electron is captured onto the 
excited ion. Because $\Delta E$ and $E_b$ are quantized, DR is a resonant process..''

For the  purposes of computation,  DR is generally treated as an independent process from 
RR (although a unified approach is discussed later in this section).  
The calculation of rate coefficients divides into two basic parts: determination of the energy structure of the 
doubly excited levels which mediate the process, and the rate coefficients or branching ratios for the stabilizing 
decays.   The first discussion of the importance of this process is that of \citet{Mass42}.
A more detailed summary of the history of the importance of DR is given by \citet{Seat76}.
Historically, the understanding of this process has progressed at a rate determined primarily by the atomic
structure calculations and measurements.  This is particularly true for DR at low temperatures, where the excitation 
energy of the initially bound core electron in the  presence of the captured electron 
(i.e. the quantity $\Delta E$ defined above)  is very small for 
each of the resonances which contribute to the rate coefficient.  
This increases the need for accurate calculations of these resonance energies, since the 
Maxwellian distribution of the recombining electron kinetic energies will be sharply peaked at
low energy if the temperature is low.  An accurate calculation of the extent to which a given resonance 
overlaps with this distribution requires values for $\Delta E$ which are accurate to $\ll kT_e$, or $\ll$1 eV
at $T_e \sim 10^4$ K.

An outline of the rate calculation was provided by \citet{Bate62}, which we reproduce here.  
If the autoionization of the capture state is much more probable than stabilization, 
then the capture state can be  regarded 
as being in LTE with the continuum to a good approximation.  Then the DR rate is simply the equilibrium abundance of the capture
state, given by the Saha-Boltzmann equation, times a (small) branching ratio expressing the fraction of ions 
which stabilize.   That is, the capture and autoionization reactions will maintain a quasi-equilibrium 
between the forward and reverse reactions of

\begin{equation}
{\rm X}_i^{q+} + e \rightarrow {\rm X}_d^{(q-1)+}.
\label{eq1}
\end{equation}

\noindent State $d$ is an excited state of the recombined 
ion ${\rm X}^{(q-1)+}$ that lies above
the ionization potential of ion ${\rm X}^{(q-1)+}$, and $i$ is the ground state of the recombining
ion X$^{q+}$.  The population of state $d$ can be derived from the local
balance between dielectronic capture and autoionization.  The  DR rate 
coefficient is then the density of the capture state times the stabilization rate:

\begin{equation}
\alpha^{dr}_i(T)=\frac{n({\rm X}_d^{(q-1)+})}{n_en({\rm X}_i^{q+})b({\rm X}_d^{(q-1)+})} \Gamma_s
\label{eq2}
\end{equation}

\noindent where $n({\rm X}_d^{(q-1)+})$, $n({\rm X}_i^{q+})$ are respectively 
the number densities of
state $d$ and state $i$, $n_e$ is the electronic density, and $\Gamma_s$ is the 
damping constant for radiative stabilization.
The factor $b({\rm X}_d^{(q-1)+})$ is a departure coefficient which reflects the fact that the
true number in state $d$ will be reduced from the Saha value by the branching ratio for 
autoionization, which is very close to unity:

\begin{equation}
b({\rm X}_d^{(q-1)+})=\frac{\Gamma_a}{\Gamma_a+\Gamma_s}.
\label{eq3}
\end{equation}

\noindent In this equation $\Gamma_a$ is the damping constant for the autoionizing transition
and $n({\rm X}_d^{(q-1)+})$ is given by the Saha equation:

\begin{equation}
\frac{n({\rm X}_d^{(q-1)+})}{n_en({\rm X}_i^{q+})}=\frac{\omega_d}{2\omega_i} \frac{h^3}{(2\pi mkT)^{3/2}}
{\rm e}^{-\varepsilon_i/kT}
\label{eq4}
\end{equation}

\noindent where $\omega_d$ and $\omega_i$ are 
respectively  the statistical weights
of states $d$ and $i$, and $\varepsilon_i$ is the energy difference between 
states $d$ and $i$. So the rate
coefficient may be written

\begin{equation}
\alpha^{dr}_i(T)=\frac{\Gamma_a \Gamma_s}{\Gamma_a+\Gamma_s} \frac{\omega_d}{2\omega_i} 
\frac{h^3}{(2\pi mkT)^{3/2}}
{\rm e}^{-\varepsilon_i/kT}
\end{equation}
\begin{equation}
=\frac{1}{\tau_a+\tau_s} \frac{\omega_d}{2\omega_i} \frac{h^3}{(2\pi mkT)^{3/2}}
{\rm e}^{-\varepsilon_i/kT}
\label{eq5}
\end{equation}

\noindent where 
$\tau_s$ = $\Gamma^{-1}_s$ and $\tau_a$ = $\Gamma^{-1}_a$ are respectively the
lifetimes for radiative stabilization and autoionization.
Since the lifetime for stabilization is generally much longer, this can be written:

\begin{equation}
\alpha^{dr}_i(T)={\Gamma_s} \frac{\omega_d}{2\omega_i} \frac{h^3}{(2\pi mkT)^{3/2}}
{\rm e}^{-\varepsilon_i/kT}
\label{eq6}
\end{equation}

\noindent 
(note that modern calculations do not need to make use of this approximation). $\Gamma_s$ can be 
expressed in terms of
the absorption oscillator strength $f_{bd}$ for transition $b \rightarrow d$, $b$ is an 
excited state (or the ground state)
of the recombined ion X$_b^{(q-1)+}$ that cannot further autoionize  and to which 
the autoionizing state $d$ can decay. 

\begin{equation}
\alpha^{dr}_i=C T^{-3/2} \frac{\omega_b}{\omega_i} \nu_{bd}^2 f_{bd}
{\rm e}^{-\varepsilon_i/kT}
\label{eq7}
\end{equation}

\noindent where

\begin{equation}
C=\frac{(2 \pi)^{1/2}e^2h^3}{ck^{3/2}m^{5/2}}
\label{eq8}
\end{equation}

\noindent and $\nu_{bd}$ is the wave number of the emitted radiation.  
%It is worth pointing out that
%after stabilization the ion is in an excited state, possibly doubly excited, with the 
%recombining electron at high $n$ and/or $l$.

The DR rate for an ion is obtained by summing over all the levels $b$ and $d$.
\citet{Burg64} pointed out the fact that the summation over statistical weights and 
Boltzmann factors can diverge formally, or can be very large if many states 
participate in the DR process.  This is reduced by the fact that the approximation 
used to derive Eq. (\ref{eq6}) breaks down for large $n$, i.e. that the 
autoionization no longer dominates over stabilization.  \citet{Burg64} calculated the rate coefficient for 
recombination onto He$^+$ using values for $\Gamma_s$ calculated by extrapolating the collisional
excitation cross section to below threshold.
%To estimate the magnitude of dielectronic 
%recombination, they consider the case ${\rm exp}\left(-\frac{\varepsilon_i}{kT}\right)=1$, 
%$\Sigma_{b,d} \frac{\omega_b}{\omega_i} f_{bd}=0.3$, $\nu_{bd}=10^5 {\rm cm}^{-1}$, which gives a rate
%coefficient of $1.3 \times 10^{-11}$ cm$^3$ s$^{-1}$ for recombination onto H at 1000 K.  This can be
%compared with the radiative rate at the comparable temperature of $2.0 \times 10^{-12}$ cm$^3$ s$^{-1}$.
A prescription for calculating DR rate coefficients  was presented in the classic work
of \citet{Burg65}, in which he demonstrated by numerical experiment that when the 
sum over $b$ and $d$ is performed on equation (\ref{eq7}) the
dependence on the nuclear charge and level energy separate, and the rate 
can be written as a sum over the 
oscillator strengths of
the stabilizing radiative transitions and a polynomial in the energies 
of the same stabilizing
transitions.  These polynomials, known together  as the Burgess General Formula (GF),  
allow the rate coefficient for DR to be calculated using 
an analytic function of nuclear charge, atomic number, and temperature.  This remains the 
standard set of rate coefficients against which others are compared, and which are in widespread 
use in various calculations of ionization balance.

The GF  is meant to be used 
for DR where the channels which dominate the total rate  are into high $n$ levels near the 
DR series limit.  This allows all these high $n$ levels to be treated 
approximately as a single resonance at the energy of the series limit.  
It is most accurate when the temperature
is comparable to the dominant core excitation, i.e. $\varepsilon_i/kT\sim 1$. 
It is not meant to be used where the dominant recombination channels are 
those at low energies far away from the series limit.
The GF models DR as a dipole core excitation followed by the inverse 
decay. It does not allow for autoionization into excited states, 
non-dipole core excitation, or alternative radiative stabilization 
pathways by either the core or Rydberg electron.  
Comparison with experiment \citep{Savi99d} shows 
the GF, and modifications by Merts and by \citet{Burg76}, to be reliable to approximately a factor of 2.

The importance of using accurate term energies rather than 
configuration-average energies in the calculation of  DR rate coefficients was pointed out by 
\citet{Shor69}.  This is because of the Boltzmann factor in Eq. (\ref{eq7}), 
and also because of the effect on the overlap of the resonance energies.
\citet{Shor69} calculated DR rate coefficients for  various ions of the 
H-like, Li-like, Na-like isoelectronic sequences, in addition to C$^+$ and Ca$^+$ ions using mono-configurational
screened hydrogenic wave functions including the effect of finite stabilization.  These results demonstrate
a qualitative difference between recombination onto ions where $\Delta n=0$ transitions are 
allowed in the core excitation, such as Li- or Be-like ions, and those in which they are 
not, such as H-like ions.  In the former case, capture states are those with $n > 100$, while in the 
latter case capture occurs primarily to states with $n < 50$.  The \citet{Burg65} formula 
provides results which are most accurate for ions dominated by the high-$n$ capture states. 
Larger errors can result for
ions where the low-$n$ states dominate.  For high $Z$ the dielectronic capture and stabilization 
can occur predominantly through states with $n \leq 4$.  The stabilization transition therefore 
can correspond to a spectroscopically resolvable emission line, although with an energy 
which is shifted from the corresponding line in the parent ion due to the partial screening 
provided by the recombining electron.   These satellite lines have diagnostic value and are 
discussed in Section \ref{satsec}.

The calculations of \citet{Shor69} neglected exchange effects and channel coupling, 
both of which are expected to reduce the autoionization rates.
The effects of stabilization and resonance overlap, in addition to exchange and channel coupling, were
examined by \citet{Burg76} using the Coulomb-Born approximation for recombination onto 
H-like ions with $1 \leq Z \leq 40$.  Comparison with the GF shows agreement to within 30$\%$ 
or better.  Compilations of DR recombination rate coefficients calculated using the GF have been made by 
\citet{Aldr73, Aldr76}, and remain in widespread use.

\noindent{\bf Experiment}

Experimental treatments of DR divide into those involving plasma measurement and those involving 
direct measurement of reaction yield using a beam or trap.  In plasma measurements ion fractions are 
measured or inferred from spectra and then the recombination rate is derived under assumptions about the 
ionization rate.  This procedure was carried out by \citet{Bret78}, who observed time variability of 
spectra from a tokamak in which the time variability comes from sawtooth heating due to an MHD instability.
By measuring the spectra of two ions simultaneously (Mg-like and Na-like Mo$^{30+}$ and Mo$^{31+}$), 
whose abundances peak near 2 keV in equilibrium, the ionization and DR rate 
coefficients could be disentangled. \citet{Broo78} measured DR  at $kT=100$ eV for Fe$^{8+}$-Fe$^{10+}$ by 
adding Fe to a theta-pinch plasma.  The measured rate coefficients are approximately  50$\%$ of those calculated by 
\citet{Jaco77}, and are much less than those predicted  by the \citet{Burg65} GF.
\citet{Isle82} measured relative abundances of iron ions in a tokamak where 
coronal equilibrium is achieved.  DR rate coefficients were inferred by assuming collisional ionization rates.  
The results lead to  inferred DR rate coefficients which are approximately $\sim$10 $\%$ of those predicted 
by the \citet{Burg65} GF.

A related technique is the observation of DR satellites from plasma experiments.  Such 
a measurement was carried out by \citet{Bitt79} for He-like Fe in tokamak.  These were found to 
agree with calculations of \citet{Jord69} and \citet{Summ74}. 
\citet{Deca91} measured satellite spectra of Fe$^{25+}$ and compared with 
various available calculations, showing agreement in the satellite 
intensity factors to within $\simeq 10 \%$.
\citet{Deca91b} compared experiment with HF  calculations using the technique  of \citet{Kari91} 
for $n=3-8$ DR satellites of the Fe XXV K$\alpha$ resonance line, 
and discussed the diagnostic use of
the DR satellites for deriving the plasma temperature.

Measurements using traps and related techniques include those of 
\citet{Bria84}, who studied the energies of the K shell  resonance in Ar$^{12+}$ -- Ar$^{15+}$
using an electron beam ion source (EBIS).
Rates for DR onto He-like argon were measured by \citet{Ali90} also using an EBIS.
\citet{Knap89} measured DR satellites from recombination into Ni$^{27+}$ using an 
electron beam ion trap (EBIT), and inferred 
cross sections at the resonance energies.
\citet{Beie92} measured DR satellites for He-like iron using an EBIT, along with 
extensive comparison with theoretical predictions for the strengths of these features.  Good agreement 
was found between experiment and theory for the strongest lines, within the 20$\%$ experimental error, although 
for weak lines much greater discrepancies  were found.

Beam experiments divide into those performed at high energy, using highly stripped 
ions, and those performed at energies similar to those expected in thermal equilibrium.  An 
example of the former is the measurement of \citet{Tani81}, who measured the  K$\alpha$ radiation
produced following electron capture by highly stripped S$^{13+}$ -- S$^{16+}$ with neutral 
argon at 13 MeV.  This fluorescence process is the high energy analog of DR.  
Similar techniques were used by \citet{Clar85} in a study of K shell excitation in Si$^{11+}$ + He at 95 MeV,
and by \citet{Schu87} who measured cross sections for the correlated emission of two K X-rays 
following the collision of S$^{15+}$ ions with H$_2$ in the energy 
range between 70 and 160 MeV.

Merged beam experiments allow reactions to be measured at low energies,  close to what is expected for
thermal equilibrium. Key to merged beam measurements is reduction of background ions in the interaction region, 
which can be an important contaminant at low beam 
energies.  Experimental measurements of DR were carried out for C$^+$ recombining to C using merged beams 
by \citet{Mitch83}.  This resulted in a lower limit which exceeds that calculated by  \citet{Laga82}.  
Also experimental measurement of DR for Mg$^+$ using merged beams was carried out by 
\citet{Beli83}, with measured cross sections found to slightly exceed those calculated by \citet{Laga82}.
\citet{Ditt83} carried out measurement of DR in  Li-like C and B using merged beams, using a high 
energy ion beam from a tandem VandeGraf accelerator.  This has the advantage of 
low background due to the high beam energy, and showed good agreement with 
\citet{Laga82} in the region of the $2s-2p$ resonance for B.
In C$^{2+}$ a discrepancy with theory was found near threshold, in which the measurement is lower, possibly 
due to errors in the theoretical calculation of the $2p4d$ and $2p4f$ levels.
Merged beam experimental measurement of DR in  Boron-like N, O, F were carried out by \citet{Ditt88}  which  
resolved the  $^2D$, $^2P$, $^2S$ states of the $2s2p^2$ electrons.

\noindent{\bf Theory}

As mentioned above, the GF provides DR rate coefficients which are easily applied and 
accurate to within a factor $\sim$2 for temperatures $kT \sim \varepsilon_i$.  Improvements to these rates 
divide into several categories:  improved term energies and bound-state wavefunctions, 
associated with more accurate structure 
calculations; inclusion of other types of transitions such as  autoionization into 
excited states, non-dipole core excitation and stabilization by decay to excited levels; 
and examination of the computational formalism, such as the isolated resonance approximation and 
the use of perturbation theory to calculate matrix elements.
Improvements to the structure initially made use of single configuration non-relativistic 
Hartree-Fock wavefunctions in $LS$ coupling, in order to study the 
systematic behavior of DR and related processes for many ions.

Study of processes  which affect DR rate coefficients, including Auger and 
radiative transition probabilities, and inner shell excitation, were carried out 
by Hahn and coworkers.  For example, inner shell excitation in electron-ion collisions in which the incident electron 
energy exceeds the first ionization threshold is the high energy analog of 
DR.  The relation between this process and DR was explored by 
\citet{Hahn77}, who pointed out the importance of inner shell excitation followed by autoionization.
%in the ion Mo$^{24+}$.  This process exceeds the cross section for resonant excitation due to the  
%filled 3p shell and the  excitation of $2p-3d$  which has a large Auger yield.   This is 
%not the case for some other ions: Mo$^{14+}$ and Mo$^{32+}$.
%Hahn77:  topic F, J
The importance of  Auger ionization to the total 
ionization cross section for highly charged ions was also pointed out by \citet{Hahn78}.
Excitation probabilities, both to discrete and continuum states, were 
calculated for the inner- and outer-shell electrons, using an improvement to the 
Bethe approximation. The result showed a decrease 
in the relative transition strength to the continuum as the degree of ionization 
$Z_I$ increases. The branching ratios for the Auger ionization and fluorescence 
decay were fitted as functions of $Z_I$ for ionized targets. The Auger ionization
and electron fluorescence  cross sections were compared with the corresponding direct processes.
%Hahn78:  topic F, J
Calculations of a related process, excitation followed by double autoionization of the ion 
Fe$^{15+}$, were carried out by \citet{Laga81}  using single configuration bound state 
and distorted wave continuum orbitals.
%Laga81 topic F, J
\citet{Mcla82} calculated cross sections for the resonant-excitation 
of the $1s$ electrons accompanied by the capture of an incident electron for the 
target ions Si$^{11+}$ and S$^{13+}$. 
%Mcla82 topics C
%\citet{Mcla88b} calculated resonant-transfer-excitation (RTE) cross sections for K-shell 
%excitation in Ca$^{17+}$+He, Ca$^{17+}$+H$_2$, and S$^{13+}$+He collision systems, and 
%compared the results with experiments. RTE  cross sections were obtained 
%in the impulse approximation by averaging the corresponding  DR
%cross sections over the momentum distribution of the target electrons. 
%Explicit calculations are made up to $n$=8, $l$=4.  
%Agreement was found for S$^{13+}$, but there are discrepancies in the Ca$^{17+}$ case.
%For $n > 5$, $\Gamma_a$ is constant and the DR cross section depends on $\Gamma_s$, 
%which scales like $n^{-3}$.  
%According to the impulse approximation, RTE is equivalent to DR with suitable average over 
%target momentums.  The discrepancy with experiment is not due to field effects, and 
%is more likely a breakdown of the impulse approximation.
%Mcla88b topics F, L, J, G
A systematic study of the dependence of the Auger and radiative 
transition probabilities of high Rydberg states on their 
principal and orbital quantum numbers was carried out by \citet{Gau78}, 
leading to a simple empirical $l$-dependence.  
%Gau78:  topic F, J, G
The scaling behavior of the transition probabilities associated with DR were studied by \citet{Rett78}, 
showing that $\Gamma_s\propto Z^4$, $\Gamma_a\sim$ const.  These were tested against 
numerical computations using single configuration Hartree-Fock wavefunctions, for the 
Be and Ne isosequences.  This simple scaling was found to break down at high Z, and a 
polynomial expression in Z  was given which is better approximation.  Relativistic effects were
shown to be important for Fe.
%Gau78:  topic J, G

Calculations of DR using single configuration non-relativistic Hartree-Fock wavefunctions in $LS$ coupling
include those of \citet{Hahn80}, who studies the scaling of $\alpha_{DR}$ with principal quantum number 
and with nuclear charge.  A correction to 
\citet{Burg65} general formula was suggested based on these results, which were 
carried out only for the Be and Ne isosequences.
%Gau78:  topic C,K
The importance of the contributions of high Rydberg states to DR of Ar$^{7+}$, Fe$^{15+}$ and Mo$^{31+}$
were studied by \citet{Laga81}, leading to derivation of approximate Auger rate coefficients and fluorescence yields 
for such states.  
%They showed that, although the $n$-dependence of the
%Auger and radiative probabilities is usually of the $n^3$ type, the DR rate may have a much weaker
%$n$-dependence, if the two active electrons are in high Rydberg states or in a light ion. 
%Laga81 topic C, G, J
\citet{Laga81b} calculated the DR rate for Na-like Mo$^{31+}$  and pointed out the 
importance of cascades to all excited levels, not just the ground level, which reduces the total rate, 
and of including doubly excited states which increases the total rate.  
%Laga81b topic C, L
\citet{Laga82b}   demonstrated the importance of $2p-3d$ core excitations and  Auger decays from 
excited states using a calculation of DR for Cl$^{7+}$ (Ne-like).  These effects cause 
departures from $n^{-3}$ scaling for the transition probabilities.
%Laga82b topics C, G, J, I, L
\citet{Laga82} calculated DR of Mg$^+$ and pointed  out the importance of  cascades to 
autoionizing levels, which leads to a reduction of the net rate.
%A fitting formula was provided to take into account this process as a function of $n$.  
%The effect is to provide cutoff at high $n$ of the fluorescence yield. 
%Laga82 topics C L J
%DR capture cross sections for B$^{2+}$ and  C$^{3+}$ were calculated by \citet{Laga83}.
%Laga83 topics C
\citet{Laga83} calculated DR for C$^+$, for which the dominant excitation is $2s-2p$.  
The capture states are at high $n > 100$, and so are densely packed in energy 
close to threshold.  
%The overlap of these states may be important, and cannot be accurately 
%calculated.  Experimental measurements of DR for C$^+$ and Mg$^+$ exceed those calculated by 
%factors of several.
%Laga83 topics C, O, H, G
%\citet{Laga83b}  explored this discrepancy, which is  a factor $\simeq$7 for Mg$^+$.  They attributed this to the 
%two effects of electric field:  ionization and $l$ mixing.  Field ionization can be important 
%since $\Delta n$=0 DR involves high $n$ states, but this  goes the wrong way
%as it will reduce the DR rate. When $l$ mixing is included in the calculation it reduces the  
%discrepancy and reproduces the approximate maximum cross section.
%Laga83 topics C, O, H, G, D
\citet{Laga83c} calculated DR for Ar$^{14+}$, examining the influence of $n^{-3}$ scaling and pointing 
out the effects of $1s$ excitation at high energy.  
%Laga83c topics C I G 
\citet{Mcla83} calculated DR for C$^{3+}$, and 
%Mcla83 topic C
\citet{Mcla83b} calculated DR for B$^{2+}$ with an improved treatment of high $n$ states.  
%Where previous calculations by these authors included only states with $n \leq 2l$, these included 
%asymptotic treatment of high $n$ states.  They find $\sigma$(C$^{3+}$)/$\sigma$(B$^{2+}$)=1.5, 
%while the experimental ratio is  $>$2. 
%Mcla83 topic C, G
\citet{Mcla83c} calculated DR for O$^{5+}$ and discussed  
scaling of $\Gamma_a$ and $\Gamma_s$ vs. Z for various ions in the isosequence. 
\citet{Laga86} discussed the effect on the DR cross section of electric-field-induced 
mixing of high Rydberg state levels, for both Mg$^{+}$ and 
Ca$^{+}$ target ions. 
%DR was calculated in the isolated resonance approximation 
%as a function of the principal quantum number of the high Rydberg electron.  The 
%agreement between theory and experiment for DR is good in the case of Mg$^{+}$, 
%while a factor of 2 discrepancy exists for Ca$^{+}$.
%Laga86 topics C, D
%\citet{Laga87b} extended the investigation of the enhancing effects of static 
%applied electric fields on the DR cross section to Ca$^+$ (K-like) and O$^{4+}$ 
%(Be-like) ions, each undergoing an intrashell dipole 
%excitation.  
%Laga87b topics C, D
%Hahn87 topics C
\citet{Omar87} calculated DR for Ca$^{12+}$, Ca$^{11+}$, Ca$^{10+}$.  
%They discussed the relation between DR and RTE in ion-neutral collisions.  
%They studied K shell excitations which decay by KLL, KLM and KMM transitions, and showed that cascade 
%effects are important, particularly for KMM.  This can affect the DR rate by a factor of 
%10 for some ions.
%Omar87 topics C, F, L
\citet{Mous88} calculated DR into Ne-like Mg$^{2+}$, P$^{3+}$, Cl$^{7+}$, in which 
DR involves $\Delta n \ne $0 core excitation, and results were compared with the  experiment
of \citet{Ditt83}.
%, a comparison which is complicated by multiple Auger channels.
% Previous work concentrated on $\Delta n $=0.
%The autoionizing levels which are accessible are sensitive to incident electron 
%energy, particularly at low energies, and the effect of this uncertainty on the DR rate was pointed out.
%Mous88 topics C, I, J
\citet{Nass89} calculate DR for N$^{2+}$, O$^{3+}$ in order to  compare  
with  the experiments of \citet{Ditt88}.
%Auger channels for the core states $^2$P, $^2$D, $^2$S, and $^4$P were explicitly 
%included.  An approximate field-enhancement factor for the maximum mixing were given.
%Nass89 topics C, D
\citet{Rama89} calculated DR for the B-like ions of C, O, Ar and Fe.  
%These  point 
%out the importance of $\Delta n=0$ transitions,  in which the 
%$2s \rightarrow 2p$ excitation is involved during the initial capture. A complete set 
%of rate coefficients was generated for all ions by extrapolation. The rate coefficients increase 
%rapidly with nuclear core charge $Z_c$.
They pointed out the importance of accurate energies for $\Delta n =0$.  Owing to 
shortcomings of the single configuration Hartree-Fock treatment for this purpose, 
these authors use the HFR code.  
They  estimate field effect enhancement based on state counting arguments, and derive an 
expression for the reduction:
$r_F\sim n_f/(1+l_{max})$, and $n_f~=~3.2 \times 10^8 /F^{1/4}$, in which $F$ is the field
strength in V cm$^{-1}$ and $l_{max}$ is the  maximum $l$ which contributes to DR.
Agreement is obtained  with the experiment by \citet{Ditt88} using $r_F=2$  for C$^+$, but the experimental
uncertainty is large. 
% An  improved semianalytic treatment of HRS is included as an appendix.
%Rama89 topics C, H, D, N
Metastables in the recombining ion can affect experimental results from beams, and these 
were calculated by  \citet{Hahn89} and by \citet{Hahn89b}, who calculated DR cross 
sections for metastable O$^{6+}$ and found them 
to be large, $\sim$10$^{-15}$ cm$^2$ for resonances, leading to rate coefficients  
which are $\sim 10^{-9}$ cm$^3$ s$^{-1}$.  
With some field enhancement and assuming that the initial 
ion beam is a mixture of metastable $1s2s$ ($^1S$ and $^3S$) states, the overall 
feature of the experimental data of \citet{Ande92} was reproduced, 
including the broad peaks at the incident-electron kinetic energies of 4.7 
and 12.5 eV, and partially also at 2.5 and 6.8 eV.  With a  different 
mixture, the C$^{4+}$ data were reproduced. 
%Hahn89 topics C, Q, R
\citet{Janj89} calculated DR rate coefficients and cross sections 
for the O$^{3+}$ ion where the $2s$ and $2p$ electrons of the initial state 
are excited to higher, $n\geq$3 states ( $\Delta n \ne 0$ transitions).
The $2p$ electron excitation dominates the  $\Delta n \ne$ 0 process, while the $2s$ excitation 
is suppressed by the cascade corrections that strongly affect the intermediate 
states $1s^22s2p3snl$ and $1s^22s2p3dnl$. The $2s$ contribution is approximately 
15$\%$ of the  $\Delta n \ne$ 0 mode, while the total  $\Delta n \ne$ 0 
contribution is roughly 10$\%$ of the  $\Delta n =$ 0 cross section.
%Janj89 topics C, I, L
\citet{Bell89} calculated DR cross sections and rate coefficients for the 
H-like and He-like C and O ions.  The DR cross sections 
for the initial metastable states of the He-like ions were also estimated 
for a few low-lying resonance states near the DR threshold. 
%The validity of LS coupling was explored and deemed to be adequate  for Z$<20$.
Comparison between various electron coupling schemes was discussed and it was pointed out that
configuration interaction (CI) will reduce the contrast between various coupling choices.  
%Bell89 topics C, Q, M, R, H, I
\citet{Hahn93} obtained rate formulas for DR 
by fitting all of the existing DR data for ions with core charges ($Z_c$) less 
than 50 and the number of electrons in the target ions less than 13. 
%The fitting procedure divides the total rate coefficients into five different excitation 
%modes rather than fitting the rate coefficients for each isoelectronic sequence. This 
%method allows better interpolation among the ions for which no explicit data 
%are available and also incorporates detailed information on the contributions 
%from the different shell excitations. The  formulas are estimated 
%to be reliable to +/- 50 $\%$ or better. Sample rate coefficients are generated for 
%C, O, Mg, Ar, Fe, Se, and Mo ions.
%Hahn93 topics C
A review of the physics of RR and DR was presented by \citet{Hahn97}, in addition pointing out 
more exotic modes of recombination.  These include off-shell dielectronic recombination (radiative DR = RDR), in 
which an electron capture is accompanied by simultaneous radiative emission 
and excitation of the target ion. 
%hahn97 topics S
 
Similar techniques were used by \citet{Rosz79}, who calculated the total rate of DR  for Mo$^{32+}$  
(Ne-like) using radial orbitals obtained from central field solution and a Hartree-Fock 
calculation for ground states. The structure was $LS$ term resolved, not configuration averaged.
Agreement with the GF for the total rate is good in the temperature 
range 1.0-6.0 keV, but poor below 1.0 keV, which is attributed to the GF use of hydrogenic 
Coulomb wavefunctions for continuum and hydrogenic wavefunctions for bound states.
\citet{Rosz87} calculated DR for members of the fluorine isoelectronic sequence, Ar$^{9+}$, 
Fe$^{17+}$, and others using the single-configuration, LS-coupled, frozen-core approximation.
These calculations employed a different treatment for the ground state orbitals, in which the 
Hartree-Fock exchange is replaced with Cowan HXR exchange potential, 
and continuum exchange is treated using a semi-classical exchange potential. A significant discrepancy is found with 
the only other available calculation, that of \citet{Jaco77}.  
This may be due to the overestimate of the decays to autoionizing levels by \citet{Jaco77}, as 
will be discussed below.  The results also differ from the rate coefficients calculated using the GF.
Similar results were found by \citet{Rosz87b} from calculations of DR for O-like
Ar$^{10+}$ and Fe$^{18+}$ using similar techniques.  
\citet{Rosz87c} calculated DR for ions of the Li isoelectronic sequence: 
 Ne$^{7+}$, Ar$^{15+}$, Fe$^{23+}$, and Kr$^{33+}$.  
When compared  with \citet{Mcla83} a large discrepancy is found, which is attributed to 
an incorrect assumption by \citet{Mcla83}  that all transitions from states $1s3pn'l'$ and $1s4pn'l'$ to $1s2pn''l''$
are treated as stabilizing, while in fact many are not since, for $n'' > n_0$, they 
are not below the first ionization limit. 
\citet{Rosz89} used the same technique to calculate rate coefficients for O$^{5+}$ and O$^{2+}$.
Comparison  with \citet{Badn88} and the \citet{Burg65} general formula shows 
differences of $\simeq$40$\%$.
The effect of density dependent correction formulas for DR as proposed by 
Jordan, Burgess, and Shore (all unpublished) are examined by \citet{Rosz89b}, 
along with the effects of metastables.
Another single-configuration Hartree-Fock calculation is that of
\citet{Youn83}, who calculated DR for He-like ions of C, Al, Ar, and Fe using DW and including CI for 
a limited set of states.  The results compare well with those of \citet{Bely79}.
 
\citet{Jaco77} pointed out that autoionization can 
occur to another level besides the ground level of the recombining ion, and that this may 
be more probable than the inverse of the initial capture. This would lead to a greater 
autoionization probability and a smaller net recombination cross section. Also, the autoionizing 
level can be collisionally ionized as well, if the density is high, leading to a net 
reduction in the recombination cross section. 
These processes were incorporated into DR rate coefficients for: 
Fe$^{8+}$ -- Fe$^{24+}$ \citep{Jaco77}, and all ions of Si  \citep{Jaco77b}, S \citep{Jaco79},
Ca, and Ni \citep{Jaco80b}.  
These calculations were widely adopted and were later shown to be inaccurate owing to 
inclusion of autoionization into excited states which are energetically inaccessible
\citep{Badn86}.   For example, for Fe$^{22+}$ \citet{Jaco77} included dipole autoionization of 
$1s^22s3p(^1P)nl \rightarrow 1s^22s3s(^1S)E_cl_c$, but \citet{Badn86} showed that this 
is only energetically allowed for $n > 19$.  At T=$10^7$~ K this translates into 
a factor 2.5 difference in the total rate.  \citet{Badn86} also showed that 
non-dipole autoionizing transitions are important for this case.

The effects of fine structure  were included in single configuration 
intermediate coupling calculations of DR rate coefficients for excited configurations 
of iron ions  by  \citet{Dasg95}.  Comparison with \citet{Savi99} shows adequate agreement, 
although their work leaves out important autoionizing 
channels by not explicitly calculating capture to states with $n~\geq~15$. Scaled rate coefficients for 
O and F-like ions were calculated by  \citet{Dasg90} and \citet{Dasg95}.

MCDF orbitals were used for DR calculations by Chen and coworkers.  
\citet{Chen86c} carried out such calculations for the He isoelectronic sequence. 
Relativistic effects drastically 
alter the satellite structure of Fe$^{24+}$, but only affect the total rate by 
20$\%$.  Relativistic effects can alter the rate coefficients 
by as much as a factor of 3 by altering the Auger energies for higher Z elements (e. g. Mo$^{40+}$). 
In non-relativistic treatments and $LS$ 
coupling the total rate is dominated by  a few capture levels, while relativistic effects 
redistributes the rate to more levels.   Calculations were carried out for 
Ne-like ions ($Z$=18, 26 and greater) by \citet{Chen86b} and for  F-like ions ($Z$=26 and greater) by \citet{Chen88}. 
Coster-Kronig channels are those in which an ion with an inner shell vacancy decays by 
autoionization of an electron with the same principal quantum number as the initial vacancy.
The influence of these channels on  DR rate coefficients were 
examined in the calculations of \citet{Chen88b}.
The total DR coefficients for B$^{3+}$, N$^{5+}$, and F$^{7+}$ ions are reduced by 60$\%$, 
13$\%$, and 4$\%$, respectively, due to the inclusion of Coster-Kronig channels. 
These effects are found to be  negligible for Z$>10$.
\citet{Chen88c} carried out DR calculations for  Be-like ions
with atomic numbers Z = 30, 34, 36, 42, 47, and 54.
Effects of relativity and configuration interaction on 
the DR satellite spectra and rate coefficients were
studied by comparing the theoretical results from nonrelativistic and relativistic 
single-configuration and multiconfiguration Hartree-Fock calculations with 
and without Breit interaction by \citet{Chen88e}.  Explicit calculations  for 
H-like Ne, Cr, Mo, and Xe show that relativistic and CI effects are 
important in calculating satellite spectra. For the light ions, nonrelativistic calculations in intermediate 
coupling with configuration interaction may be sufficient. For medium heavy 
and heavy ions, however, {\em ab initio} relativistic calculations in intermediate 
coupling with CI including the Breit interaction are necessary. 
DW calculations using MCDF orbitals were carried out for  Li-like ions by \citet{Chen91}.
It was shown that the total rate coefficients 
for the fine-structure states for ions with Z$\geq$4 can differ by as much as 
one order of magnitude at low temperatures due to the effects of relativity 
and intermediate coupling. Comparison with \citet{Rosz87}, finds $\Delta n$=0  rate coefficients 
20$\%$ larger for Ne$^{7+}$ at low temperature, which is likely due to \citet{Rosz87} neglect of $1s2pnl$
for $l > $8 and the use of non-relativistic $2s - 2p$ energies.  For $\Delta n >$0, computed rate coefficients 
are lower at low T and  higher at high T, than those of \citet{Rosz87}.  The 
discrepancy at low  temperatures is likely due to discrepancy in Auger energies, 
and at high temperatures it is due to inclusion of K excitation channels in the work of Chen.
DW calculations using MCDF orbitals have been carried out for B-like ions by \citep{Chen98}. 
Comparison with experiment \citep{Savi99} shows that the MCDF rate coefficients agree with experiment to within 
$\simeq 30 \%$.  The discrepancy  may be due to the neglect of $l \geq 8$ Rydberg states, and possibly 
also due to overestimate of resonance energies. 

Badnell showed the importance of the intermediate coupling and
CI effects on DR rate coefficients in a series of papers reporting calculations using his 
{\sc autostructure} code in Fe$^{22+}$ and Fe$^{21+}$ .
This enables the relatively rapid calculation of large numbers of radiative and autoionization transition 
rates for arbitrary atomic configurations which are needed to calculate DR rate coefficients.
\citet{Badn86} calculated configuration-mixing $LS$ coupling or intermediate coupling 
autoionization rate coefficients for Fe$^{24+}$, and pointed out 
omissions in the state contribution in work on the same ion by \citet{Bely79}.
Calculations using the same techniques for  Fe$^{21+}$ and Fe$^{22+}$ were carried out by \citet{Badn86b}.
This was also done for Be-like ions \citep{Badn87}, B-like ions \citep{Badn91b}
S$^+$ -- S$^{5+}$ , \citep{Badn91} and  oxygen ions \citep{Badn89, Badn92}. 
Intermediate coupling calculations were carried out by \citet{Badn89b} for the boron 
isoelectronic sequence, by \citet{Badn89c} for oxygen and by \citet{Badn89d} for Na-like
P and Cl.  The  effects of coupling are expected to be less important for 
highly charged ions, i.e., charge greater than 20, since the change from $LS$ coupling
does not open significant additional radiative channels.
DR for the ground and excited states of He-like C and O were calculated by \citet{Badn90b},
and for Li-like Al by \citet{Badn90}.  These were compared with R-matrix calculations by \citet{Tera90}, 
and the importance of high angular momentum states and stabilization were pointed out. 
The importance of fine structure and CI were examined in detail for He-like ions by \citet{Pind90},
and compared with storage ring experimental results.
\citet{Gorc96} demonstrated the importance of CI in Na-like ions, including Fe$^{15+}$.

Calculations of DR cross sections are most often done perturbatively, using bound state wavefunctions which 
are calculated separately from the continuum. 
\citet{Gorc96b} compared calculations using R-matrix with an optical potential 
against perturbative methods for calculating 
DR in the Ar$^{15+}$ ion, showing agreement between the methods.  This provides validation for the use of 
R-matrix methods for cases where the isolated resonance approximation is likely to be inaccurate.  
Rate coefficients which use continuum wavefunctions calculated in the R-matrix method  and which 
combines both RR and DR into a unified rate coefficient have been calculated by \citet{Naha94} and 
\citet{Naha95, Naha96}.
Total recombination rate coefficients were calculated using R-matrix  for DR into 
Si, Si$^+$, S$^+$, S$^{2+}$, C$^+$, 
C, N$^+$, O$^{2+}$, F$^{3+}$, Ne$^{4+}$, Na$^{5+}$, Mg$^{6+}$, Al$^{7+}$, Si$^{8+}$, and S$^{10+}$,
\citep{Naha95}  and for iron ions \citep{Naha96}.

Important to any close-coupling treatment of DR is the effect of radiation damping.  This 
corresponds physically to the effect of radiative decays on the resonant capture state, and 
which formally determines the width of the resonance and therefore its effect on the rate coefficient.
The importance of this effect, and the limitations of R-matrix calculations of DR, were
discussed by \citet{Gorc02}, by comparing computational techniques applied to recombination into
Fe$^{16+}$.  They showed the importance of accurate treatment of  resonance damping, 
adequate numerical resolution of resonances and inclusion of radiative decays to autoionizing states,
and that these effects can be treated more efficiently and accurately using 
perturbative calculations.  

\noindent{\bf Recent Developments}

In a series of papers beginning with \citet{Savi97}, measurements were made of the resonance 
strengths and energies for several  iron ions in the Li-Ne isoelectronic sequences using the 
heavy-ion Test Storage Ring in Heidelberg, Germany.   This apparatus has the combined 
advantages of low background and negligible metastable content. 
\citet{Savi99} measured the energy dependent 
cross section for Fe $^{17+} \rightarrow$ Fe$^{16+}$ and Fe$^{18+} \rightarrow$ Fe$^{17+}$ $\Delta n$=0 
DR and calculated DR rate coefficients. 
They found significant discrepancies between rate coefficients inferred from their measured  
cross sections and those of other published calculations. 
In a comparison with the calculations of \citet{Jaco77} and \citet{Rosz87}, who only published 
Maxwellian-averaged rate coefficients, \citet{Savi99}
demonstrated that such comparison can be used to identify discrepant calculations, but cannot 
clearly identify the approximation responsible for the error, and that agreement 
between measured and calculated rate coefficients is not a reliable test of the validity of a calculation.
The accuracy of DR calculations can best be evaluated by comparison of resonance 
strengths and energies in the cross section.  \citet{Savi99} show that single configuration $LS$ coupling 
calculations overestimate Fe$^{18+} \rightarrow$ Fe$^{17+}$ by a 
factor 1.6. This may be due to the use of $LS$ coupling which leaves out 
important autoionizing channels, or it could be due to 
inaccurate resonance energies.  

\begin{figure*}
\includegraphics*[angle=270, scale=0.3]{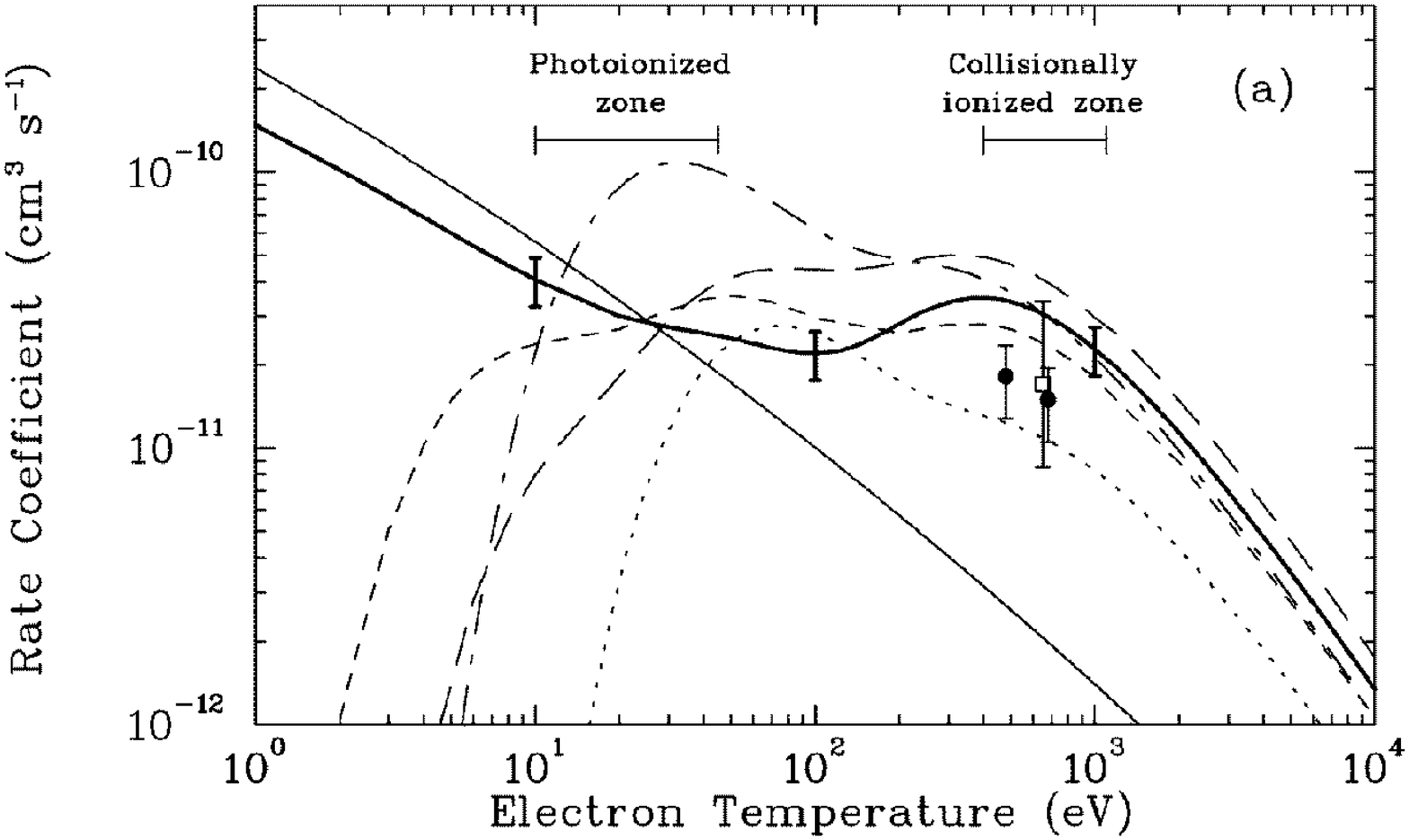}% Here is how to import EPS art
\includegraphics*[angle=270, scale=0.3]{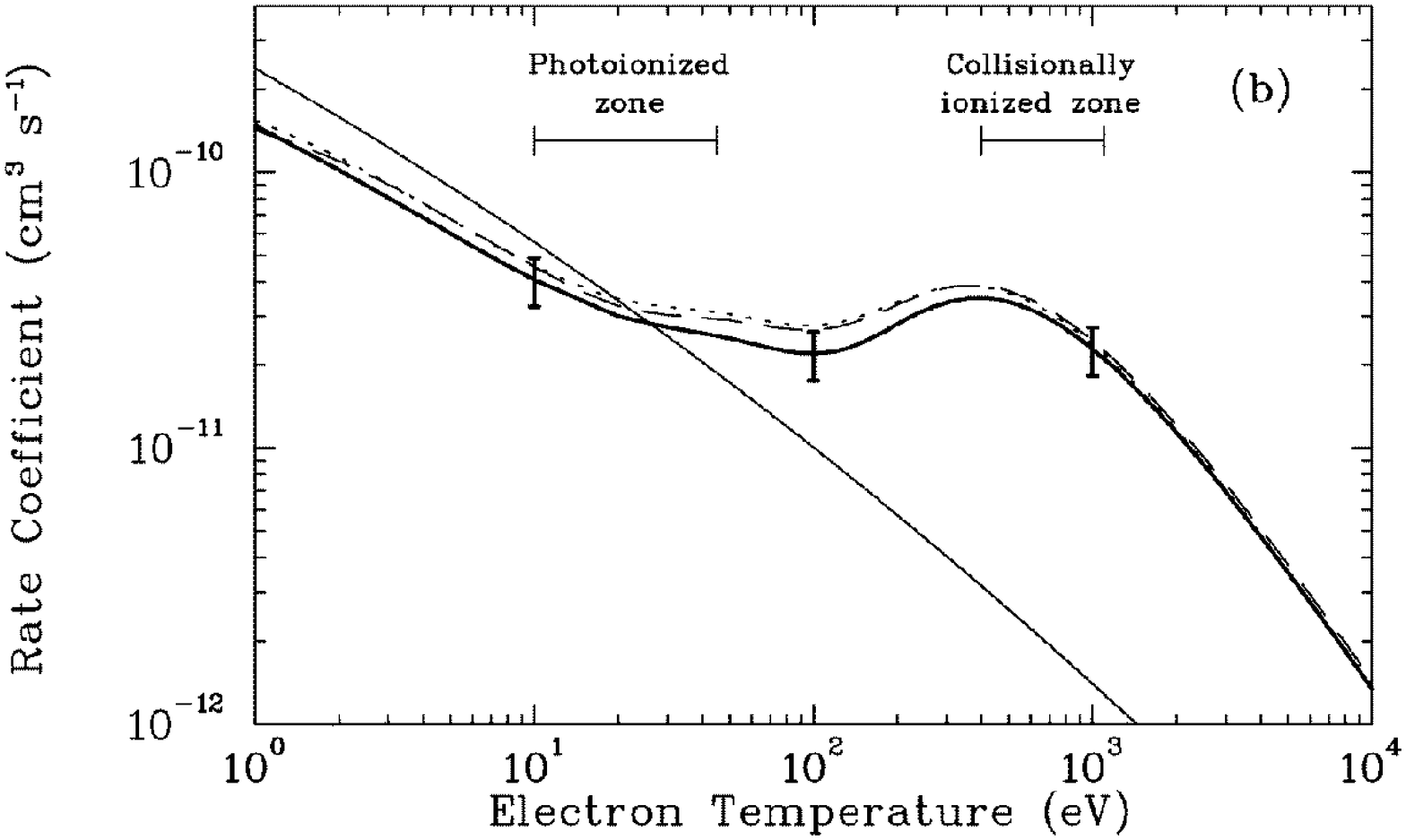}% Here is how to import EPS art
\caption{\label{savindr} Left panel (a): 
Fe$^{18+}$ -- Fe$^{17+}$  rate coefficient for recombination. 
The thick solid curve represents the experimentally derived rate coefficient from \citet{Savi02b}.
The thick error bars show the estimated experimental uncertainty of 20$\%$. 
Symbols show tokamak results of \citet{Isle82} (filled circles) and \citet{Wang88b} (open squares). 
Curves are DR rate coefficients calculated by  \citet{Jaco77} as fitted by \citet{Shul82b,Shul82c}
(dotted curve), \citet{Rosz87} (short-dashed curve), and of \citet{Dasg94} (long-dashed curve), 
and recommended DR rate coefficient of \citet{Mazz98} (dot-long-dashed curve), and 
recommended RR rate coefficient of \citet{Arna92} (thin solid curve).  
None of the experimental or theoretical DR rate coefficients include RR.  The range of temperatures 
at which these processes are likely to occur in equilibrium are denoted for 
photoionized and collisionally ionized plasmas.
Right Panel (b): As in (a) the thick solid curve represents the rate 
coefficient derived from the storage ring experiment of \citet{Savi02}, 
and the thin solid curve shows the recommended RR rate. The dotted curve shows the
MCBP calculations and the dashed curve shows the MCDF calculations of the DR rate coefficents calculated
by \citet{Savi02}. 
Both calculations are for capture into states including capture states for all values of $nl$ (i.e. $n_{max}=\infty$)
and both include DR via 1 $\rightarrow$ 2, 2 $\rightarrow$ 2, and 2 $\rightarrow$ 3 core excitations.   
None of the experimental 
or theoretical DR rate coefficients in (a) or (b) include RR. From \citet{Savi02}. }
\end{figure*}

\citet{Savi02} measured resonance 
strengths and energies for DR of Fe$^{18+} \rightarrow$ Fe$^{17+}$ 
via $n = 2 \rightarrow n'=2$ and $n=2 \rightarrow n'=3$ core excitations. 
They have also calculated these resonance strengths and energies using two independent techniques:
the perturbative multiconfiguration Breit-Pauli (MCBP) and 
multiconfiguration Dirac-Fock (MCDF) methods, finding reasonable agreement  
between experimental results and theoretical calculations. 
The left panel (a) of figure \ref{savindr} shows the results of  measurements
made using tokamaks  by \citet{Isle82} and \citet{Wang88b}, theoretical calculations by \citet{Jaco77}, 
\citet{Rosz87}, and \citet{Dasg94} and those adopted in the compilation of  
\citet{Mazz98} and the RR rate coefficients calculated by \citet{Arna92}.  This shows the dispersion
in rate coefficients obtained by various workers is a factor $\sim$10 at low temperatures, and 
a factor $\sim$3 at coronal temperatures, far greater than the estimated 
experimental uncertainties.
The right panel (b) of Figure \ref{savindr} shows the storage ring measurements of 
\citet{Savi02}, along with calculations using MCBP ({\sc autostructure}) and MCDF.  
\citet{Savi02b} measured the resonance strengths and energies for  
DR of Fe$^{19+} \rightarrow$ Fe$^{18+}$  $\Delta n$=0 core excitations. They have 
also calculated the DR resonance strengths and energies using  four different 
theoretical techniques: {\sc autostructure}, {\sc hullac}, 
{\sc mcdf}, and R-matrix methods. On average the theoretical resonance 
strengths agree to within $\leq 10\%$. However,  the 1 $\sigma$ scatter in the comparison is 30$\%$, 
so that calculations of individual emission lines due to DR will 
have a scatter of approximately this magnitude, although the ensemble
will be more accurate.

\citet{Savi03} measured DR resonance strengths and energies for Fe$^{20+} \rightarrow$ Fe$^{19+}$ 
for Fe$^{21+} \rightarrow$ Fe$^{20+}$ via $\Delta n=0$ core
excitations. They have also calculated these resonance strengths and energies using three 
techniques: multiconfiguration Breit-Pauli 
(MCBP) method using {\sc autostructure}, and
{\sc mcdf} and  {\sc fac}.  Although there is general 
agreement between experiment and theoretical calculations, 
discrepancies occur for collision energies $\leq$3 eV.   Nonetheless, the storage ring 
experiments and the careful comparison with various computational platforms provide
crucial benchmarks.  These make it clear which computational techniques and approximations 
are most reliable, and the level of accuracy which can be expected.  These techniques 
can then be applied on a large scale to provide DR cross sections and rate coefficients for 
astrophysical applications.

Other storage ring measurements include those of \citet{Dewi95} for He$^{+}$, \citet{Zong97} for Ar$^{15+}$, 
\citet{Mann98} for C$^{3+}$,  \citet{Glan99} for F$^{6+}$, \citet{Glan01} for N$^{4+}$, 
\citet{Bohm02, Bohm03} for O$^{5+}$, \citet{Bohm01} for Ne$^{7+}$,
\citet{Fogl03} for Ni$^{17+}$, and \citet{Niko04} for Na$^{8+}$; these are summarized by \citet{Glan04}.
Many of these include comparison with calculations which utilize relativistic many-body perturbation theory, 
and take into account QED effects in the position and strengths of resonances.
\citet{Kenn95} measured DR for Li-like Cl and Si and \citet{Fogl05} has measured DR for Be-like C, N and O
using similar apparatus.
Measurements of DR have been done using a single-pass merged-beam  
by \citet{Ande92, Ande92b} for 
He-like and Li-like ions of nitrogen, fluorine, and silicon.  These authors  point out the importance
of resonances associated with the $1s2s(^3S)$ metastable state in the He-like case.
They also point out the importance of intermediate coupling in determining the resonance 
structure and hence the DR rate coefficient in situations when the electron velocities
sample resonances close to the ionization threshold for the metastable state.  
The use of merged beams for DR and for electron collisional
excitation and for charge transfer has been reviewed by \citet{Phan99}, 
and the use of storage rings for measurements of recombination is reviewed by \citet{Schi99}.

Building on the experimental results of Savin and coworkers, large scale calculations of state-selected 
cross sections and rate coefficients for DR in intermediate coupling using 
the  {\sc autostructure} \citep{Badn03} code have been carried out by 
Badnell, Gorczyca, and coworkers.   These include calculations for the 
various  isoelectronic sequences H-Na 
\citep{Altu04, Altu05, Altu06, Badn06, Colg03, Colg04, Colg05, Mitn04, Zats03,Zats04,Zats04b,Zats05,Zats05b,Zats06}.
In a complementary effort \citet{Gu03b} 
has used {\sc fac} to calculate total DR rate coefficients for the H-like through Ne-like 
isoelectronic sequences for the 7 elements Mg, Si, S, Ar, Ca, Fe and Ni.  
Together these remove the shortcomings of
early calculations, namely that of $LS$ coupling and the choice of energetically allowed channels for 
applications at coronal temperatures.  For applications at temperatures characteristic 
of photoionized plasmas, large uncertainties remain  owing to the 
use of  {\em ab initio} energy level structure intrinsic in this work.
This has been shown by the storage ring measurements of \citet{Schi04}, who demonstrate 
the existence of strong resonances in the recombination cross section 
for Mg$^{8+}$ at electron energies 20-70 meV.  They point out that errors in the 
location of these resonances of only 100 meV can result in 
changes of a factor up to a factor $\sim$3 in the total recombination rate coefficient
at low temperatures.
However, with these calculations and their experimental validation,
the rate coefficients for DR at coronal temperatures for  many ions of astrophysical 
interest have reached a level accuracy previously unattained.  

One remaining area of uncertainty is the role of external fields on DR. 
According to \citet{Badn03},
this  renders pointless efforts to compute field-free rate coefficients to an accuracy of better than 20$\%$.   
It has long been known that the high Rydberg states that frequently 
dominate the DR process can be Stark-mixed by weak 
electric fields \citep{Burg69}, such as  the plasma microfield 
\citep{Jaco76}, and so increase the partial rate coefficients by factors 
of 2 or more over a wide range of $n$. Recently, the picture has been 
further complicated by the discovery that magnetic fields, when crossed with 
an electric field, strongly affect the electric field enhancement, by reducing 
it in most cases \citep{Robi97,Bart99}. 
This suppression of the electric field enhancement improves the 
applicability of field-free DR rates, but does not remove the uncertainty 
due to field effects in modeling of real plasmas.
The importance of field effects, and the field strengths themselves, can be 
derived through collisional radiative modeling of the dynamic part of the plasma 
microfield \citep{Badn03}.
Previously it appeared that a reasonable approach would be to use the values 
of the plasma microfield for the electric field strength for use in the generation of field dependent 
data as input to plasma modeling, but the added sensitivity to  the magnetic 
field makes this impractical.  

\subsubsection{Collisional Excitation} 

\noindent{\bf Background}

Collisional excitation or deexcitation, by electrons or protons, is 
closely associated with discrete diagnostics,
since it is the dominant mechanism affecting the level populations associated 
with many diagnostic features.  However, it also is key to the calculation of 
synthetic spectra and for cooling.   For spectrum synthesis, comprehensiveness 
is important in order to accurately calculate cooling rate coefficients and pseudo-continuum emission
due to large arrays of blended weak lines.  

In tabulating collisional excitation data it is customary to work with the 
collision strength \citep{Hebb40}, rather than the cross section.  The collision strength 
between a given pair of levels $\Omega_{ij}$ is defined such that 
$\Omega_{ij}=\Omega_{ji}$, and the excitation cross section  is given by

\begin{equation}
\sigma_{ij}={\frac{\pi a_0^2}{\omega_i k_i^2}}\Omega_{ij}
\end{equation}

\noindent where $\omega_i$ is the statistical  weight of 
level $i$.  The Maxwellian averaged collision strength is denoted $\Upsilon_{ij}$ 
and preserves this symmetry.  

Although, calculations are of greater practical importance to collisional excitation than experiments
owing to the large number of transitions which may be of interest, 
measurement of plasma spectra can be used to infer collision rates.  This makes use of assumptions 
about the excitation mechanism and with independent measurements of the gas density
and temperature.   Examples include the measurements of \citet{Datl76}, who deduced ionization rate 
coefficients for He-like B and C  from 
the time histories of the lines emitted by these ions ions in a theta-pinch 
plasma and compared  theoretical estimates using the semi-classical formula 
of Burgess with experimental results, corrected for the effects of metastables.   
\citet{John71}  determined  electron collision-excitation rate coefficients experimentally 
for $n$=2, 3, and 4 levels of  Be-like N$^{3+}$,  O$^{4+}$, Ne$^{6+}$, and Si$^{8+}$ 
using plasma  produced in a theta 
pinch device. \citet{Tond71}  measured excitation rate coefficients for eighteen transitions 
in Ne$^{7+}$.  A  review of methods and some results  was presented  by \citet{Kunz72}.

A general review of experimental techniques and results is given by \citet{Dunn95}.
The use of merged beams for  electron collisional
excitation and for charge transfer has been reviewed by \citet{Phan99}.
Recent measurements of $2s- 2p$ excitation in Li-like ions C$^{3+}$ and O$^{5+}$ 
have been carried out using merged beams by
\citet{Gree99} and \citet{Loza01}, and for S$^{3+}$ by \citet{Smit00}.
Measurements of selected transitions have provided crucial checks on calculations
\citep{Beie02} for spectra such as Fe$^{16+}$, and the EBIT apparatus has been 
used to benchmark cross sections and density dependent line ratios for
N$^{5+}$, Ar$^{13+}$ and Fe$^{21+}$ \citep{Chen04}.

An early review of computational techniques for 
collisional excitation is that of \citet{Bely66}.  These include
the Born approximation, in addition to other computational 
methods, which require the consideration of many partial waves and so are slow when 
carried out on older computers.  For this reason the Bethe approximation was 
widely used in early work.  This is based on the fact that at very high energy 
distant encounters are most important and the projectile remains outside the 
atom most of the time.  Then the cross section can be expressed 
as a proportionality with the oscillator strength of  the transition.  The 
proportionality is expressed as a Gaunt factor, and values of effective 
$\bar g$ were derived by \citet{Vanr62}.  
This allowed the use of radiative transition probabilities directly in 
modeling of collisional plasmas.  
This formula was used, for example, in calculation of structure and collision strengths 
for Fe$^{7+}$ by \citet{Czyz66} and compared with and results from Hartree-Fock with exchange 
self-consistent wavefunctions.  This approximation is no longer 
in widespread use, since more accurate methods can now be conveniently applied to 
large numbers of transitions.  

 Coulomb Born Oppenheimer (CBO)
calculations of  collisional excitation in H-like ions
were performed by \citet{Burg70} and \citet{Gold71}.
Screened hydrogenic calculations of collisional excitation for atoms with less than 4 electrons
were performed by \citet{Samp74}.
\citet{Gold81} calculated scaled
 collision strengths for hydrogenic ions using the CBO method, 
and \citet{Clar82} calculated scaled collision strengths for excitation of 
highly charged ions.

An early mention of the importance of CI for collisional excitation rate coefficients was by \citet{Layz51}.  
\citet{Jone70} extended the Eissner and Nussbaumer CI code to include relativity, and 
illustrated that drastic changes in $A$ values can occur depending on the treatment of CI.
\citet{Ermo72} explored the importance of CI in He-like ions and mixing between 
$^1P$ and $^3P$, especially for $Z>$2 using  both Breit-Pauli and $Z$-expansion techniques.
\citet{Nuss72} discussed the importance of CI for the ion C$^{2+}$.
\citet{Flow72} estimated collision strengths and showed that  Li-like lines are useful 
temperature diagnostics in solar corona if 2-2 and 2-3 lines can both be measured.
\citet{Bely73} and \citet{Flow73} performed CI calculations for  Fe$^{12+}$.
The pitfalls of CI calculations which have deficient sets of configurations 
and can lead to errors were examined by \citet{Nuss73}, with the ion Fe$^{12+}$ \citep{Flow73}  
used as an example. 
\citet{Loul73} demonstrated the effects of blending on the 17.06 \AA\ line in a 
CI calculation of the structure and excitation rate coefficients for Fe$^{16+}$.
%A mono-configurational Hartree-Fock formulation for electron 
%collisional excitation for ions in configurations 
%$1s^22s^22p^q$ and $1s^22s^22p^63s^23p^q$ is given by \citet{Sara69}.
\citet{Jone74} calculated collision strengths for He-like ions of Si, Ca, Fe  comparing  
the results of $LS$ coupling and intermediate coupling using a DW method.
The effects of relativity were examined by \citet{Walk74b}, who  
calculated electron impact  excitation of the $n~=~1$ and $n~=~2$ states of 
hydrogenic ions in the  Coulomb-Born approximation using Dirac wavefunctions.

A comparison of  distorted wave and close coupling calculations of collision 
strengths for transitions in C$^{2+}$ excited by electron collisions \citep{Flow72b} indicates 
that the distorted wave method is sufficiently accurate for many astrophysical 
and laboratory applications.
DW calculations for many ions have been carried out 
by Bhatia, Mason and coworkers, cited in Section \ref{discretediagnostics}. 
Calculations using similar techniques have been made for  S$^{10+}$ \citep{Land03}, Ca$^{6+}$ \citep{Land03b},
Fe$^{17+}$ \citep{Corn92},  Fe$^{20+}$ \citep{Phil96}, Ne$^{2+}$ \citep{Land05c}, Ca$^{13+}$ \citep{Land05f}, 
Ca$^{12+}$ \citep{Land05}, and Ar$^{11+}$ \citep{Eiss05}.
\citet{Fawc91} calculated collision strengths and oscillator strengths for Fe$^{8+}$. 
Collisional excitation rate coefficients for H-like and He-like ions calculated using a relativistic
DW method have been calculated by \citet{Samp83} and \citet{Zhan87}. 
This  has also been applied to the F \citep{Samp91}, 
Na \citep{Samp90}, B \citep{Samp86,Zhan94,Zhan94b}, Li \citep{Zhan90}, Be \citep{Zhan92}, and C \citep{Zhan96b}
isoelectronic sequences.
Collections of references to calculations of collisional excitation have been 
provided by \citet{Kato76}, \citet{Raym77}, \citet{Mewe81},  and  \citet{Mewe85, Mewe86}.
Reviews of calculations of available electron excitation cross sections for many ions of interest were presented as 
part of a 1994 conference on collisions by \citet{Foss94}, \citet{Corn94}, \citet{Maso94}, \citet{Prad94}, 
\citet{Duft94}, \citet{Bhat94}, \citet{Lang94}, \citet{Kato94}, \citet{Samp94}, 
\citet{Berr94}, \citet{Mcwh94}, \citet{Duba94}, and \citet{Call94}.
%from Lang 1994 ATNDT 57 1.
%The need for J resolved data, for target wavefunctions which include a sufficient number of configurations,
%enough levels so that collisional-radiative modeling can be dons, temperature range should include region 
%where abundance is expected to peak in equilibrium, when only presenting collisions strength data enough 
%energy points should be given to allow accurate thermal averaging, the need for treatment of resonances.

\noindent{\bf Recent Developments}

Close coupling calculations are available for an increasing fraction of 
transitions needed to model coronal plasmas.  
R-matrix calculations of collisional excitation have been carried out for hydrogenic
ions of C \citep{Agga91a}, Ne  \citep{Agga91b}, 
Si \citep{Agga92a}, Ca \citep{Agga92b}, Fe \citep{Agga93}.
R-matrix intermediate coupling calculations have been made for He-like and Li-like Ar and Fe by \citet{Whit01, Whit02}.
Calculations for Be-like ions have been reviewed by \citet{Berr94}.
The breakdown of the isolated resonance approximation and the 
importance of accurate treatment of radiation damping in calculations
of collisional excitation, in contrast with the situation for DR, was pointed out by \citet{Badn93b}.
The point is that two or more Rydberg series of resonances will interfere 
only if they overlap because coupling through the
background is weak, or absent. In contrast, for electron-impact excitation there is strong 
coupling through the background.  There is no dipole selection rule and 
consequently number of resonances increase greatly, so that 
close-coupling technique or a perturbative approach to overlapping 
resonances are needed.
 Such a situation is seldom met in photoionization and DR cross 
section because the number of accessible resonances is limited by 
dipole selection rules.

A comparison of relativistic DW and R-matrix calculations for Fe$^{21+}$ was carried out by \citet{Gu04b}.
In order to test for the importance of channel coupling and breadown of the isolated resonance approximation,
the two calculations were compared using the same energy level structure, and 
good agreement between the two methods was found for most transitions, for both resonance structure and 
for the background cross section.  This is in contrast to \citet{Badn93b}, who demonstrated factor $\sim$2 
errors in the dominant resonant contributions to the excitation of Mg-like ions.
Thus uncertainties introduced by the DW + isolated resonance approximation can be  significant.
R-matrix calculation of excitation of many X-ray transitions have been carried out by the Iron 
project \citep{Humm93}.   This work is continuing and 
the results are contained in the TIPTOPbase database \citep{Cunto92}.  
A sample of publications include those for  
Fe$^{20+}$ \citep{Butl00}, Fe$^{19+}$ \citep{Butl01a}, Fe$^{18+}$ \citep{Butl01b}, 
Fe$^{11+}$ \citep{Binn98a, Binn98b}, the  oxygen isosequence \citep{Butl94}, 
Ca$^{7+}$ \citep{Land04}, Fe$^{15+}$ \citep{Eiss99},
the Cl isosequence \citep{Pela95}, Fe$^{13+}$ fine structure \citep{Stor96},  the B isosequence
\citep{Zhan94c} and for Fe$^{2+}$ \citep{Zhan96}.
R-matrix calculations have been made for the fine structure levels of Fe$^{12+}$  and for Fe$^{10+}$
\citep{Gupt98,Gupt99}.
Most of these do not include levels in the 
$n=4$ manifold, but do account for relativistic effects 
using the Breit-Pauli Hamiltonian and intermediate coupling.  
Exceptions include the R-matrix calculations for He-like O by \citet{Dela02} and
He-like Ne by \citet{Baut03}.  
The mission of the Iron Project has been extended into the X-ray region by 
the UK RmaX Network (http://amdpp.phys.strath.ac.uk/UK$\_$RmaX). 

\begin{figure}
\includegraphics*[angle=270, scale=0.4]{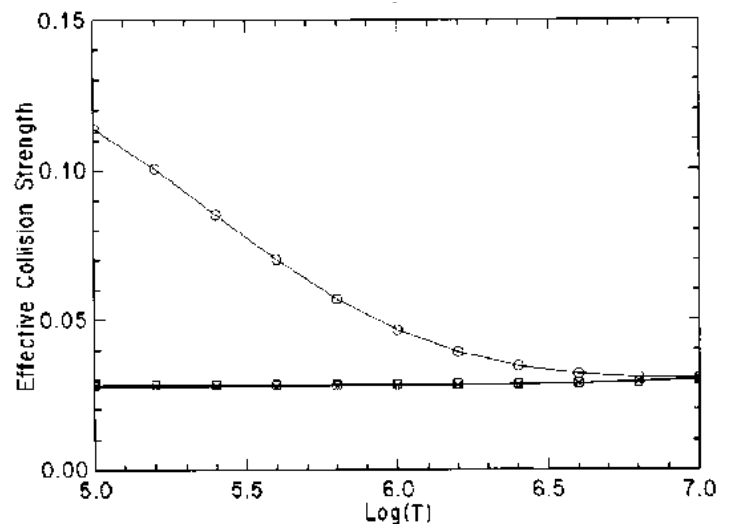}
\caption{\label{Eissfig} The effective 
collision strength for the $3s^2\,^1S0\rightarrow3s3p^3P^{\rm o}_1$ transition in Fe$^{14+}$
as a function of temperature for close-coupling (circles) and DW calculations (squares).  
from \citet{Eiss99}}
\end{figure}

\begin{figure}
\includegraphics*[angle=270, scale=0.4]{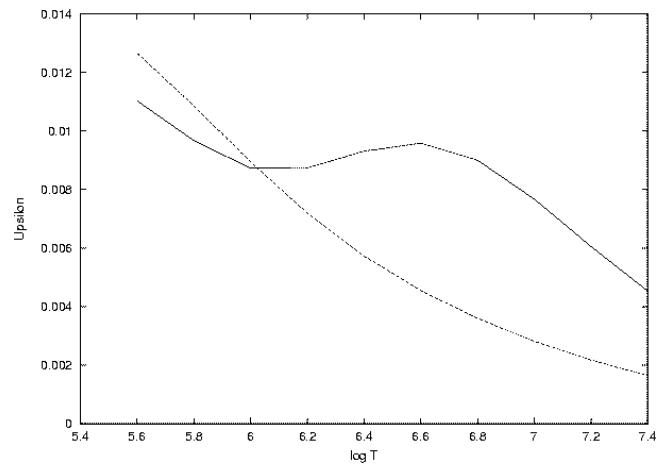}
\caption{\label{Butl33fig} Comparison
of the collision strength for the  $2s^22p^2 (^3P ^{\rm e}_0) \rightarrow 2s^2 2p^2 (^1D ^{\rm e}_2)$ transition in Fe$^{20+}$
calculated using BPRM (solid curve; \cite{Butl00}) with calculated using MCDF by \citet{Agga91c} (dashed)
as a function of temperature
(from \citet{Butl00})}
\end{figure}

Close-coupling calculations which 
include $n=4$ and intermediate coupling have been carried out for 
Fe$^{20+}$ by \citet{Badn01} and for Fe$^{21+}$ by \citet{Badn01b}, and show 
general consistency with the Iron Project results.
The difference between collision strengths calculated using the close-coupling approximation 
from those calculated using DW is illustrated in Figure \ref{Eissfig}, taken from \citet{Eiss99}.  
This shows the effective 
collision strength for the  $3s^2~^1S_0 \rightarrow 3s3p~^3P^o_1$ transition in Fe$^{14+}$
as a function of temperature for close-coupling (circles) and DW calculations (squares).  The
difference at low temperatures is due to the cumulative effect of resonances, which are 
not accounted for by DW method.  At the temperature corresponding to the peak abundance of this 
ion in coronal equilibrium, log(T)$\simeq$6.4, the two methods give very nearly the 
same result.
The role of CI in collisional calculations is illustrated in Figure \ref{Butl33fig}.  This shows comparison
of the effective collision strength for the 
$2s^22p^2 (^3P ^{\rm e}_0) \rightarrow 2s^2 2p^2 (^1D ^{\rm e}_2)$ transition in Fe$^{20+}$
as a function of temperature calculated using BPRM \citep{Butl00} with those of \citet{Agga91c} which 
used MCDF.  The latter, although it includes a more complete treatment of relativistic interactions,
omits the $n$=3 configurations and so under predicts the resonance structure at high temperature, where these 
resonances can be excited.

\begin{figure}
\includegraphics*[angle=0, scale=0.4]{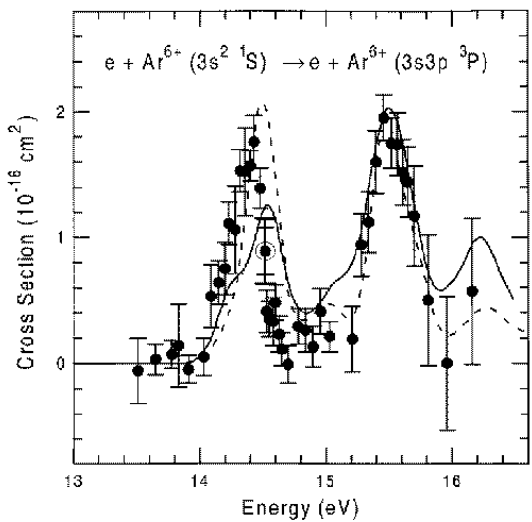}
\caption{\label{Smithfig} Comparison
of the collision strength for the  $3s^2~^1P_0 \rightarrow 3s3p~^3D$ transition in Ar$^{6+}$
calculated using an R matrix calculation (solid curve) \citep{Griff93}, an isolated-resonance DW calculate (dashed curve), and
experimental merged beam measurements \citep{Smit03}.
from \citet{Smit03}}
\end{figure}

Measured line ratios and absolute rate coefficient 
are important for validating calculations.  Absolute cross 
sections have been measured for electron impact excitation of  a 
few ions, for example C$^{3+}$ $2s(^2S_{1/2}) \rightarrow 2p(^2P_{1/2,3/2})$ 
for energies near threshold  by \citet{Savi95} (see also \citet{Janz99} for 
further work and references to other measurements), for Fe$^{16+}$ by
\citet{Brow01b}, and \citet{Beie02, Beie04}, and for Fe$^{20+}$ -- Fe$^{23+}$ by \citet{Chen05}.  
These serve as important tests for close-coupling 
calculations such as those of \citet{Chen02}.
An example is shown in figure \ref{Smithfig}, which shows
a comparison of the collision strength for the  $3s^2 (^1P_0) \rightarrow 3s3p (^3D)$ transition in Ar$^{6+}$
calculated using an R matrix calculation  \citep{Griff93}; an isolated-resonance DW calculations  
and experimental merged beam measurements \citep{Smit03}.  These 
illustrate the importance of resonances, which are more
prominent in the spin-forbidden transition shown here, and 
the difficulty in accurately calculating resonance strength.  An added 
uncertainty in experimental measurements comes from the fact that experiments such as those shown in 
Figure \ref{Smithfig} must detect the scattered electrons in order to discriminate 
between various excitation channels, and the angular scattering distribution must be 
understood in order to correct for the electrons which miss the detectors.  
Similar comparisons have been performed by \citet{Djur02}.

Bibliographies of the available data for collisional excitation were published by 
\citet{Itik84} and \citet{Itik91,Itik96,Itik02}, all designed primarily for 
fusion applications, and by \citet{Prad92}.  Reviews  organized by isosequence include
the aforementioned conference proceedings from 1994 (e.g. \citet{Berr94} etc.).
Much  recent work relevant to X-ray astronomy is contained in 
the {\sc chianti} database.  These include papers by \citet{Dere97b},
\citet{Land99}, \citet{Dere01}, \citet{Youn03}, 
  \citet{Land04} \citet{Land05b}, and \citet{Land05d}, and papers describing  
the Arcetri spectral code \citep{Land98, Land02}.

\subsubsection{Radiative Transition Probabilities} 

\noindent{\bf Background}

Traditional modeling of coronal plasmas assumed that radiative decay to the ground level is more 
rapid than excitation or decay to excited levels.  As a consequence, all dipole-allowed line emission 
and cooling could be calculated without explicit reference to the rate coefficients for 
radiative decay.  Recent plasma models, such as {\sc apec} \citep{Smit01}, calculate all level 
populations explicitly, and therefore require transition probabilities for 
transitions in all decay paths of excited levels.  

Radiative transition probabilities are conveniently parameterized in terms of 
the oscillator strength, which has a value of unity by definition for a classical 
point charge.  Non-relativistic quantum mechanical oscillator strengths for hydrogen
have been tabulated by \citet{Menz35}.  It is also customary to define the oscillator 
strength for hydrogen in terms of the semi-classical Kramers expression:

\begin{equation}
f_K(n,n')=\frac{32}{3 \pi \sqrt{3}}\left(\frac{1}{n'^2}-\frac{1}{n^2}\right)^{-3}
\left(\frac{1}{n^3 n'^5}\right) g_I(n\prime,n)
\end{equation}

\noindent along with a Gaunt factor $g_I(n\prime,n)$.  These are tabulated in,  e. g., \citet{Bake38}.

Classic tabulations of these include the 
H-like oscillator strengths of \citet{Wies66}, along with the compilations of 
NIST \citep{Fuhr99}. For hydrogen-like ions without relativity exact calculation is possible
\citep{Betsal72}.  Treatment of 
non-dipole allowed processes such as  2 photon decay  \citep{Shap59} 
and magnetic dipole rate coefficients \citep{Parp72} requires treatment of the effects of QED and nuclear size.

Examples of DW calculations include: transition probabilities for lines from 
Al-like ions \citep{Nuss77}, transition probabilities within 
the $2s^2 - 2s2p - 2p^2$ manifold in the Be isosequence \citep{Muet76}, 
radiative data for the Mg isosequence \citep{Chris86}, 
transition probabilities for ground configuration of S$^{3+}$ \citep{John86}, 
transition probabilities for transitions within the ground configuration of S$^{2+}$ and other Si-like ions 
\citep{Huan85},  and oscillator strengths for He-like allowed lines and collision 
strengths among fine structure levels \citep{Zhan87}.   It is 
worth noting that, although ground state fine structure transitions 
often fall outside of what is typically considered the X-ray band,
the structure and transition probabilities are relevant to the X-ray spectrum
from these ions seen in absorption, and they may also be useful for study of 
higher energy transitions which belong to the same isoelectronic sequence.
\citet{Mart93} calculated transition probabilities in the lithium sequence. 
DW calculations of oscillator strengths have been carried out for $2s-2p$ transitions 
Be, B, C, N, and O \citep{Fawc78}, and 
for  $\Delta n=$0, 1  transitions in the O,  S, P \citep{Fawc86, Fawc86b, Fawc86c}
and C \citep{Fawc87} isosequences.  

\citet{Cowa84} presented theoretical calculation of wavelengths and oscillator strengths for Fe$^{9+}$.
They pointed out the importance of spin-orbit and configuration-mixing effects in the 
calculation of wavelengths and oscillator strengths for heavy-element moderately charged ions.
This technique involves calculation of the energy level structure and line spectrum for a variety of ions
by collecting available experimental data from laboratory and solar measurements, adjusting in order to 
obtain smooth variation along isoelectronic sequence, and then using these levels to calculate the structure of 
model ions using a semi-empirical code \citep{Cowa81}.  This was done for  Fe$^{11+}$ and Fe$^{12+}$ $3p-3d$ and 
$3s-3p$ transitions \citep{Brom78}, Be-like and B-like iron  \citep{Brom78b}, the 
$2p^2-2p3d$ transition array of Fe$^{20+}$ and isoelectronic spectra \citep{Brom77},
the $2s^22p^n-2s^22p^{n-1}4d$ Fe$^{17+}$ and Fe$^{18+}$ lines \citep{Brom77b},
much of which was updated by \citet{Corn92},
the $2p^3-2p^23d$ transition array in Fe$^{19+}$ and isoelectronic ions \citep{Brom77c}, and the
$3s^23p^n-3s^23p^{n-1}3d$ transitions in Fe$^{9+}$ and Fe$^{10+}$ \citep{Brom77d}.
A compilation of collisional excitation and radiative decay rate coefficients for 
lines of $2s^22p^k$, $2s2p^{k+1}$, and $2p^{k+2}$ configurations, and for the 
O, N, C, B, and Be isoelectronic sequences of Ti, Cr, Fe, Ni  was produced by \citet{Feld82}.
A compilation of  transition probabilities for Cr, Fe, Ni ions in the 
B, C, N, O, and F isosequences was produced by \citet{Feld80}.

\noindent{\bf Recent Developments}

Close-coupling transition probabilities for many dipole-allowed transitions of interest have 
been calculated by the Opacity Project \citep{Seat87} and are contained in the 
the TOPbase database \citep{Cunto92}.  These use LS coupling and are tabulated 
with theoretical wavelengths, and therefore are not directly applicable to 
synthesis of spectra and do not include forbidden transitions.  
Their application to observations has been 
aided by the work of \citet{Fuhr99} who have matched these with 
experimental wavelengths where possible and redistributed the oscillator strength 
among fine structure levels.  Other CI calculations of transition probabilities 
include oscillator strengths of F-like ions for 18$< Z <$33 using CIV3 \citep{Blac94}. 
Iron Project calculations include intermediate coupling and 
relativistic effects in the Breit-Pauli approximation.  These include 
extensive work on forbidden transitions within the ground configuration 
of complex ions, whose wavelengths generally place them outside the scope 
of this review \citep{Gala97}. 
An annotated bibliography of transition probabilities for allowed and forbidden transitions with some 
overlap to the X-ray and EUV band  is  presented by \citet{Biem96}.  
\citet{Quin00} calculated  the wavelengths and oscillator strengths using 
{\sc hfr} for the $3p4s$, $3p4d$, $3p5s$ and $3p5d$ transitions in Fe$^{8+}$--Fe$^{13+}$ 
appearing in the soft X-ray region. 
Close coupling calculations of transition probabilities for fine structure transitions 
in S$^{3+}$ have been calculated by \citet{Taya99}. 
\citet{Stor00} presented  transition probabilities for transitions within the ground configuration of 
the carbon  and oxygen iso-electronic sequences. 
Opacity Project line strengths have been incorporated into the 
line list for use in absorption line studies by 
\citet{Vern96b}.  These used a compilation of experimental energy levels similar to that used by NIST 
and Opacity Project wavelengths derived using LS coupling rules 
for lines originating from ground-term multiplets  in the spectral region 
1--200 \AA\  .
Experimental measurements of radiative lifetimes for excited levels of 
many ions have been reviewed by \citet{Trab02}.  

\subsection{Photoionized Plasmas} 

In gases exposed to strong ionizing radiation, or in the absence of strong mechanical or non-radiative
heating, the ionization balance can be determined by the effects of photoionization and recombination.
In this case the gas temperature is determined by a balance between heating and cooling due to photon 
interactions.  Heating processes include slowing down of fast photoelectrons, and Compton scattering,
while cooling is due to emission processes analogous to those in a coronal plasma.
Early discussion of this was by \citet{Tart69} and \citet{Tart69b}.  
Traditional application for photoionization models was HII regions and planetary nebulae
\citep{Flow68,Flow83}, but it is now apparent that photoionization is dominant in many 
X-ray sources as well, such as active galaxies (e. g., Figures \ref{3783fig} and \ref{ngc1068}).

The condition for photoionization to dominate over collisions depends on the rate of mechanical heating, if 
any,  and on the ratio of the ionizing flux to the gas density, $n$, which is 
called the photoionization parameter. Various definitions are in use, those most widely 
quoted are $U=N/({\rm n}c)$ where $N$ is the number flux of ionizing photons (i.e., photons with energies 
greater than 1 Ry), and 
$\xi=4 \pi F/{\rm n}$, where $F$ is the energy flux of ionizing photons, typically 
in the 1-1000 Ry energy range.  

The choice of convention for ionization parameter definition is arbitrary.  Any 
accurate calculation of photoionization rate coefficients or heating  must take into account the 
shape of the ionizing spectrum in detail; the ionization parameter serves as a constant 
of proportionality for use in describing model results.  Certain choices
of ionization parameter are more directly related to the problem to be solved, which 
explains in part the origins of the differing conventions.  
Photoionization rate integrals can be written

\begin{equation}
{\rm rate}=\int_{\varepsilon_{th}}^\infty F_\varepsilon \sigma(\varepsilon) \frac{d \varepsilon}{\varepsilon}
\end{equation}

\noindent where $\varepsilon_{th}$ is the ionization threshold energy for the bound level in question,
$F_{\varepsilon}$ is the ionizing flux in units erg cm$^{-2}$ s$^{-1}$ erg$^{-1}$, 
and the  photoionization cross section is $\sigma(\varepsilon)$.  This quantity scales with energy above threshold as 
$\sigma(\varepsilon)\propto \varepsilon^{-\gamma}$ where $\gamma \sim$ 3.  So the photoionization rate and 
heating integrals are always dominated by the behavior at threshold if  
$d\log(F_\varepsilon)/d\log(\varepsilon) \leq \gamma$.
In classical photoionized nebulae, such as HII regions or planetary nebulae, the behavior of the model 
depends most sensitively on the ionizing photon flux at 1 Ry.  The total number of ionizing photons 
$N$ is also weighted toward the flux at 1 Ry unless $F_{\varepsilon}$ increases faster than $\propto\varepsilon^1$. 
However, photons at 1 Ry
have less influence on the ionization of ions typically observed in the X-ray band than do photons 
at energies greater than $\sim$0.25 keV.  Study of X-ray photoionized plasmas differs from classical 
photoionized nebulae in that it is often possible to directly observe the photons responsible for 
the ionization of gas responsible for line emission or absorption, which is motivation 
for  an ionization parameter which is more closely coupled to the continuum in the X-ray band than
is $N$.  This has led to the suggestion by \citet{Netz96} of the use of the parameter 
$U_X=N_X/({\rm n}c)$ where $N_X$ is the number flux of ionizing photons above 0.1 keV.  
The choice of $\xi$ is motivated by the fact that it is more heavily weighted toward 
the X-ray band by using ionizing energy flux rather than number flux.  Also, heating by Compton 
scattering is proportional to the photon energy flux.  $\xi$ has units implied
(typically erg s cm$^{-1}$), while $U$ is dimensionless.  
The study of photoionized gases in which the pressure is 
prescribed rather than the density has led to the further definition of an ionization parameter
$\Xi=F/(cP)$, where $P$ is the gas pressure \citep{Krol81}.  This quantity is 
dimensionless.
It is important to note that no matter what definition of ionization parameter is 
adopted, it serves only as a convenient scaling quantity.  Real calculations take into 
account the full frequency dependence of the radiation spectrum, and so do not 
depend ultimately on the convention which is used.

Since the ionizing radiation spectrum may not, in general,
be a pure thermal spectrum, photoionization is not conveniently parameterized by a 
rate coefficient depending on temperature. Rather, it is customary to consider the full
energy dependent cross section, and perform the calculation of the ionization rate
by integrating this over the photon flux distribution. Photoionization cross sections are also 
important for calculations of the inverse process, RR, 
via the Milne relation \citep{Oste06}.  In the case of RR,
cross sections to excited levels are important. In many cases of astrophysical interest
the equilibrium populations of excited levels are negligible, so that for the purpose of 
calculating the ionization rates, it is only necessary to consider photoionization from the 
ground level or term. For this reason many traditional calculations were applied only 
to the ground term.  Recent calculations have addressed rate coefficients for excited levels, 
which are increasingly being self-consistently included in level population calculations 
for most ions. An added feature of photoionization is that, owing to the non-thermal 
nature of the radiation field, ionization from inner electronic shells can be important 
for ions with 3 or more electrons.  In contrast, EII generally results in 
a smaller contribution from inner shell ionization.  A thermal electron energy distribution
has too few electrons at the high energies needed for inner shell ionization 
(at least 5 -- 10 times the valence threshold),  although
excitation-autoionization is important for many isoelectronic sequences.

\subsubsection{Photoionization} 

\noindent{\bf Background} 

Analytic expressions for the photoionization cross section of hydrogen 
can be obtained in the same way as for bound-bound transition oscillator 
strengths if the upper level is allowed to have an imaginary principle 
quantum number, $ik$.  The oscillator strength can then be written

\begin{equation}
f_{nk}=\frac{32}{3 \pi \sqrt{3}}\left(\frac{1}{n^2}+\frac{1}{k^2}\right)^{-3}
\left(\frac{1}{k^3 n^5}\right) g_{II}(n,k)
\end{equation}

\noindent where $g_{II}(n,k)$ is the bound-free Gaunt factor \citep{Menz35, Karz61}.  
The cross section involves the density of states, and if this is calculated 
in the limit of large $n$ then a reasonably accurate approximation is

\begin{equation}
\sigma(E)=(2.815 \times 10^{29}) \frac{g_{II}(n,k)}{n^5 \nu^3}
\end{equation}

\noindent 
This is a hydrogenic approximation and is independent of angular momentum. 
The Gaunt factor can take angular momentum into account;  the 
assymptotic dependence of the cross section on energy is $\propto\varepsilon^{-l-7/2}$
\citep{Fano68}. 
The exact analytic non-relativistic cross section for $l$=0,1 is given by \citet{Betsal72}.

A widely used simple parameterization of photoionization cross sections is that of 
\citet{Seat58}.   This has been fitted to  monoconfigurational 
Hartree-Fock results for the moderately ionized ions important to planetary nebulae 
by \citet{Henr68}, \citet{Henr70}, \citet{Chap71} and \citet{Chap72}.
Photoionization cross sections for He-like ions were  calculated  by \citet{Brow71}.
Photoionization cross sections using Hartree-Slater wavefunctions were 
calculated by \citet{Barf79}, who derived
a simple scaling behavior along isonuclear sequences which could be used to 
relate  ionic cross sections to those for neutrals.  \citet{Reil78, Reil79} 
performed the first large scale calculations of photoionization cross 
sections using Hartree-Slater wavefunctions, resulting in tabulations
of cross sections for all ions of the elements with $Z$ up to 30, including inner shells. 
Cross sections for excited levels of many ions based on a hydrogenic approximation
were calculated by \citet{Clar86}. 

The behavior of photoionization cross sections near the thresholds in multi-electron ions
is affected by resonance structure due to quasi-bound states.  Accurate models for these 
require calculations using the close-coupling approximation.  Near inner shell  edges 
these take the form of photo-excited core (PEC) resonances below the threshold, and the
net effect is to fill in the cross section below threshold and smooth the inner shell edge.

\noindent{\bf Recent Developments}

\begin{figure}
\includegraphics*[angle=270, scale=0.3]{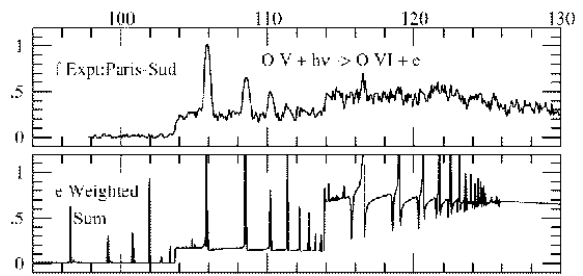}% Here is how to import EPS art
\caption{\label{naharfig} Comparison of BPRM calculation of 
the photoionization of O$^{4+}$ with experiment \citep{Cham03} in the energy range 90-130 eV.
Below 103 eV the cross section is due to ionization from the $2s2p$ excited configuration,
and the $2s^2$ above,  from \citet{Naha04}.} 
\end{figure}

A large scale compilation of close coupling calculations is included in the Opacity Project 
\citep{Seat87}, with results contained in the TOPbase, which is in turn part of TIPTOPbase \citep{Cunto92}  
http://cdsweb.u-strasbg.fr/topbase/home.html.  Also included as part of TIPTOPbase 
is the Iron Project database  and the OP server, which can calculate 
Rosseland mean opacities \citep{Miha78} for various elemental mixtures using the Opacity Project cross 
sections.  These are LS coupling calculations including channel couplings, 
carried out for ground states and excited states for all ions of elements 
with $Z$ up to 30. A widely used compilation  
was presented by \citet{Vern95}, who calculated central field cross sections 
including inner shells based on the Dirac-Slater potential, for most astrophysically 
important ions. These were scaled to  fit the behavior of the $R$-matrix calculations near threshold, 
while also having the correct behavior in 
the high energy asymptotic limit.  These are convenient to use, and provide an approximation 
to the resonance structure near threshold, and therefore can be used to calculate 
integral quantities such as photoionization rates.  However they do not contain the  
detailed resonance structure near threshold, and so should not be applied to calculations 
of monochromatic opacity or synthetic spectra.  
Opacity Project  opacities have been recently revised to include
inner-shell contributions, using the {\sc autostructure} package, and these 
are available through the Opacity Project on line database TIPTOPbase \citep{Badn05}.  
Figure \ref{naharfig} shows a comparison of BPRM calculations 
for O$^{4+}$ with experiment \citep{Cham03} in the vicinity of the threshold for ground state ionization,
showing the extensive resonance structure and general consistency between the measured 
and calculated cross section.

Even when the total ionization rate is so low that the gas is neutral, 
or if photoionization is not the dominant ionization mechanism, the spectrum of the 
radiation field transmitted in the X-ray band is affected by the photoionization 
cross section.   Thus a knowledge of the cross section is needed to interpret observations 
of X-ray absorption. In this case, spectroscopic accuracy can be important, since 
features in the cross section can be used to diagnose the conditions in the absorbing 
gas \citep{Paer01}. Examples of such features include the line features due to K shell photoexcitation
in oxygen and its ions, which is abundant in the interstellar medium.
The importance of this process has been emphasized by \citet{Prad00} and \citet{Prad03}.
An early detection of interstellar oxygen K absorption was by \citet{Scha86}.

\begin{figure}
\includegraphics*[angle=0, scale=0.4]{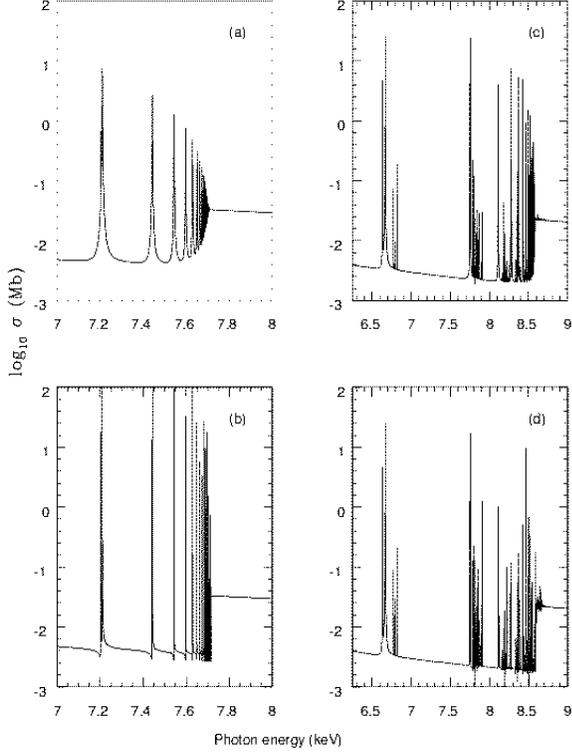}% Here is how to import EPS art
\caption{\label{bautistafig}High-energy total photoabsorption cross sections of the ground
level of:
(a) Fe$^{16+}$ including radiative and Auger damping effects;
(b) Fe$^{16+}$ excluding damping effects for resonances with $n>2$;
(c) Fe$^{22+}$ including damping;
(d) Fe$^{22+}$ excluding damping for resonances with $n>2$.
This demonstrates that when damping is included the resonance widths
are constant for high $n$ leading to a smearing of the K edge.  From \citet{Baut04}.} 
\end{figure}

The calculation of photoionization from inner shells is affected by 
the  resonance structure  associated with excitation of states with one $np$ excited
electron and a K shell vacancy.  They decay predominantly by spectator Auger 
transitions, in which the $np$ electron does not participate, and have
widths nearly independent of principle quantum number $n$.  This leads to a series of 
resonances with constant width close to the threshold for photoionization,
thereby smearing and lowering the location of the edge.  This effect, called Auger damping, has 
also been observed in the laboratory \citep{Farh97}.  
It was emphasized and pointed out earlier
in  calculations of inner shell ionization by \citet{Gorc00b} for oxygen (c. f.,   
\citet{Gorc99}), and Ne \citep{Gorc00}.  Computation of these effects 
using an $R$-matrix technique relies on the use of an optical potential to 
mimic the decays to spectator channels not explicitly included in the close 
coupling expansion.  {\sc BPRM}  photoionization cross sections including Auger damping 
have been carried out for inner shells of all ions of iron \citep{Baut03, Baut04, Kall04, Mend04, Palm02, 
Palm03, Palm03b},  and for oxygen \citep{Garc05}.  
An example of these effects is shown in figure \ref{bautistafig}, which shows the photoionization 
cross sections of Fe$^{16+}$ and Fe$^{22+}$ near the threshold for the  K shell, taken from \citet{Baut04}.
The left panels show the resonance structure with the effects of  damping by resonances with $n>2$ 
included, while the right panels show the resonance structure with damping by resonances with $n>2$  
excluded.

The vacancies created by inner shell photoionization
are filled by fluorescence or Auger decays.  The rates for 
these processes have been calculated by  \citet{Jaco86} using configuration-average energy levels 
and by \citet{Kaas93} based on 
isoelectronic scaling from the calculations of \citet{Mcgu69, Mcgu70, Mcgu71a, Mcgu71b, Mcgu72}.
The validity of these approximations has been discussed  by \citet{Gorc03} and \citet{Gorc06}.  Level resolved intermediate 
coupling calculations of K Auger and fluorescence rates have been calculated for iron 
by \citet{Baut03, Baut04}, \citet{Mend04}, and \citet{Palm02, Palm03, Palm03b}.   

\subsubsection{Radiative Recombination (RR)} 

Radiative recombination (RR) is  more important as a line emission mechanism in 
photoionized plasmas than in coronal plasmas and therefore receives more discussion in this context.
Rate coefficients can be calculated using detailed balance arguments
via the Milne relation \citep{Oste06}.  The rate coefficient can be expressed in the form defined by
\citet{Seat59} for hydrogenic ions:

\begin{eqnarray}
\alpha_n(Z,T)=&\frac{1}{c^2}\left(\frac{2}{\pi}\right)^{1/2} (mkT)^{-3/2} 2 n^2 e^{I_n/kT}
\nonumber\\ &\int_{I_n}^{\infty}(h\nu)^2 e^{-h\nu/kT}\sigma_n(Z,h\nu)d(h\nu)
\end{eqnarray}

\noindent where $I_n$ is the ionization potential and the photoionization cross section can be written

\begin{equation}
\sigma_n(Z,h\nu)=\frac{2^6\alpha\pi a_0^2}{3 \sqrt{3}}\frac{n}{Z^2}(1+n^2\epsilon)^{-3}g_{II}(n,\epsilon)
\end{equation}

\noindent where $\alpha$ is the fine structure constant, $a_0$ is the Bohr radius, and energy conservation 
requires that $h\nu=h R c Z^2(1/n^2+\epsilon)$.  The Kramers-Gaunt factor can be expanded as a polynomial
in $u=n^2 \epsilon$, and the rate coefficient can be expressed as:

\begin{equation}
\alpha_n(Z,T)=D Z \frac{\lambda^{1/2}}{n} x_n S_n(\lambda)
\end{equation}

\noindent where 
$D=\frac{2^6}{3}\left(\frac{\pi}{3}\right)^{1/2} \alpha^4 c a_0^2=5.197 \times 10^{14} {\rm cm}^3 ~{\rm s}^{-1}$
and $\lambda=hRcZ^2/(kT)=157890 Z^2/T$, $x_n=\lambda/n^2$,

\begin{equation}
S_n(\lambda)=\int_0^\infty \frac{g_{II}(n,\epsilon)e^{-x_n u}}{1+u}du
\end{equation}

\noindent and $u=n^2\epsilon$.  \citet{Seat59} uses the asymptotic expansion for $g_{II}(n,\epsilon)$ to 
evaluate $S_n(\lambda)$.  Numerical tables for $S_n(\lambda)$ and for the 
total recombination rate coefficients for 
$\alpha_\Sigma(Z,T)=\Sigma_{n=1}^{\infty}\alpha_n(Z,T)$
and for the spectrum
of recombining electrons have been published by \citet{Seat59}.  
When $R Z^2/n^2 \gg kT$ then $\alpha_n(1,T) \propto n^{-1} T^{-1/2}$
and when $R Z^2/n^2 \ll kT$ then, for $Z=1$, 

\begin{eqnarray}
\alpha_n(1,T) \propto &n^{-3} T^{-3/2} ({\rm ln}\left(\frac{n^2T}{157890}\right)-0.5772
\nonumber\\ & +8.56 \times 10^{-3} T^{1/3}-2.3 \times 10^{-5} T^{2/3} )
\end{eqnarray}

\noindent The first term in the brackets becomes less 
important at high temperatures, corresponding to the fact that  
excited levels contribute more at low temperatures than at 
high temperatures.  

Such calculations are typically tabulated as total recombination 
rates, in which all the possible radiative transitions from the lowest continuum state
(i.e., the ground state of the next highest ion stage)
to the bound states of a given ion are summed.  This requires 
a sum over all the rate coefficients into excited levels of an ion, and implicitly assumes 
that all recombinations to excited levels decay to the ground level.  Such sums traditionally 
use ground state photoionization cross sections appropriate to the ion, together 
with hydrogenic photoionization cross sections for the excited levels.
\citet{Tart71} calculated RR rate coefficients for many 
ions of astrophysical interest, as did  \citet{Goul78}. 
A widely used compilation is by \citet{Aldr73, Aldr76} who calculated RR and 
DR rate coefficients for many ions 
of astrophysical interest, calculated using the \citet{Burg65} general formula for DR
 and the Milne relation.  
\citet{Pequ91} calculated total and effective RR  coefficients for all ions with Z$\leq$10 
for important optical and UV transitions of these ions.
\citet{Wood81} calculated RR rate coefficients for iron, using the cross sections of \citet{Reil79}.
\citet{Prad83} calculated the RR for the ground states 
of Li-like ions. \citet{Vern96} calculated rate 
coefficients for RR based on Opacity Project photoionization cross sections. 
\citet{Gu03c} has provided total RR rate coefficients for the H-like through Ne-like 
isoelectronic sequences for the 7 elements Mg, Si, S, Ar, Ca, Fe and Ni, and \citet{Gu03d}
has provided X-ray line emission rate coefficients for Fe$^{17+}$ -- Fe$^{23+}$ due to both RR cascades and DR. 
The need for tabulations of total rate coefficients or of effective rate coefficients is reduced by advances in computer 
speed, which allow plasma modeling codes to calculate or make use of 
state-specific recombination rates.

\subsubsection{Dielectronic Recombination (DR)} 

Dielectronic recombination (DR) in photoionized plasmas is generally less important than 
in coronal plasmas, but it is not negligible.  As shown in figure \ref{savindr}, 
the contributions of radiative 
and DR are comparable for Fe$^{18+}$ DR at temperatures
characteristic of photoionized plasmas.  For Fe$^{19+}$ -- Fe$^{22+}$ DR dominates RR by a factor 
of 2 or more \citep{Savi02, Savi02b, Savi03}. Theoretical rate coefficients are less certain
at low temperatures, owing to the fact that the 
electron velocity distribution samples low energy states.  
Theory cannot calculate the resonance 
energies for the relevant DR resonances at low energy with sufficient precision.
\citet{Netz04} and \citet{Krae04} provide examples of the sensitivity of 
astrophysical results to DR rates in photoionized plasmas, and also point out 
the consequences of errors in the rate coefficients which are in widespread use
for astrophysical modeling.

The importance of low temperature DR was pointed out by \citet{Stor81}, 
who showed that the C$^{2+}$ $2p^2~^3D \rightarrow 2s2p~^1P_0 \lambda$2296 line 
observed from planetary nebulae is likely excited by DR.
This was extended to ions of C, N, and O by \citet{Nuss83}, who calculated 
rates for these ions using LS coupling. The associated satellite spectra 
were calculated by \citet{Nuss84}.  These were extended to Mg, Al, Si by \citet{Nuss86} 
and to Ne by \citet{Nuss87}.
Calculations of low temperature DR are most affected by 
the uncertainties in the low energy resonance structure, 
and this in turn 
is affected by the assumption of LS coupling in many calculations.
LS coupling can be applied when the fine structure splitting is small compared to the 
term separation which is caused by the Coulomb interaction. 

Recent experimental  measurements using storage rings \citep{Savi00,Savi02b}
and calculations using {\sc autostructure} \citep{Altu04, Altu05, Altu06, Badn06,
Colg03, Colg04, Colg05, Mitn04,
Zats03,Zats04,Zats04b,Zats05,Zats05b, Zats06} of DR include intermediate coupling 
structure calculations, and so are likely to be more accurate for low temperature DR
than previous calculations.  Nonetheless, as pointed out by \citet{Schi04} and by 
\citet{Gu03b}, DR at low temperatures 
continues to be a major source of uncertainty in calculations of ionization balance under 
photoionization conditions.

\subsubsection{Charge Transfer} 

\noindent{\bf Background}

Charge transfer is of potential importance in photoionized plasmas owing to the 
fact that the charge transfer rate coefficient for an ion with neutral H or He can exceed that 
for electron recombination by a factor $\geq 10^4$  \citep{King96}.  Photoionized plasmas can 
have a hydrogen neutral fraction as great as $\sim 0.1$ coexisting with 
several times ionized metals, depending on the shape 
of the ionizing spectrum.  
Another application for charge transfer rate coefficients was pointed out by \citet{Savi04}, who
calculated the rate coefficients for  H$^+$+H$_2$ $\rightarrow$ H+H$_2^+$ 
using recently published theoretical cross sections, and showed that uncertainties 
in these rate coefficients can affect the cooling and formation of primordial 
structures such as stars and galaxies.

The importance of charge transfer in the 
interstellar medium was pointed out by \citet{Fiel71} and \citet{Stei71} 
who also calculated rate coefficients for  C$^{2+}$ + He $\rightarrow$ C$^+$ + He$^+$  \citep{Stei75}
and discussed  the influence of charge transfer on the O$^+$ 
and N$^+$ ionization balance.
The potential importance of charge transfer and, for the first time, charge transfer
involving ions more than singly ionized on the 
structure of planetary nebula was pointed out  by \citet{Pequ78} and \citet{Pequ80}. 
This led to efforts to calculate cross sections and 
rates, including calculations of the charge transfer of N$^+$ with H, and of 
C$^+$ and S$^+$ with H and He using a 
distorted wave calculation  by \citet{Butl79, Butl80}.
\citet{Butl80b} calculated the charge transfer of a variety of multiply 
charged ions with H and He using a Landau-Zener calculation. 
\citet{Wats79} calculated 
charge transfer of C$^{3+}$ +H and N$^{3+}$+H using close-coupling approximation. 
A review and compilation by \citet{King96} provides a collection of 
rate coefficients in a form useful for modeling.

\noindent{\bf Recent Developments}

The most accurate calculations at the low energies of interest to astrophysics  ($\leq$ 1 eV/amu) 
are techniques similar to molecular calculations, using the molecular-orbital close-coupling (MOCC)
technique \citep{Heil85}.  
Recent MOCC calculations include  O$^{3+}$ + H  \citep{Wang03}, 
Si$^{4+}$ + He \citep{Stan97},   N$^{4+}$ + H \citep{Zyge97}, 
S$^{4+}$ + H \citep{Stan01},  and S$^{4+}$ + He \citet{Wang03}.
These show large ($\geq 10^2$) differences with Landau-Zener calculations in some cases.
Recent measurements include  Si$^{3+}$, Si$^{4+}$, Si$^{5+}$ + He \citep{Tawa01},
Ne$^{3+}$ + H \citep{Rejo04},  Ne$^{2+}$ + H \citep{Mroc03},
C$^{4+}$ + H \citep{Blie97},  N$^{2+}$ + H \citep{Piek97},
C$^+$ + H \citep{Stan98,Stan98b}. 

A review of the charge transfer process is given by \citet{Jane85}.
Recent reviews of charge transfer data for photoionized plasmas are given
in \citet{Stan01r} for theory and in \citet{Have01} for experiment.
Although quantum mechanical MOCC calculations provide the most reliable
results when no experimental data are available \citep{Stan01r}, still large
discrepancies between theory and experiment at collision energies of less than
a few hundreds eV/amu
point to the difficulty of theory at this energy range and the need for
benchmark measurements \citep{Have01}.

\subsection{Charge Transfer in Cometary and Planetary Atmospheres} 

The discovery of X-ray emission from comets by the ROSAT X-ray astronomy satellite
\citep{Liss96}
led to the appreciation of the importance of  charge transfer of solar 
wind nuclei with neutrals in comets and gaseous planets.  This process
has a distinct spectral signature associated with the cascade of the captured
electron from the excited state where capture initially occurs.
The most probable states are determined by the energetics of the 
collision and the level structure of the reactants.  

\citet{Habe97} extended this idea to look at individual spectral lines 
produced from C, O, and Ne ions. The excited state of the ion following 
capture was assumed to have a principal quantum number $n$ which is 
the nearest integer to the quantity $q^{0.75}$, 
where $q$ is the charge on the ion, followed 
by a cascading decay of $\Delta n = 1$.  This assumption leads to X-ray spectra
which  fits to the observed X-ray spectra in a subset of the cases.
A recent review of the physics of cometary X-ray emission was presented by \citet{Crav02}.  
EBIT simulations of cometary charge exchange were performed by \citet{Beie03}. 
Additional models for the cascade, level populations and X-ray line emission  associated with cometary
charge transfer have been calculated by \citet{Khar00}.  

Measurements for various species have been carried out using beam techniques.
Total cross sections for C, N and O are tabulated by \citet{Phan87}, 
\citet{Have89}, and \citet{Huq89}.  For interpreting X-ray spectra, 
and for testing theoretical calculations, state-selective measurements are 
needed.  This need, along with  the relatively low collision energy, 
make the X-ray astronomy data needs distinct from those for fusion plasmas.
Techniques for measuring state-selected charge transfer include energy 
loss spectroscopy and photon emission spectroscopy.  Energy 
loss spectropscopy provides greater counting rates,  while 
photon emission spectroscopy which allows higher energy resolution. 
Measurements have been made using energy loss spectroscopy of  
O$^{2+}$ in H, H$_2$ and He \citep{Mcla90}, C$^{4+}$ ions in collisions with 
H$_2$ and O$_2$ \citep{Mcla92},  
Fe$^{3+}$ and Fe$^{4+}$ ions with H and He atoms \citep{Mcla93},
 S$^{2+}$ with H and H$_2$ \citep{Wils90b}, and  S$^{3+}$ with  H, H$_2$ and He  \citep{Wils90}.
Using similar techniques, \citet{Kimu87} have studied highly 
stripped Ne, O, N and C in collisions with H and H$_2$, and
\citet{Kamb96} have studied N$^{3+}$ colliding with H$_2$, He, Ne, and Ar.
Beam measurements have been made using photon emission spectroscopy 
to obtain state selective cross sections of single electron charge 
transfer of   C$^{4+}$ on H and H$_2$ \citep{Hoek90}, O$^{3+}$ on H and H$_2$  \citep{Beij96}, 
C$^{4+}$, N$^{5+}$, O$^{6+}$ with H, H$_2$ and He \citep{Dijk85a}, C$ ^{6+}$, N$^{6+}$, O$^{6+}$, 
and Ne$^{6+}$ onto H$_2$, He, Ar \citep{Dijk85b}.
Similar techniques have been used to study  C$^{5+}$ and N$^{6+}$ with He and H$_2$ 
\citep{Sura91}, and  C$^{5+}$ and N$^{6+}$ with He and H$_2$ \citep{Ciri85}.
Double charge exchange can be an important loss process for highly charged ions 
incident on neutrals, but in many cases it leads to the production of autoionizing 
states, and therefore does not contribute to X-ray emission.
Cross sections with molecular targets other than H$_2$ have been measured 
for ions of C, N, O, and Ne colliding with He, H$_2$, CO$_2$, and H$_2$O 
by \citet{Gree01}.   Some of these experimental 
data have been collected at the ORNL/UGA charge transfer database website,
http://cfadc.phy.ornl.gov/astro/ps/data/home.html, including both total 
and state-specific cross sections and rates, along with fits to many of these.

Computational techniques for charge exchange collisions 
include the classical trajectory Monte-Carlo technique \citep{Corn00},
in which the motion of the projectile, target and electron are calculated 
by integrating Hamilton's equations including only the Coulomb interactions.
This technique is of limited use at energies below 1 keV/amu, where the structure 
of the projectile ion is important.  Another method is the
semi-classical impact parameter close coupling method, in which the 
wavefunction of the electrons is calculated from a  set of orbitals
appropriate to the projectile ion.  This technique is most 
suitable to systems with a small number of electrons 
\citep{Saha95, Kuma98},  and so has not been applied to the study of 
many ions.  The Landau -Zener model has also been widely applied, and 
involves the use of avoided level crossings in order to 
determine the cross sections for various processes.  
It turns out that especially for energies below 200 eV/amu there 
are considerable differences between experiment and theory both in relative 
and state-selective cross sections.

\section{\label{discussion}Discussion}

Basic needs and physical processes important to atomic data for X-ray astronomy 
have been known since the advent of solar X-ray astronomy.  This has led to steady 
production of rate coefficients and cross sections for use in modeling and interpreting 
observed astrophysical X-ray spectra, and with considerable overlap with the 
needs for fusion plasmas.  Prior to the launch of $Chandra$ and 
$XMM-Newton$ a great deal of work had been done toward the goal of developing an atomic 
database for coronal plasmas.  Much of this was devoted to the study of discrete diagnostics,
since the spatial and temporal structure of the X-ray emission limits the accuracy of 
global modeling for the Sun.

The launch of instruments capable of observing spectra from extrasolar objects 
with comparable resolution and good counting statistics has added  
motivation for data production and it has changed the emphasis 
somewhat from that of studies of solar X-rays.  
Extrasolar objects often cannot be spatially resolved and the 
sensitivity to temporal variability is limited.  Arguments based on 
other knowledge of their properties motivates attempts to construct global models 
based on a single set of physical conditions, such as the assumption of a single temperature
or a cooling flow for the coronal emission from a cluster of galaxies.  For some objects a simple distribution of 
conditions is assumed, such as a differential emission measure analysis performed for 
a cool star.  In addition, the spectra of distant objects often do not have
sufficient statistical accuracy to allow the application of detailed discrete diagnostics.  
This creates a greater need for global modeling, in which the ionization balance
and spectrum can be calculated based on simple assumptions about the conditions, such 
as temperature and density.  In addition extrasolar data have broadened the range of 
physical processes of interest to include charge exchange, photoionized plasmas, 
inner shell processes, and opacities associated with interstellar gas.  

Significant advances in the calculation and measurement of atomic cross sections and rate coefficients needed 
for X-ray astronomy have occurred in parallel with the launch of the new observatories.
Notable among these are the energetic application of experimental apparatus such as 
the EBIT and storage rings, and the improvements in computer technology and campaigns 
to calculate large quantities of data such as the Iron Project and Opacity Project.
At the same time, long term efforts to compile reliable data from more traditional 
laboratory sources have continued to yield results, and the advent of free on line databases,
such as the NIST database, has aided in their use.   Other important databases 
are those developed primarily for the fusion energy program at ORNL (http://www-cfadc.phy.ornl.gov/) 
and ALADDIN and AMBDIS at the IAEA (http://www-amdis.iaea.org/).
As a result, there now exist accurate 
experimental datasets for line wavelengths and  cross sections or rate coefficients for 
some key processes, along with computations using the most accurate known algorithms 
for many quantities.   Laboratory measurements have been made for DR
of many ions, and  comparison with calculations allows benchmarking of computational techniques.
Comprehensive calculations of state-specific DR, which have been 
benchmarked against the measurements, are available for many ions of interest.
Measurements of EII exist for most ions of astrophysical interest, 
although verification is needed for many of these.  Calculations, including 
close-coupling calculations with adequate treatment of intermediate coupling, 
relativistic effects, CI and  radiation damping, 
have been made for radiative transition probabilities and electron impact 
collision strengths of many ions of interest.   Measurements of absolute electron 
impact excitation cross sections 
have proven to be crucial in benchmarking these calculations. 
Beam measurements have yielded state-selective 
cross sections for charge exchange for many ions of interest to solar system X-ray 
studies.  

In addition to the databases discussed above, a great deal of useful data has been collected and 
made publicly available as part of the databases associated with analysis packages such as {\sc chianti}
\citep{Land05}, {\sc apec} \citep{Smit01}, and {\sc mekal/spex} \citep{Kaas96}.  It is important 
to point out that the ultimate source of  all databases and compilations is extensive 
computational and experimental work, and that the credit for this work is often neglected 
when the compilation or database is used.  Astrophysicists and modelers should, whenever possible, 
attempt to cite original sources even when using the compilation or database as a guide or 
repository for atomic data.

As a result of accuracy demanded by the new instruments for X-ray astronomy 
some of the atomic calculations which were in widespread use for 
modeling astrophysical plasmas are no longer adequate for application to many observations.
The approximations necessitated by early computers or analytic work provide
valuable insight and constraints on the more recent work, and therefore have been 
crucial in the development of modern tools.  However, there are few remaining 
processes or ions for which rate coefficients calculated using simple approximations are 
all that is available.  These include the use of the Born approximation, or CBO, 
in collisional ionization cross sections, the Bethe approximation for 
collisional excitation, {\em ab initio} wavelengths for strong lines 
in the X-ray band, the Burgess general formula  for DR, 
hydrogenic or central field cross sections for photoionization or radiative 
recombination, and Landau-Zener rate coefficients for charge exchange.  An added consequence of 
the capabilities of new computers is that it is no longer necessary to use total 
recombination rates when calculating ionization balance.  Rather, collisional-radiative 
models (eg. \cite{Summ06}) can be quickly calculated which take into account state-specific recombination
and ionization rates, and so are applicable to a wide range of gas densities 
and radiation environments. 

Areas where there are still critical needs include the accumulation of energy level 
structures and transition wavelengths which are of sufficient accuracy.  These are needed 
for applications including line identification, where the observations have an accuracy of 
10$^{-3}$ in many cases, and for calculations of DR 
at low temperature.  Experiments have the most promise for useful work in this area, but 
new theoretical techniques are needed for this challenge as well.   
There are few experimental measurements of inner shell photoabsorption
for ions of interest to astronomy.  Similarly, many lines in observed spectra such as that 
of NGC 3783 remain unidentified, and these may be associated with inner shell transitions 
not typically observed from coronal plasmas.  X-ray processes involving molecules, including 
detailed spectra associated with charge transfer, and inner shell opacities, have not 
been studied for many species.  Experimental work is needed to continue the campaign to
measure all the DR and CI rate coefficients needed to benchmark the calculation of coronal ionization
balance in both electron ionization and photoionization driven
plasmas, and in cometary and planetary atmospheres.

The most probable choices for future X-ray instrumentation will likely be 
only a partial continuation of the trends of the recent past.
Both technological challenges and astronomical priorities suggest that 
X-ray astronomy satellites following $Chandra$ and $XMM-Newton$ will 
have significantly greater sensitivity, so that the number of objects which 
can be observed spectroscopically, the statistical quality of the spectra
and the ability to study time variability will all be greatly enhanced.
However, it is less likely that either the spectral resolution or the 
spectral bandpass will be improved in such instruments.  Thus, spectral 
resolution comparable to that available from the best 
optical or UV instruments will not be attained.
In view of this it seems likely that the interpretation of data from such future 
missions will continue to rely on tools similar to those currently in use, 
tools which attempt to calculate ionization, excitation 
and synthesize the spectrum over a wide range of wavelengths.  
New targets for these observations will likely include 
galaxies, clusters of galaxies, and intergalactic gas 
in order to study the formation of 
structure, nucleosynthesis, and cosmological parameters.   More sensitive instruments 
will also study spectra of fainter nearby  objects such as stars and 
supernova remnants, and allow improved statistics and more detailed study of 
time variability of the brightest objects.  This is in contrast to 
what might be expected if the trend toward both greater spectral 
resolution and greater sensitivity were to occur, which might lead 
to more extensive application of discrete diagnostics and a reduced reliance on
global modeling.  If so, the atomic data needed for the foreseeable future
will not be greatly altered, in the sense that the precision required will 
be comparable to the best currently available, and physical processes of interest
will span those covered in this review.  Spectra with improved statistical 
accuracy, or time resolution, are likely to reveal physical effects which have not 
been incorporated into the available atomic database.  These may include departures from 
ionization equilibrium, non-stationary processes, optical depth effects, the effects 
of magnetic fields, and interactions with energetic particles and 
non-Maxwellian electron energy distributions.  This implies the need for 
cross sections at energies both significantly above and below the traditional range 
of energies prescribed by coronal equilibrium.  In addition, it is likely that X-ray 
astronomers will further explore molecular, or solid matter in astrophysics using charge transfer
and also absorption features near edges or inner shell lines.  

\begin{acknowledgments}
We thank Manuel Bautista, Nigel Badnell, Greg Madejski, Alfred M\"uller, 
Daniel Savin for careful reading many valuable comments and suggestions.
We also thank Prof. Claude Canizares and the Center for Space Research for access to the 
MIT libraries.  This work was supported in part by a grant from NASA though the Astrophysics
Theory program. 
\end{acknowledgments}

\begingroup
\squeezetable
\begin{table}
\caption{\label{Glossary}Glossary}
\begin{ruledtabular}
\begin{tabular}{ccc}
Acronym&Meaning&Section\\
\hline
AMA&Angular Momentum Average&\ref{disec}\\
APEC&Atomic Physics Emission Code&\ref{coronal}\\
BP&Breit-Pauli&\ref{theory}\\
BPRM&Breit-Pauli R-Matrix Code Package&\ref{theory}\\
CB&Coulomb Born Approximation&\ref{theory}\\
CBE&Coulomb Born with Exchange&\ref{theory}\\
CBO&Coulomb Born Oppenheimer &\ref{theory}\\
CCD&Charge Coupled Device&\ref{need}\\
CI&Configuration Interaction&\ref{theory}\\
DI&Direct Ionization&\ref{eiion}\\
DR&Dielectronic Recombination&\ref{disec}\\
DW&Distorted Wave Aprroximation&\ref{theory}\\
EA&Excitation-Autoionization&\ref{disec}\\
EBIS&Electron Beam Ion Source&\ref{exp}\\
EBIT&Electron Beam Ion Trap&\ref{exp}\\
ECIP&Exchange Classical Impact Parameter&\ref{eiion}\\
EII&Electron Impact Ionization&\ref{eiion}\\
FAC&Flexible Atomic Code&\ref{theory}\\
GF&Burgess General Formula&\ref{disec}\\
HETG&High Energy Transmission Grating&\ref{need}\\
HFR&Hartree-Fock Relativistic Code&\ref{theory}\\
HRS&High Rydberg States&\ref{disec}\\
HULLAC&Hebrew University LLNL Code&\ref{theory}\\
HXR&Hartree-Fock Exchange Potential\ref{theory}\\
LLNL&Lawrence Livermore Nat'l Lab&\ref{theory}\\
LTE&Local Thermodynamic Equilibrium&\ref{theory}\\
MBPT&Many Body Perturbation Theory&\ref{theory}\\
MC&Multi Configuration&\ref{theory}\\
MCBP&Multi Configuration Breit Pauli&\ref{theory}\\
MCDF&Multi Configuration Dirac Fock&\ref{theory}\\
MCHF&Multi Configuration Hartree Fock&\ref{theory}\\
ORNL&Oak Ridge National Laboratory&\ref{theory}\\
QED&Quantum Electrodynamic&\ref{theory}\\
READI&Resonant Excitation Auto-Double-ionization&\ref{eiion}\\
REDA&Resonant Excitation Double Autoionization&\ref{eiion}\\
RGS&Reflection Grating Spectrograph&\ref{need}\\
RTE&Resonant Transfer of Excitation&\ref{disec}\\
SCF&Self Consistent Field&\ref{theory}\\
\end{tabular}
\end{ruledtabular}

\end{table}
\endgroup

\end{document}